\documentclass[twocolumn, switch]{article} 

\usepackage{preprint}

\usepackage{enumitem}
\usepackage{amsmath, amsthm, amssymb, amsfonts}
\usepackage{bm}
\usepackage{amsmath}
\usepackage{graphicx}
\usepackage[algoruled,algo2e]{algorithm2e}
\usepackage{subfig}

\usepackage{setspace}
\usepackage{amssymb}
\usepackage{booktabs}
\usepackage{array}
\usepackage{color}
\usepackage{multirow}
\usepackage{multicol}
\usepackage{threeparttable}
\usepackage{url}
\usepackage[page]{appendix}

\usepackage[numbers,square]{natbib}
\usepackage[utf8]{inputenc}	
\usepackage[T1]{fontenc}	
\usepackage{xcolor}		
\usepackage[colorlinks = true,
            linkcolor = blue,
            urlcolor  = blue,
            citecolor = blue,
            anchorcolor = black]{hyperref}	
\usepackage{booktabs} 		
\usepackage{nicefrac}		
\usepackage{microtype}		
\usepackage{lineno}		
\usepackage{float}			

\usepackage{newfloat}
\DeclareFloatingEnvironment[name={Supplementary Figure}]{suppfigure}
\usepackage{sidecap}
\sidecaptionvpos{figure}{c}

\usepackage{titlesec}
\titlespacing\section{0pt}{12pt plus 3pt minus 3pt}{1pt plus 1pt minus 1pt}
\titlespacing\subsection{0pt}{10pt plus 3pt minus 3pt}{1pt plus 1pt minus 1pt}
\titlespacing\subsubsection{0pt}{8pt plus 3pt minus 3pt}{1pt plus 1pt minus 1pt}

\usepackage{graphics}
\usepackage{hyperref}
\usepackage{color}
\usepackage{multirow}
\usepackage{hhline}
\usepackage{lipsum}

\usepackage{fancyhdr} 
\title{JPEG Quantized Coefficient Recovery \\via DCT Domain Spatial-Frequential Transformer}

\usepackage{authblk}

\author[1]{Mingyu Ouyang}
\author[1]{Zhenzhong Chen}
\affil[1]{School of Remote Sensing and Information Engineering, Wuhan University}

\begin{document}

\twocolumn[ 
  \begin{@twocolumnfalse} 
  
\maketitle

\begin{abstract}

JPEG compression adopts the quantization of Discrete Cosine Transform (DCT) coefficients for effective bit-rate reduction, whilst the quantization could lead to a significant loss of important image details. Recovering compressed JPEG images in the frequency domain has recently garnered increasing interest, complementing the multitude of restoration techniques established in the pixel domain. However, existing DCT domain methods typically suffer from limited effectiveness in handling a wide range of compression quality factors or fall short in recovering sparse quantized coefficients and the components across different colorspaces. To address these challenges, we propose a DCT domain spatial-frequential Transformer, namely DCTransformer, for JPEG quantized coefficient recovery. Specifically, a dual-branch architecture is designed to capture both spatial and frequential correlations within the collocated DCT coefficients. Moreover, we incorporate the operation of quantization matrix embedding, which effectively allows our single model to handle a wide range of quality factors, and a luminance-chrominance alignment head that produces a unified feature map to align different-sized luminance and chrominance components. Our proposed DCTransformer outperforms the current state-of-the-art JPEG artifact removal techniques, as demonstrated by our extensive experiments.

\end{abstract}

\vspace{0.4cm}

  \end{@twocolumnfalse} 
] 

\newcommand\blfootnote[1]{%
\begingroup
\renewcommand\thefootnote{}\footnote{#1}%
\addtocounter{footnote}{-1}%
\endgroup
}

{\blfootnote{Corresponding author: Zhenzhong Chen, E-mail:zzchen@ieee.org}}

\section{INTRODUCTION}
\begin{figure}[!t]
  \centering
  \subfloat[Lossless DCT Coefficients]{
    \includegraphics[width=0.233\textwidth]{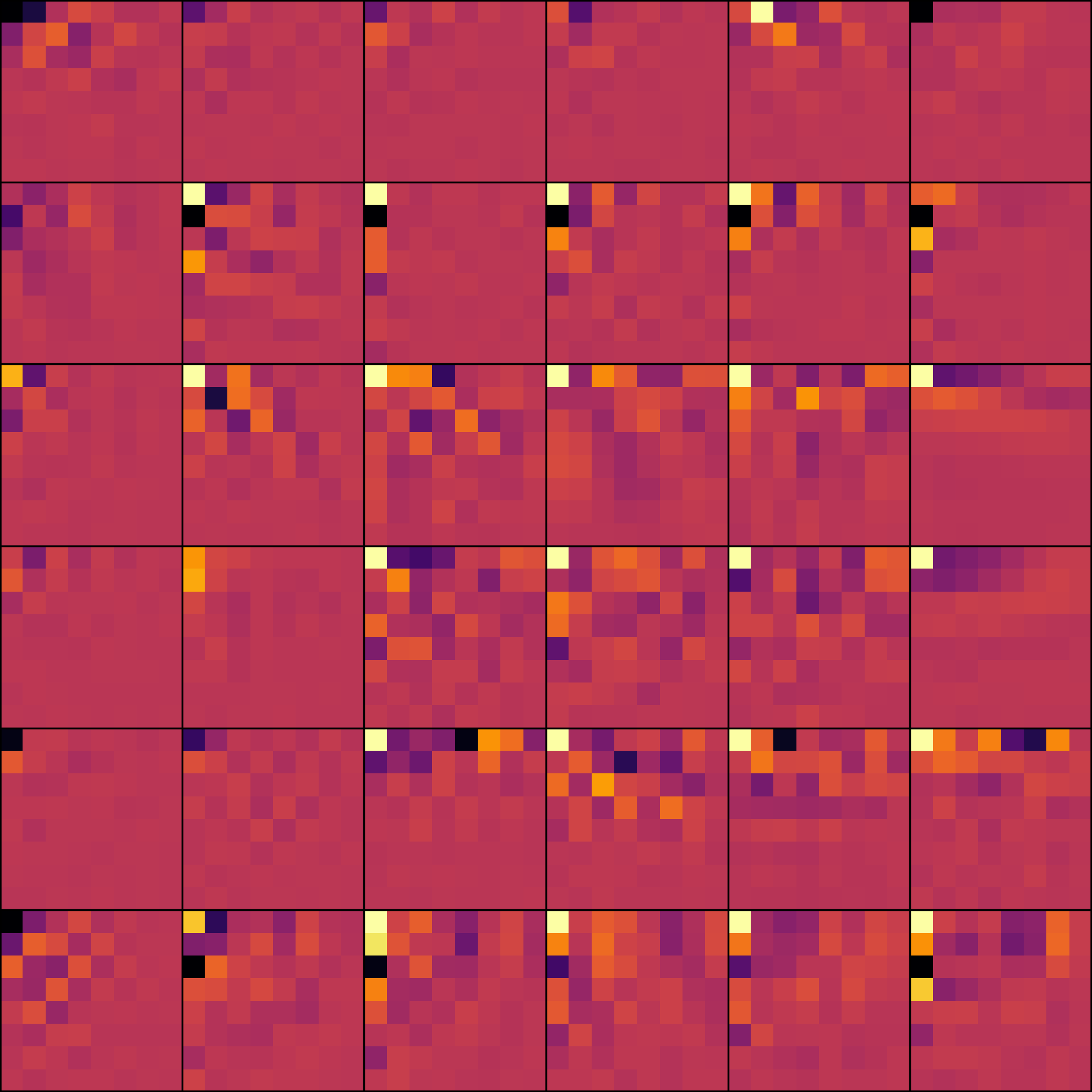}%
    \label{fig1:subfig_a}
  }
 \hspace{-0.1cm}
  \subfloat[Original Image]{
    \includegraphics[width=0.233\textwidth]{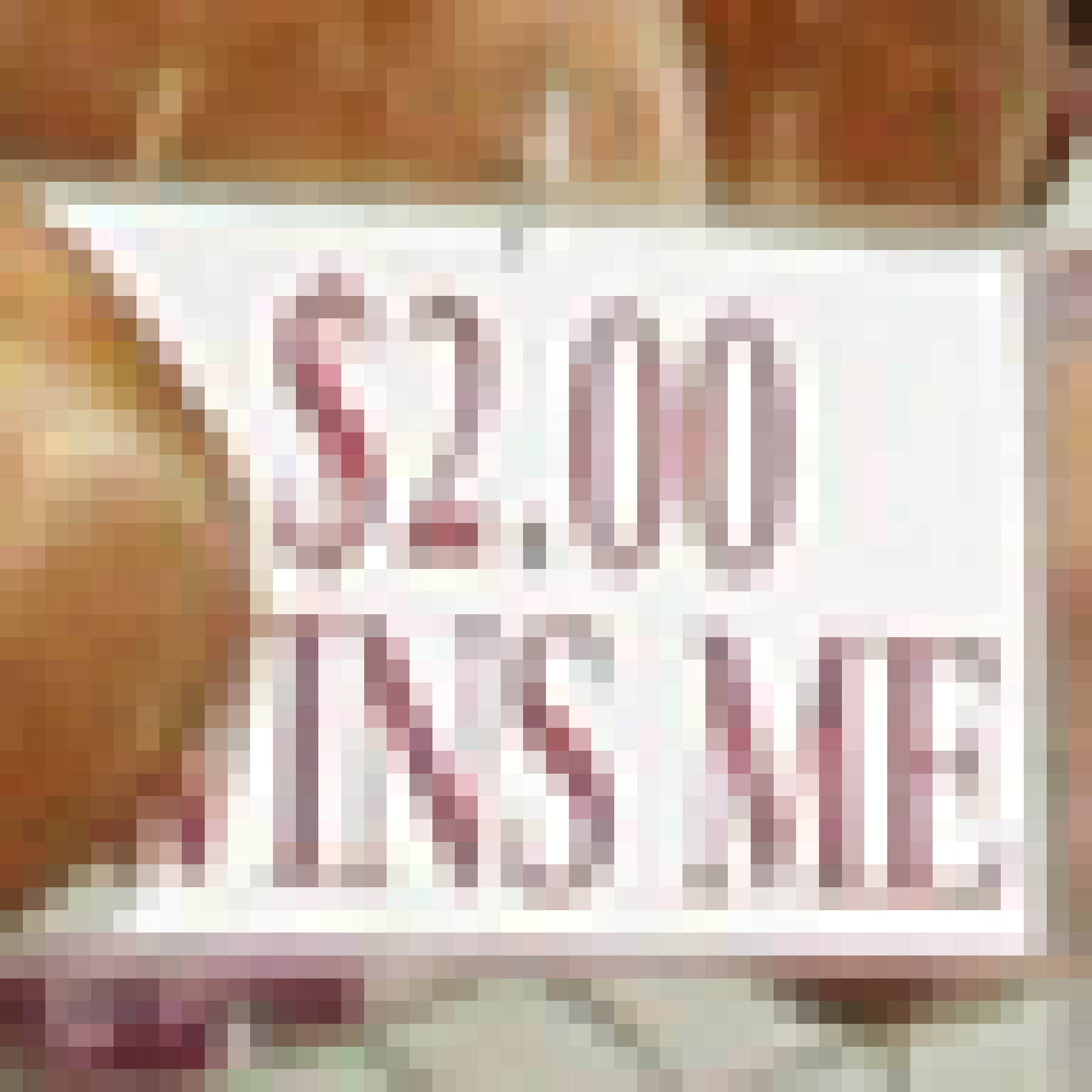}%
    \label{fig1:subfig_b}
  }
\vspace{-7.5pt}
  \subfloat[Quantized DCT Coefficients]{
    \includegraphics[width=0.233\textwidth]{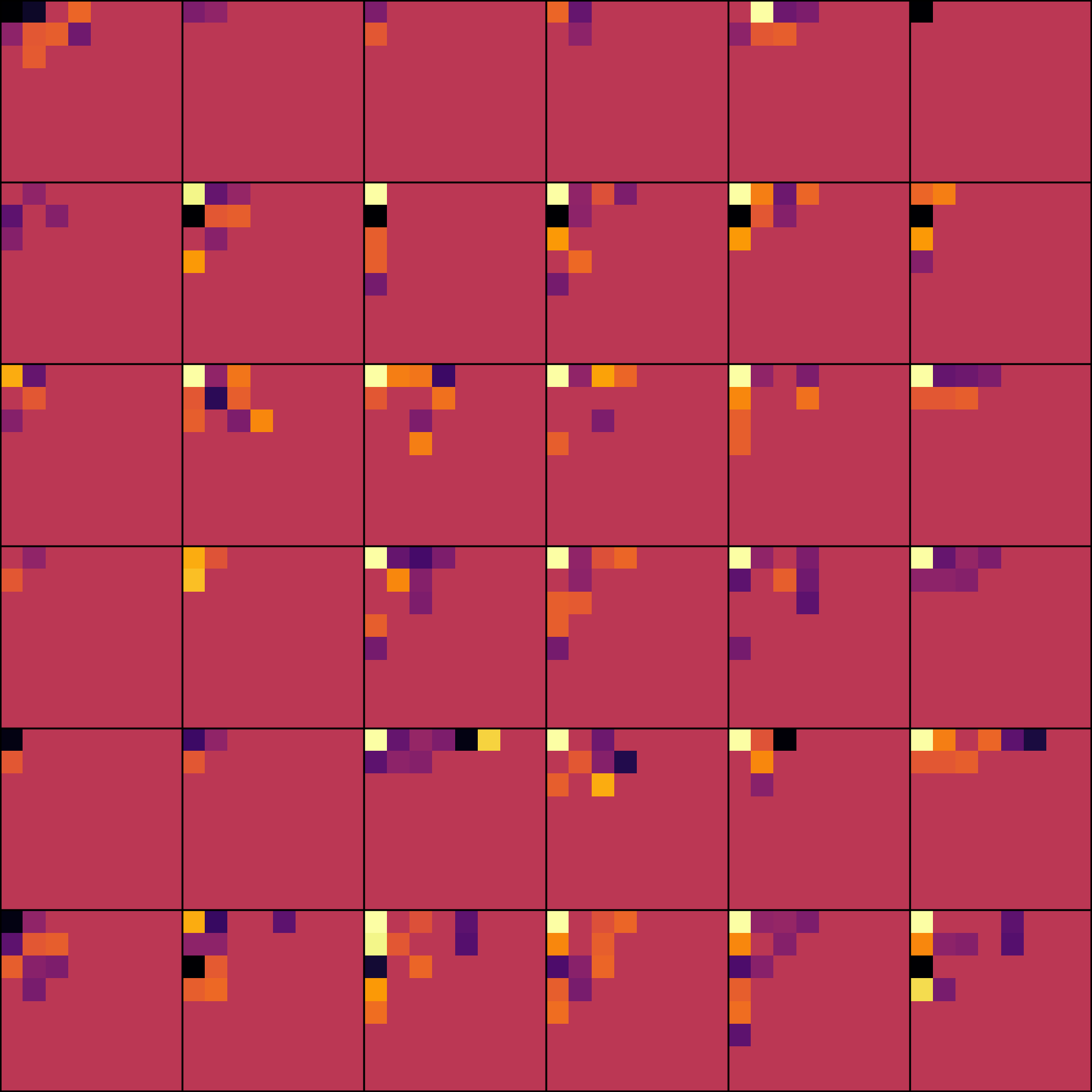}%
    \label{fig1:subfig_c}
  }
 \hspace{-0.1cm}
    \subfloat[QF=10 JPEG]{
    \includegraphics[width=0.233\textwidth]{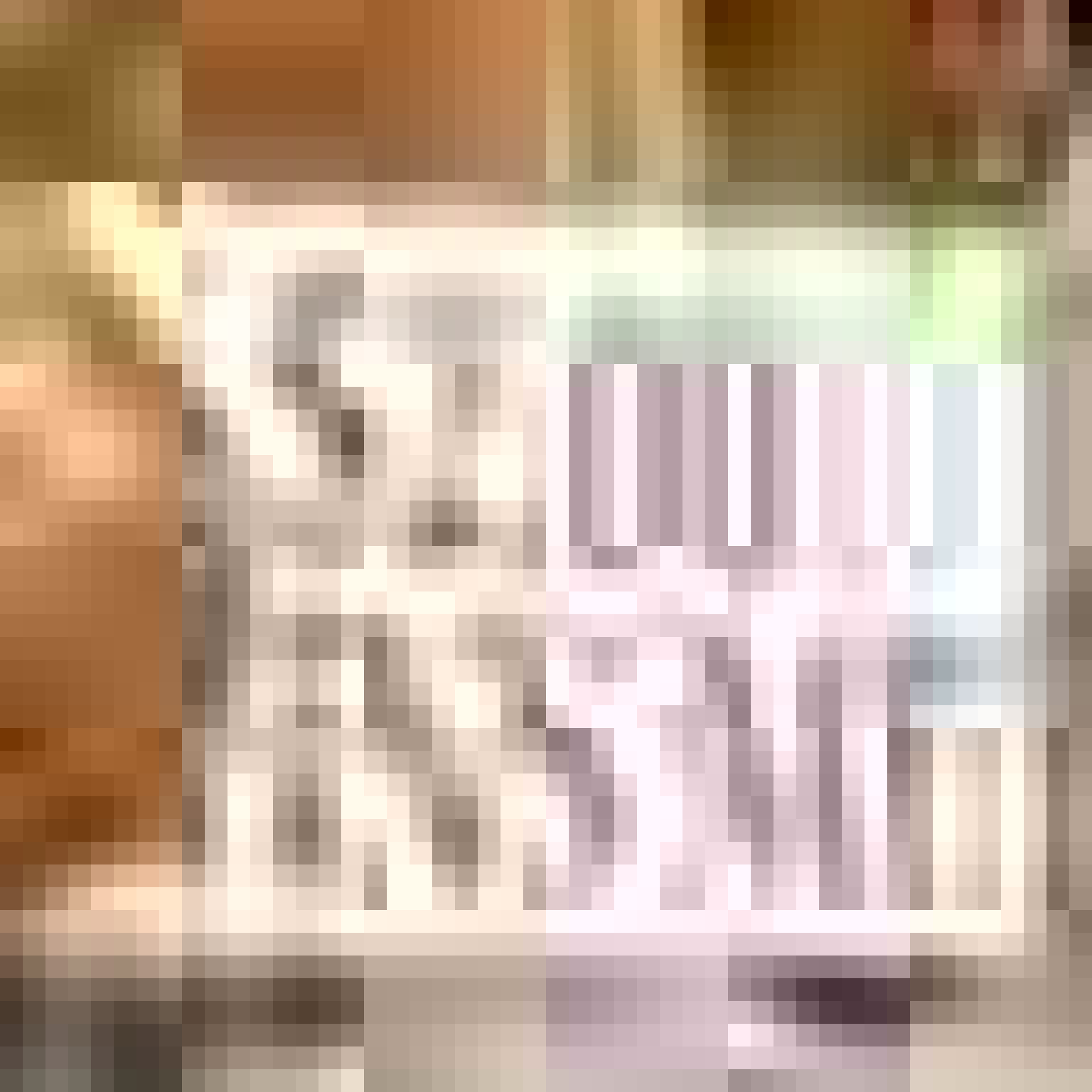}%
    \label{fig1:subfig_d}
  }
  \vspace{-7.5pt}
  \subfloat[Recovered Coefficients]{
    \includegraphics[width=0.233\textwidth]{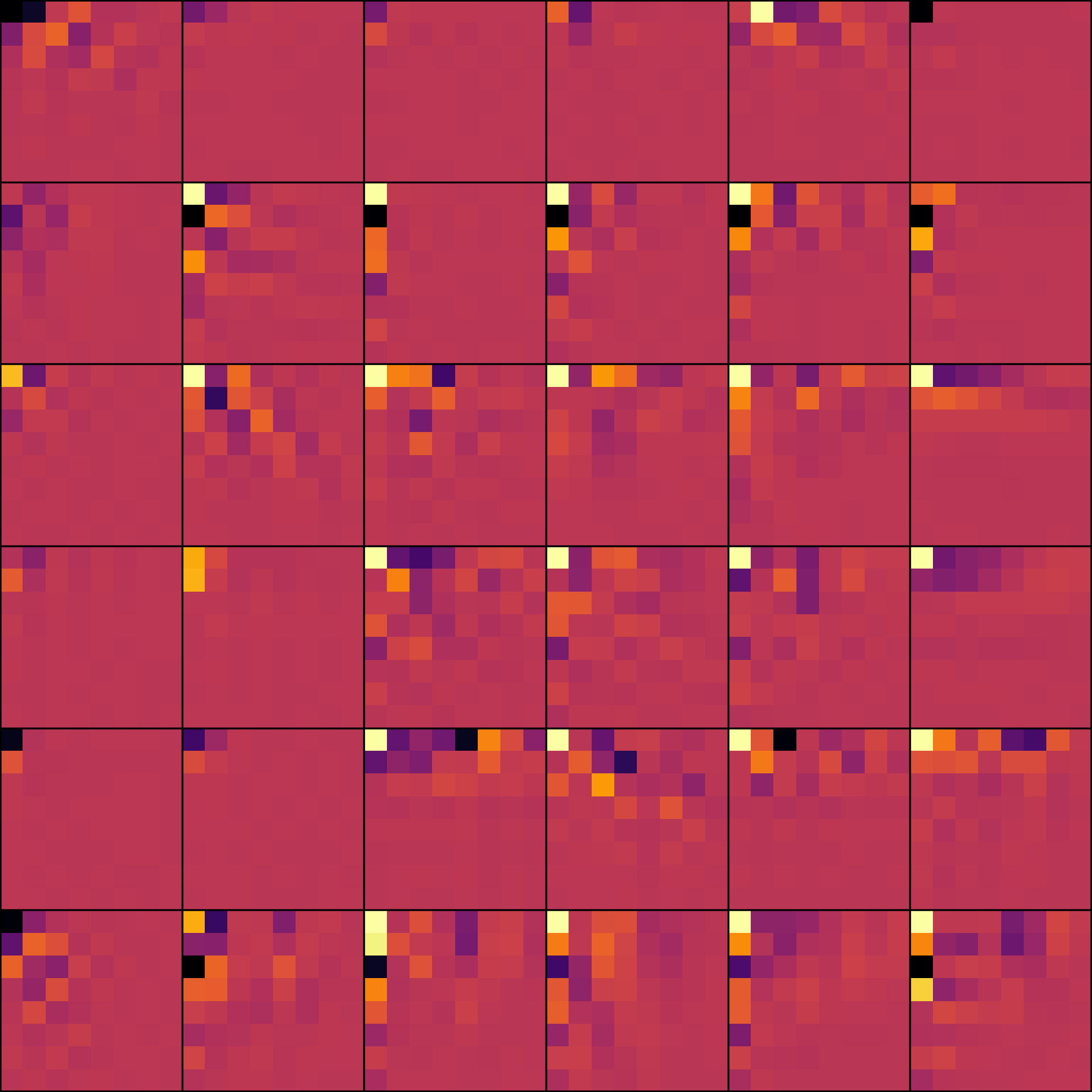}%
    \label{fig1:subfig_e}
  }
 \hspace{-0.1cm}
    \subfloat[Recovered Image]{
    \includegraphics[width=0.233\textwidth]{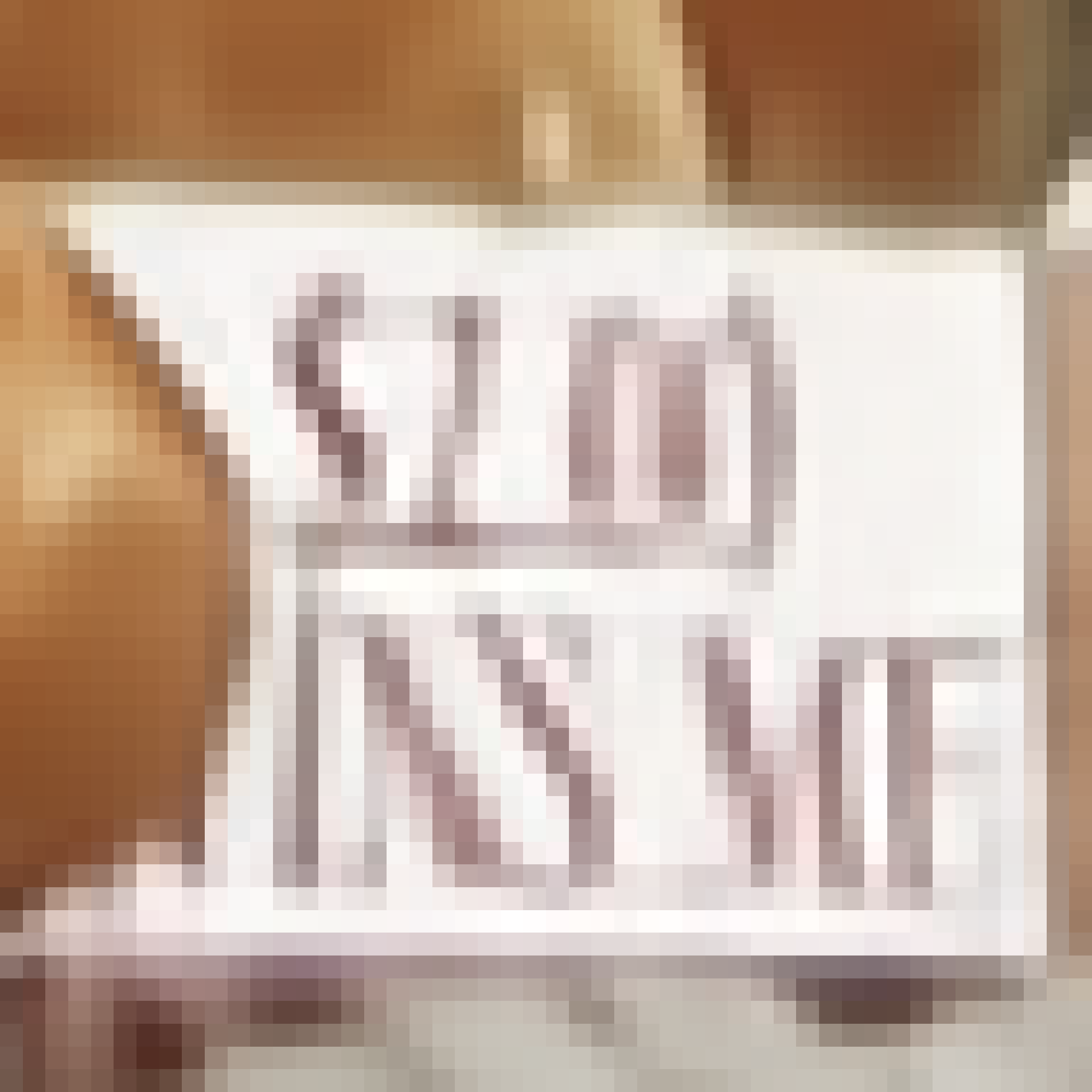}%
    \label{fig1:subfig_f}
  }
  \caption{Visualizations of the quantization and recovery of DCT coefficients and corresponding images. (a)-(b) The original image and its lossless DCT coefficients. (c)-(d) Compressed JPEG at QF = 10 and its highly sparse quantized coefficients. (e)-(f) Recovered coefficients of our DCTransformer and the reconstructed image. Note that only the coefficients of the Y channel are presented.}
  \label{fig_1}
\end{figure}

Aiming at aggregating the information and reducing data size of digital images, the JPEG (Joint Photographic Experts Group) algorithm \cite{wallace1991jpeg} is a cornerstone in image compression technology for its flexibility and balance between compression ratio and visual quality preservation. It has remained a widely used and pivotal format in digital images for decades due to its widespread compatibility and sufficient efficiency for many applications. The JPEG compression process first divides the image into $8 \times 8$ blocks and applies Discrete Cosine Transform \cite{ahmed1974discrete} (DCT) to each block (refer to Figure \ref{fig_1} (a)). The produced DCT coefficients are then divided by a quantization matrix and rounded to the nearest integer. An example of the DCT coefficients that are quantized in the JPEG compression and recovered through our model is depicted in Figure \ref{fig_1}, along with corresponding images. Since the human eye is less sensitive to high-frequency details and chrominance variants, JPEG compression relies on quantization and chroma subsampling to balance the compression ratio and preserve visual quality. The quantized coefficients are subsequently entropy-coded and packed into a compact bitstream for efficient transmission or storage. These characteristics enable JPEG images to be widely used across various devices, establishing it as one of the most critical image formats. However, as a lossy compression algorithm, JPEG compression introduces coding artifacts, such as blockiness resulting from dividing the image into blocks or ringing artifacts due to the degradation of high-frequency information (refer to Figure \ref{fig_1}. (d)), which makes a negative impact on the subjective visual quality. 

To improve the visual experience, the task of removing compression artifacts from the decoded JPEG images, also referred to as JPEG artifact removal, has received significant interest. Many traditional approaches \cite{ahumada1994smoothing, price1999biased, reeve1984reduction, list2003adaptive, 6823743, dabov2007image} have been developed to estimate the lost information from prior modeling perspectives or manually design filters. Deep learning-based methods, especially convolutional neural networks (CNNs) based ones, have also been proposed for image denoising such as some revolutionary works \cite{dong2014learning, DNCNN, ARCNN} showed significant improvements. Although CNN-based methods have achieved remarkable progress compared with traditional methods \cite{ARCNN, DNCNN, 7410430, DMCNN, MWCNN, QGAC, FBCNN}, it still suffers from some inherent limitations like weak in modeling long-range dependencies compared to Transformer-based methods \cite{dosovitskiy2021image, liu2021Swin, liang2021swinir, 10.1145/3503161.3547986}.

To sum up, existing JPEG artifact removal approaches suffer from the following limitations in the practical deployment: (1) Most of them train a specific model for each quality factor and cannot handle a wide range of quality factors. Due to the flexibility of JPEG compression, varying quality factors degrade images differently in coefficient representations, posing challenges in recovering the identical original image. (2) Few studies consider learning from their decoded frequency coefficients. Since DCT coefficients serve as the original representation of JPEG compression, direct recovery in the DCT domain would be worth further exploration in this task. (3) In existing DCT coefficient recovery methods, luminance and chrominance components are processed separately due to further subsampling of chrominance components and different quantization values.

To address the above limitations, we introduce a DCT domain spatial-frequential Transformer (DCTransformer) that employs a single frequency domain model to recover the JPEG quantized coefficients. Our core idea targets the lossy steps in JPEG compression and directly recovers decoded coefficients within the DCT frequency domain. In particular, we present a dual-branch Spatial-Frequential Transformer Block (SFTB) for capturing correlations among both spatial and frequential dimensions of collocated DCT coefficients. Furthermore, by integrating quantization matrix embedding and a luminance-chrominance alignment head, our single model effectively handles the recovery of compressed luminance and chrominance components across a wide range of compression quality factors. Besides, different from prevalent CNN-based models in quantized coefficient recovery tasks, our approach employs a Transformer-based architecture in the DCT domain. To the best of our knowledge, it is the first Transformer-based model adopted for JPEG quantized coefficient recovery in the DCT domain.

In summary, the main contributions of this paper are as follows: 

\begin{enumerate}
\item{A DCT domain spatial-frequential Transformer for JPEG quantized coefficient recovery (DCTransformer) is proposed. The dual-branch self-attention architecture is specifically designed to capture the correlations across different dimensions within collocated DCT coefficients. Additionally, the feature concatenation and residual connections facilitate handling the sparsity of quantized DCT coefficients.}

\item{Towards a single model handling luminance and chrominance components across a wide range of quality factors. By incorporating the quantization matrix embedding and a luminance-chrominance alignment head, we effectively introduce the information of the quantization matrix and unify different-sized components into our DCTransformer.}

\item{The proposed recovery scheme is fully based on learning in the DCT domain. To the best of our knowledge, this is the first research that adopts a Transformer-based model within the DCT domain to learn from coefficients directly. We also provide a DCT domain quantitative evaluation for thorough analysis. Our DCTransformer achieves state-of-the-art performance in color JPEG restoration among both the pixel and DCT domain approaches.}
\end{enumerate}

The remainder of this paper is structured as follows. Section II reviews the frequency domain learning and quantized coefficient recovery. The proposed DCT domain method is introduced in Section III. Experiments and ablation studies on benchmark datasets demonstrate the effectiveness of our proposed method in Section IV. Finally, Section V concludes our paper.

\begin{figure*}[!t]
\centering
\includegraphics[width=7in]{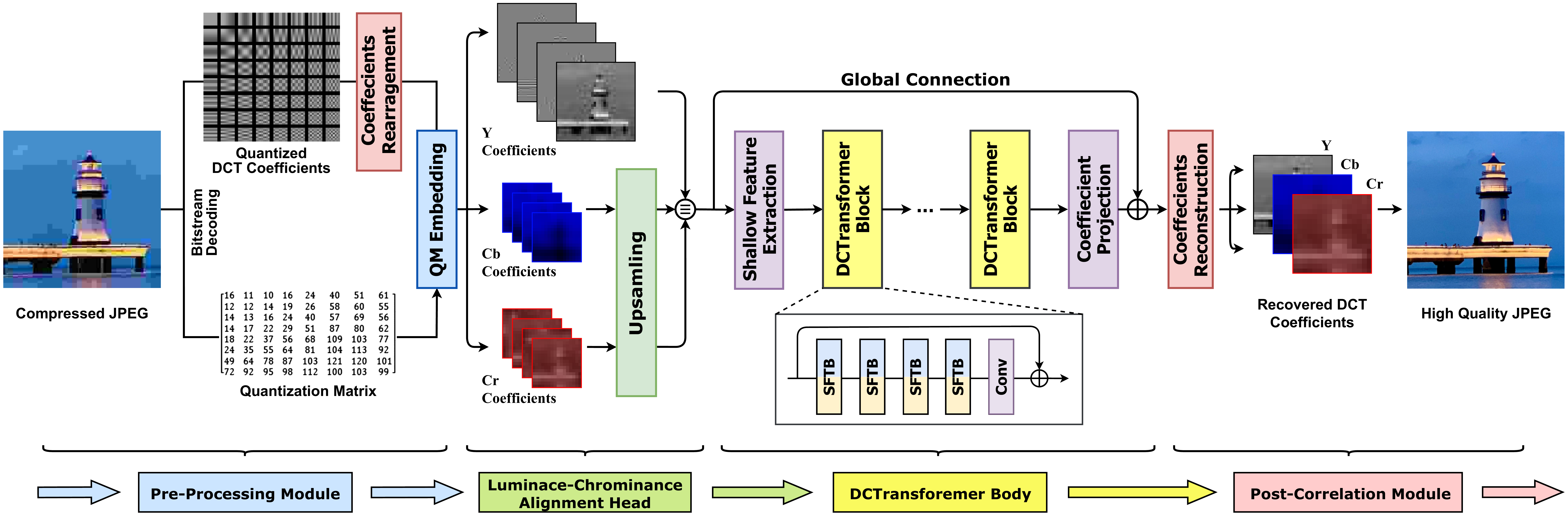}
\caption{An overview of the pipeline of the proposed DCTransformer for JPEG quantized coefficient recovery. DCTransformer consists of four modules, \textit{i.e.}, pre-processing module, luminance-chrominance alignment head, DCTransformer body, and post-correlation module. The pre-processing module includes operations of quantization matrix embedding and coefficient rearrangement to prepare collocated DCT coefficients. Then the luminance-chrominance alignment head unifies different-sized Y and CbCr coefficients and fed into the DCTransformer body. This is processed through several Spatial-Frequential Transformer Blocks (SFTB) in the DCTransformer body. Finally, the post-correlation module reconstructs the full image from the recovered coefficients.}
\label{fig_2}
\end{figure*}

\section{RELATED WORKS}

Quantized coefficient recovery, aiming to improve the quality of compressed images, shares the same objective as JPEG artifact removal but directly processes decoded coefficients as input and output. Due to the quantized DCT coefficients having excessive invalid values, this problem is reflected in the loss of image details in the pixel domain. Traditional quantized coefficient recovery methods address the degradation problem by approximating the lost information via Gaussian \cite{ahumada1994smoothing}, and Laplace distributions \cite{price1999biased}, or manually design a filter to reduce the artifact \cite{reeve1984reduction, list2003adaptive, 6823743, dabov2007image}. These estimation approaches deviate from the complexity in reality practice due to their priori assumptions, leading to critical performance limitations.

Given the powerful nonlinear mapping capability shown in various vision tasks, deep neural networks have been introduced to restore compressed JPEG images since \cite{DNCNN, ARCNN, 7410430}. However, many previous methods implement a model trained for each quality factor \cite{ARCNN, 7410430, zhang2019rnan, 8653951, MWCNN}, which is inefficient and unrealistic in the implementation. To handle the varying quality factors, \cite{QGCN} attempts to incorporate the quantization matrix by concatenating it with images as additional input channels. \cite{FBCNN} predicts the quality factor from the lossy input, and reconstructs the image by embedding the adjustable quality factor to achieve flexibility. Fu \textit{et al.} \cite{9446478} introduces an interpretable method by leveraging advantages from model-driven and data-driven approaches. \cite{10080951} combines a variational auto-encoder with a deformable offset gating network for controlling flexible qualities. More recently, \cite{wang2022jpeg} presents an unsupervised learning strategy for compression quality estimation via contrastive learning. Besides operating in the pixel domain, several dual-domain methods, such as \cite{guo2016building, DMCNN, IDCN}, integrate processing both the pixel and DCT domain information and aim to learn dual-domain corrections for JPEG artifact reduction tasks. However, one challenge is that the optimal representation in the frequency domain may not correspond to that in the pixel domain and vice versa \cite{xu2020learning}. Moreover, for more general image restoration purposes, many post-processing methods like \cite{zhang2018ffdnet, Soh2022variational, 9454311}, or Transformer-based \cite{liang2021swinir, 10.1145/3503161.3547986} prove their excellent capability when extended to JPEG artifact removal tasks.  

Despite some existing methods (e.g., QGAC \cite{QGAC}) also fully operating in the frequency domain, our method differs in three main aspects. Firstly, while traditional methods typically apply a CNN-based model, we propose a Transformer-based approach for DCT domain modeling. Secondly, existing methods tend to have over-complicated schemes. For example, QGAC employs a convolutional filter manifold to introduce the quantization information to guide the recovery. Yet we find that our quantization matrix embedding can efficiently and directly introduce varying loss information from quantization while reducing the computational cost of the network. Thirdly, in existing methods, the recovery of color JPEG is divided into two stages: first restoring the Y channel as intermediate output, and then restoring the Cb and Cr channels separately. In contrast, our method provides an alignment toward the unified feature map to recover both the luminance and chrominance components simultaneously.

\section{PROPOSED METHOD}

Given an original image $I_{gt}$ and the corresponding JPEG compressed image $I_j$, the JPEG compression process can be formulated as:
\begin{equation}
C_j = round(Q(D(I_{gt}))) + e,
\end{equation}
where $C_j$ is the quantized DCT coefficients of $I_j$, $D(\cdot)$ is the DCT operation, $Q(\cdot)$ denotes quantization controlled by the quantization matrix (QM), and $e$ encapsulates other compression losses, \textit{i.e.}, chroma subsampling. Thus, we can formulate the task of JPEG quantized coefficient recovery as:
\begin{equation}
I_{rec} = D^{-1}(Q^{-1}(C_j)),
\end{equation}
where $D^{-1}(\cdot)$ and $Q^{-1}(\cdot)$ denote inverse DCT and inverse quantization operations respectively, and $I_{rec}$ is the recovered image. Our primary objective is to design an optimal function for $Q^{-1}(\cdot)$ that minimizes the distance between $I_{gt}$ and $I_{rec}$, \textit{i.e.}, $||I_{gt} - I_{rec}||$, under constraints of recovering the coefficients in DCT domain. However, finding this $Q^{-1}(\cdot)$ is an ill-posed problem due to the lossy nature of the rounding and downsamping operation, which makes the inverse quantization process nontrivial. The recovery of quantized coefficients is not merely a mathematical inversion problem but also a perceptual one, as it should aim to maximize the perceived quality of the recovered image. This necessitates employing a learning-based approach that can incorporate perceptual capabilities into the recovery.

\subsection{The Overall Framework}
We propose a novel framework aiming at recovering the quantized coefficients in the DCT domain. Figure \ref{fig_2} shows an overview of our framework (note that some components are omitted for simplicity), which consists of four modules as follows. 

The first module, a coefficient pre-processing module, is designed to enable the single model to recover JPEG compressed images with varying quality factors. Itcomprises a quantization matrix (QM) embedding and a coefficient rearrangement operation. The second module, a luminance-chrominance alignment head, allows the model to simultaneously process luminance and chrominance components, with producing a unified feature map. The third module, which serves as the core structure for quantized coefficient recovery, employs a DCT domain spatial-frequential Transformer network, named DCTransformer. DCTransformer consists of a shallow feature extraction layer, several DCTransformer blocks, a coefficient projection layer, and a global connection. Each DCTransformer block contains several Spatial-Frequential Transformer Blocks (SFTBs) and a convolutional layer, followed by a residual connection. The last module is a post-correlation module that incorporates coefficient reconstruction and inverse DCT operation, converting the recovered quantized DCT coefficients into restored pixel domain images.

\subsection{QM Embedding and DCT Coefficient Rearrangement}
\begin{figure}[!t]
\centering
\includegraphics[width=3.5in]{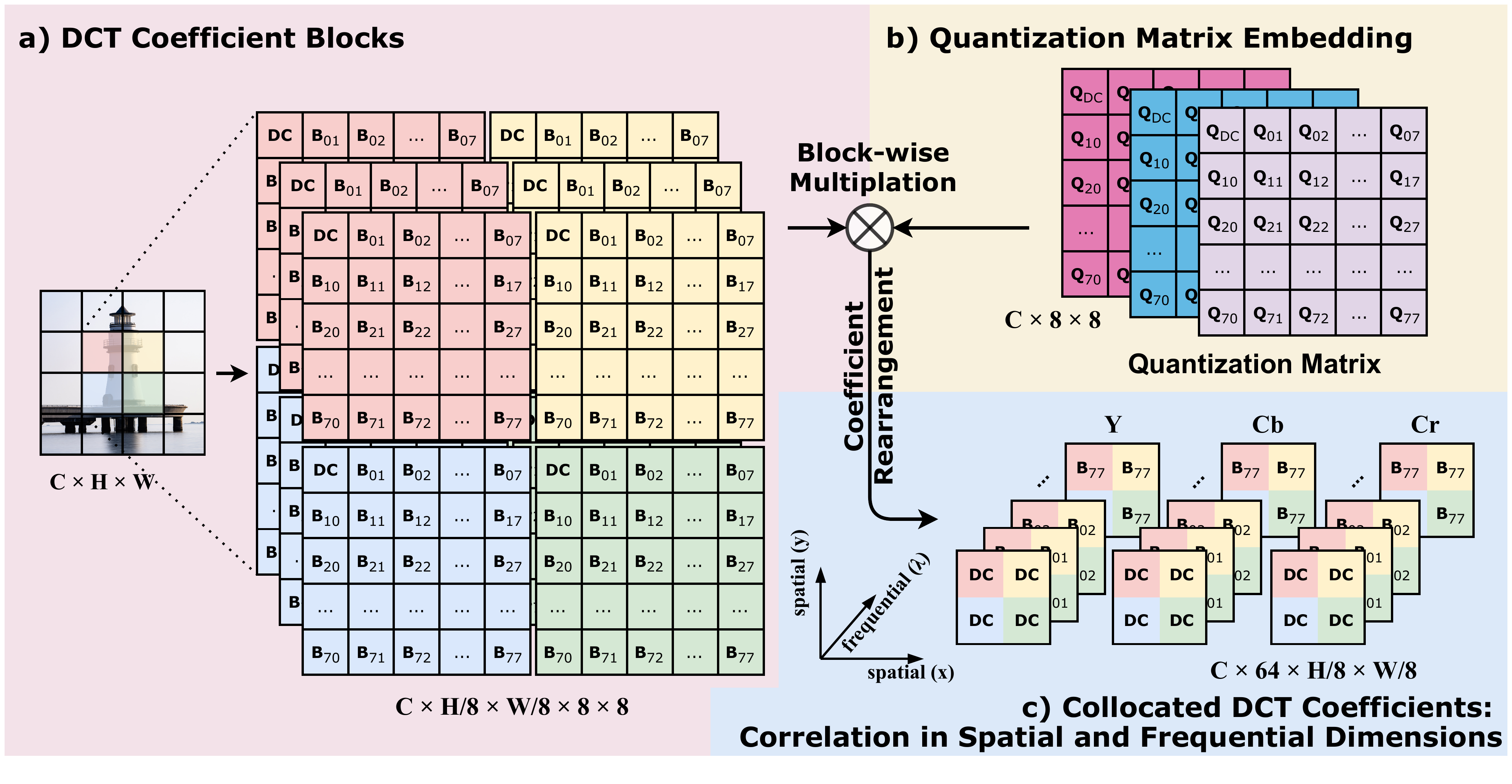}
\caption{The schematic illustrations of a) Decoded DCT coefficient blocks. b) The proposed quantization matrix embedding. c) The rearrangement of collocated DCT coefficients. The resulting collocated DCT coefficients have an intrinsic correlation in both spatial and frequential dimensions.}
\label{fig_3}
\end{figure}

Many frequency domain methods suffer from different compression quality factors since they bring differences in frequency domain representations. Thus, many previous models are trained for a specific quality factor or channel component. Otherwise, the quantization matrix is regarded as the input to the network for spatial feature extraction or predicted from the unknown to approximate the loss of information. To address these problems, we perform the operations of \textit{1) Quantization Matrix Embedding} and \textit{2) DCT Coefficient Rearrangement} in our pre-processing module. A schematic diagram of these operations is shown in Figure \ref{fig_3}.

\textit{1) Quantization Matrix Embedding:~~}For JPEG compression without lossy subsampling operation, the losses are all caused by quantization. Considering $B_{coef}$ in each of the $8 \times 8$ coefficients block and corresponding quantization matrix $Q_{mat}$, the loss $\delta_{coef}$ can be expressed as:
\begin{equation}
\delta_{coef} = round(\frac{B_{coef}}{Q_{mat}}) - round(\frac{B_{coef}}{Q_{mat}}) \cdot Q_{mat},  
\end{equation}
where $round$ is a function that rounds to the nearest integer, and the division is element-wise. This operation restricts the coefficient loss to the loss range of the rounding operation, which signifies the quantization value for the corresponding frequency in the quantization matrix. Derived from Eq. (3), the loss $\delta_{coef}$ is satisfies:
\begin{equation} 
\delta_{coef} = \theta_i \odot Q_{mat}, 
\end{equation}
where $\odot$ donates Hadamard product of two $8 \times 8$ matrix and $\theta$ is a relative factor in range:
\begin{equation} 
-0.5 < \theta < 0.5 ~~~ \forall ~ \theta_i \in \theta~.
\end{equation}

Inspired by this, we propose an intuitive approach as quantization matrix embedding. Recall that previous methods, like \cite{QGAC} employs convolutional layers to extract features from the quantization matrix, or \cite{QGCN} duplicates the quantization matrix and concatenates them with images as additional input channels. We suggest they may have drawbacks since the information of the quantization matrix should be combined with DCT coefficients that are quantized by it. Therefore, we recover the direct representation of quantization losses, as in the JPEG decoding process, the lossy coefficients are block-wise multiplied by the quantization matrix, resulting in the value range from -1024 to 1023 approximately. The proposed quantization matrix embedding can be formulated as:
\begin{equation} 
\tilde{B}_{coef} = B_{coef} \odot Q_{mat}, 
\end{equation}
where $Q_{mat}$ is the quantization matrix. This is illustrated in Figures \ref{fig_3}  a) and b). Hence, the loss of $\tilde{B}_{coef}$ in position $i, j \in N^{\{0,...,7\}}$ can be constrained by the value of the quantization matrix at position $(i,j)$:
\begin{equation} 
-0.5 * Q_{mat}(i,j) < \tilde{\delta}_{coef}(i,j) < 0.5 * Q_{mat}(i,j).
\end{equation}

\begin{figure*}[!t]
\centering
\includegraphics[width=\textwidth]{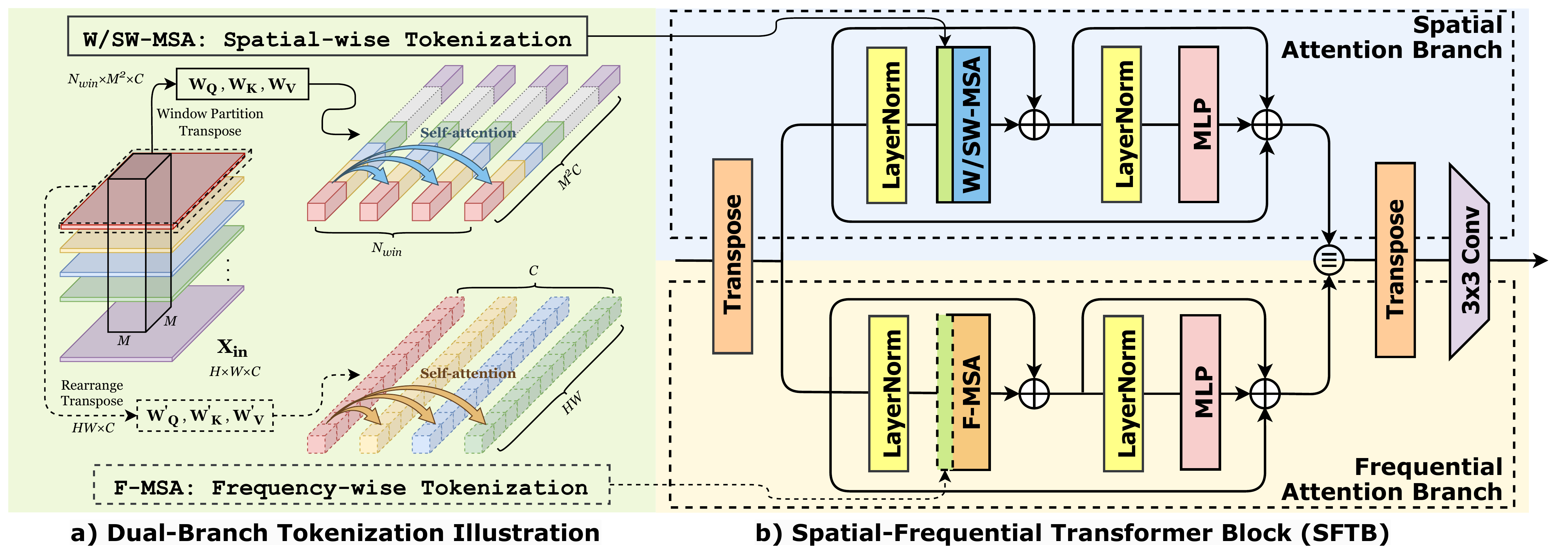}
\caption{The architecture of our Spatial-Frequential Transformer Block (SFTB). a) Illustration of the tokenization in spatial-wise and frequency-wise self-attention to extract diverse correlations. Note that multi-head design and learnable biases have been omitted for simplicity. b) The dual-branch architecture of SFTB. Each SFTB consists of two attention branches: spatial attention branch and frequential attention branch. The outputs of two branches are then channel concatenated and passed through a 3 $\times$ 3 convolutional layer.}
\label{fig_4}
\end{figure*}

\textit{2) DCT Coefficient Rearrangement:~~}Due to the $8 \times 8$ block-based DCT operation in JPEG compression, each $8 \times 8$ block contains 64 different frequency components of the original pixel domain image. To emphasize the spatial and frequential correlations within coefficients, we rearrange the embedded DCT coefficients into collocated DCT maps. For each YCbCr channel, we reshape the coefficients from the shape $1 \times H \times W$ to $64 \times \frac{H}{8} \times \frac{W}{8}$, where the $64$ represents different frequency components, and the $H$ and $W$ denote the original height and width of the Y or CbCr channel, respectively. 

To better understand the collocated DCT coefficients, we illustrate them from the spatial ($x,y$) and frequential ($\lambda$) axes, as shown in Figure \ref{fig_3} c). Most importantly, after the rearrangement, the DCT coefficients are collocated with spatial features. Analyzing along the frequential (first) dimension, 64 orthogonal frequency components within the same $8 \times 8$ block are represented by different channels at the same spatial position. Meanwhile, along the spatial (last two) dimensions, adjacent spatial positions correspond to the same frequency components in neighboring $8 \times 8$ blocks. This forms the basic shape of the input and output feature map for our frequential-spatial feature extraction process.

\subsection{Luminance-Chrominance Alignment Head}

Color JPEG images encode information in YCbCr colorspace. To achieve a higher compression ratio, the Cb and Cr channels, also known as chrominance components, are quantized to carry less information and further subsampled to half-sized. Due to the high sensitivity of prior frequency domain models, the correction process for color JPEG images in \cite{ouyang2020towards, QGAC} recovers the luminance and chrominance coefficients through separate models or different stages. These may reduce efficiency and lead to insufficient information interaction between the luminance and chrominance components.

To manage different-sized Y and CbCr components, we propose a luminance-chrominance alignment head to unify color JPEG coefficients. It takes in DCT collocated coefficients from both luminance $X_y$ and chrominance components $X_{cb}$ and $X_{cr}$ as input, produces an aligned and unified DCT feature map $X_{aligned}$ feeding to the DCTransformer body. In detail, $X_{y}$, $X_{cb}$, and $X_{cr}$ are first projected into higher dimensions by three different convolutional layers, respectively. Subsequently, $X_{cb}$ and $X_{cr}$ are upsampled twice by each of the two transpose convolutional layers to match the size of $X_y$. Finally, we channel-wise concatenate the three and use a $3 \times 3$ convolutional layer $H_{cat}(\cdot)$ to fusion the components. This alignment process can be expressed as:
\begin{equation}
X_{aligned} = H_{cat}([H_y(X_{y}),~H_{cb}(X_{cb}),~H_{cr}(X_{cr})]),
\end{equation}
where $[\cdot]$ represents the concatenation operation, $H_y$, $H_{cb}$ and $H_{cr}$ stand for convolutional heads for each component. This alignment ensures the luminance and chrominance coefficients are spatially matched and mapped to higher dimensional feature space for the following feature extraction.

\subsection{DCTransformer Body}

\textit{1) Dual-branch Spatial-frequential Transformer Architecture: }
Transformer \cite{vaswani2017attention} and its self-attention module are primarily designed for sequence modeling and extracting correlations within long-range dependencies. Compared to convolution-based methods, it also shows promising results when adapted to various computer vision tasks \cite{dosovitskiy2021image, liu2021Swin, liang2021swinir, 10.1145/3503161.3547986}. However, in JPEG compression, the quantization operation produces many invalid coefficients to shorten the coding, resulting in sparsity where coefficients are quantized to zeros. This sparsity problem is especially significant for the high-frequency components \cite{nash2021generating}, which brings a great challenge for our recovery task.

Let us revisit the feature representation of our collocated DCT coefficient map first. After the rearrangement, the intrinsic spatial and frequential correlations of collocated quantized DCT coefficients manifest along different dimensions, as shown in Figure \ref{fig_3} c). However, due to the sparsity problem issued above, we argue that the representation of quantized DCT coefficients is relatively unsuitable for CNNs to extract localized spatial features or insufficient for standard Transformers to compute the attention along the patches. Consequently, it becomes imperative for the restoration model to address this representation of quantized coefficients, aiming to simultaneously extract both spatial and frequential correlations to facilitate recovery.

Following the aforementioned motivation, we introduce a DCT domain frequential-spatial Transformer architecture, namely DCTransformer, which serves as the core module for quantized coefficient recovery. This architecture, characterized by its dual-branch design, is specifically crafted to extract correlations within both spatial and frequential dimensions, with its operations fully conducted within the DCT domain. Both branches of our DCTransformer is carefully designed to ensure the comprehensive capture of correlations across both spatial and frequential dimensions, thereby facilitating a more effective recovery.

An illustration of the dual-branch spatial-frequential self-attention structure is shown in Figure \ref{fig_4}. To be specific, in each Spatial-Frequential Transformer Block, the input feature map $X_{in}$ undergoes two parallel branches separately: the spatial attention branch and the frequential attention branch. Each involves a different type of tokenization and follows a self-attention mechanism. The operations can be represented as:
\begin{equation}
\begin{gathered}
    Y_{spat} = B_{spat}(X_{in}), \\
    Y_{freq} = B_{freq}(X_{in}),
\end{gathered}
\end{equation}  
where $Y_{spat}$ and $Y_{freq}$ are output feature maps of the spatial attention branch $B_{spat}$ and frequential attention branch $B_{freq}$, respectively.

\textit{2) Spatial Attention Branch: }
For the spatial branch in our SFTB, we utilize a Swin Transformer Block \cite{liu2021Swin} to extract spatial correlations in collocated DCT coefficients. Note that this operation is still entirely in the DCT domain, where we extract the spatial correlation of the rearranged DCT coefficients. Swin Transformer, with a modification of the standard multi-head self-attention (MSA), computes local attention within each partitioned window, thereby generally reducing computational complexity. Figure \ref{fig_4} shows a simplified version of this mechanism. The window self-attention (W-MSA) and shifted window self-attention (SW-MSA) are alternatively ordered in successive SFTBs. Therefore, the cross-window connections are facilitated by the shifting window operation. As illustrated in Figure \ref{fig_4} (b), the spatial attention branch consists of one W/SW-MSA module and a multiple-layer perception (MLP), accompanied by two layer normalizations (LN) preceding each SW-MSA and MLP.

In detail, the input is first partitioned into $M \times M$ non-overlapped windows to compute the self-attention. Given an input feature map shaped as $H \times W \times C$, the partitioned windows are reshaped to $\frac{HW}{M^2} \times M^2 \times C$, where standard self-attention is computed in each shaped $M^2 \times C$ window for W-MSA. Moreover, SW-MSA incorporates a shifting by displacing the windows by $(\frac{M}{2}, \frac{M}{2})$ to further introduce cross-window connections. Given the input feature $X_{in}$, the entire process of the spatial attention branch is expressed as follows:
\begin{equation}  
\begin{split}
\begin{aligned}
    X^{\prime}_{spat} &= \text{W/SW-MSA}(\text{LN}( X_{in})) +  X_{in}, \\
    Y_{spat} &= \text{MLP}(\text{LN}(X^{\prime}_{spat})) + X^{\prime}_{spat},
\end{aligned}
\end{split}
\end{equation}
where the W/SW-MSA donates the operation of window-based multi-head self-attention and shifted window-based multi-head self-attention; MLP regards two successive multiple layer perceptions, projecting the embedded feature into a higher dimensional space representation, supplemented with a GELU function to introduce more non-linearity.

\textit{3) Frequential Attention Branch: }
Collocated DCT coefficients are rearranged into spatially aligned, with each of 64 orthogonal frequency components in separate channels. To enhance extracting the dependencies in different frequency components in the frequential attention branch, we propose a frequency multi-head self-attention (F-MSA), in which each channel of the frequency feature map is process isolated as a token. In this approach, the attention is performed on the channel level so that the information within each frequency can interact globally with each other. Different from directly projecting each patch of the feature map as $Q$, $K$, and $V$, the input $X_{freq}$ is firstly reshaped into $HW \times C$, and linear projected through $W_Q$, $W_K$, and $W_V$ into $Q$, $K$, and $V$ as $C$ tokens in shape $HW \times 1$. In order to attend more frequency patterns, we extend it to multi-heads by splitting $Q$, $K$, and $V$ into $N=\frac{C}{d_{head}}$ heads in parallel, where in our experiments $d_{head}$ is set to 32. Subsequently, the standard attention mechanism is computed within each frequency-wise token, and outputs from all heads are concatenated. We illustrate this modified tokenization process of F-MSA in Figure \ref{fig_4} (a). Given the input feature $X_{in}$, the entire frequential branch can be formulated as:
\begin{equation}  
\begin{split}
\begin{aligned}
    X^{\prime}_{freq} &= \text{F-MSA}(\text{LN}(X_{in})) + X_{in}, \\
    Y_{freq} &= \text{MLP}(\text{LN}(X^{\prime}_{freq})) + X^{\prime}_{freq}, 
\end{aligned}
\end{split}
\end{equation}
where the F-MSA donates frequential multi-head self-attention; MLP has the same structure as in the spatial branch, which consists of two successive multiple-layer perceptions and a GELU function.

Within the F-MSA operation, we utilize two $3 \times 3$ depth-wise convolutional layers with a GELU activation in the middle as a channel-wise positional embedding. Each channel is added by a $X_{ebd}$ feature map before projecting into $Q$, $K$, and $V$. Since \cite{cordonnier2019relationship} has shown the potential of learnable positional embedding, this could enable attention heads to exploit the channel-wise information of different frequencies.

\begin{table*}[!t]
\caption{Pixel domain evaluation on \textbf{Color} JPEG images from three benchmark datasets. Represented in \textbf{PSNR(dB)/SSIM/PSNR-B(dB)} format. The best and second best performances are \textbf{Boldfaced} and \underline{Underlined}, respectively. Please note that marker $ ^*$ represents DCT domain methods.}
\centering
\setlength{\tabcolsep}{1pt} 
\begin{tabular}{cccccccc}
\toprule
\multicolumn{1}{c|}{Dataset} & \multicolumn{1}{c|}{QF} & \multicolumn{1}{c|}{JPEG} & \multicolumn{1}{c|}{QGCN\cite{QGCN}} & \multicolumn{1}{c|}{FBCNN\cite{FBCNN}} & \multicolumn{1}{c||}{ARCRL\cite{wang2022jpeg}} & \multicolumn{1}{c|}{QGAC$ ^*$\cite{QGAC}} & \multicolumn{1}{c}{DCTransformer$^*$} \\ 
\midrule
\midrule
\multicolumn{1}{c}{\multirow{4}{*}{LIVE1}} &
  \multicolumn{1}{|c|}{10} &
  \multicolumn{1}{c|}{25.69/0.743/24.20} &
  \multicolumn{1}{c|}{27.78/0.804/27.55} &
  \multicolumn{1}{c|}{27.77/0.803/27.51} &
  \multicolumn{1}{c||}{\underline{27.80}/\underline{0.805}/\underline{27.57}} &
  \multicolumn{1}{c|}{27.62/0.804/27.43} &
  \multicolumn{1}{c }{\textbf{28.06}/\textbf{0.808}/\textbf{27.87}}
   \\
 &
  \multicolumn{1}{|c|}{20} &
  \multicolumn{1}{c|}{28.06/0.826/26.49} &
  \multicolumn{1}{c|}{30.12/0.870/29.74} &
  \multicolumn{1}{c|}{30.11/0.868/29.70} &
  \multicolumn{1}{c||}{\underline{30.23}/\underline{0.872}/\underline{29.85}} &
  \multicolumn{1}{c|}{29.88/0.868/29.56} &
  \multicolumn{1}{c }{\textbf{30.32}/\textbf{0.873}/\textbf{30.00}}
   \\
 &
  \multicolumn{1}{|c|}{30} &
  \multicolumn{1}{c|}{29.37/0.861/27.84} &
  \multicolumn{1}{c|}{31.44/0.898/30.97} &
  \multicolumn{1}{c|}{31.43/0.897/30.92} &
  \multicolumn{1}{c||}{\underline{31.58}/\textbf{0.900}/\underline{31.13}} &
  \multicolumn{1}{c|}{31.17/0.896/30.77} &
  \multicolumn{1}{c}{\textbf{31.62}/\textbf{0.900}/\textbf{31.23}}
   \\
 &
  \multicolumn{1}{|c|}{40} &
  \multicolumn{1}{c|}{30.28/0.882/28.84} &
  \multicolumn{1}{c|}{32.33/0.914/31.83} &
  \multicolumn{1}{c|}{32.34/\underline{0.913}/31.80} &
  \multicolumn{1}{c||}{\textbf{32.53}/\textbf{0.916}/\underline{32.04}} &
  \multicolumn{1}{c|}{32.05/0.912/31.61} &
  \multicolumn{1}{c }{\underline{32.51}/\textbf{0.916}/\textbf{32.07}}
   \\ \midrule
\multicolumn{1}{c}{\multirow{4}{*}{BSDS500}} &
  \multicolumn{1}{|c|}{10} &
  \multicolumn{1}{c|}{25.84/0.741/24.13} &
  \multicolumn{1}{c|}{27.85/0.800/27.53} &
  \multicolumn{1}{c|}{{27.85}/0.799/{27.52}} &
  \multicolumn{1}{c||}{\underline{27.91}/\underline{0.803}/\underline{27.59}} &
  \multicolumn{1}{c|}{27.74/0.802/27.47} &
  \multicolumn{1}{c}{\textbf{27.95}/\textbf{0.803}/\textbf{27.87}}
   \\
 &
  \multicolumn{1}{|c|}{20} &
  \multicolumn{1}{c|}{28.21/0.827/26.37} &
  \multicolumn{1}{c|}{30.18/0.870/29.61} &
  \multicolumn{1}{c|}{30.14/0.867/29.56} &
  \multicolumn{1}{c||}{\textbf{30.31}/\textbf{0.872}/\underline{29.74}} &
  \multicolumn{1}{c|}{30.01/0.869/29.53} &
  \multicolumn{1}{c}{\underline{30.21}/\underline{0.871}/\textbf{30.07}}
   \\
 &
  \multicolumn{1}{|c|}{30} &
  \multicolumn{1}{c|}{29.57/0.865/27.72} &
  \multicolumn{1}{c|}{31.52/0.899/30.79} &
  \multicolumn{1}{c|}{31.45/0.897/30.72} &
  \multicolumn{1}{c||}{\textbf{31.69}/\textbf{0.901}/\underline{30.96}} &
  \multicolumn{1}{c|}{31.33/0.898/30.70} &
  \multicolumn{1}{c}{\underline{31.54}/\textbf{0.901}/\textbf{31.34}}
   \\
 &
  \multicolumn{1}{|c|}{40} &
  \multicolumn{1}{c|}{30.52/0.887/28.69} &
  \multicolumn{1}{c|}{32.44/0.916/31.62} &
  \multicolumn{1}{c|}{32.36/0.913/31.52} &
  \multicolumn{1}{c||}{\textbf{32.66}/\textbf{0.918}/\underline{31.82}} &
  \multicolumn{1}{c|}{32.25/0.915/31.50} &
  \multicolumn{1}{c}{\underline{32.46}/\underline{0.917}/\textbf{32.21}}
   \\ \midrule
\multicolumn{1}{c}{\multirow{4}{*}{ICB}} &
  \multicolumn{1}{|c|}{10} &
  \multicolumn{1}{c|}{29.44/0.757/28.53} &
  \multicolumn{1}{c|}{32.00/0.813/31.96} &
  \multicolumn{1}{c|}{\underline{32.18}/0.815/\underline{32.15}} &
  \multicolumn{1}{c||}{32.05/0.813/32.04} &
  \multicolumn{1}{c|}{32.06/\underline{0.816}/32.04} &
  \multicolumn{1}{c}{\textbf{32.73}/\textbf{0.818}/\textbf{32.69}}
   \\
 &
   \multicolumn{1}{|c|}{20} &
  \multicolumn{1}{c|}{32.01/0.806/31.11} &
  \multicolumn{1}{c|}{34.21/0.842/34.15} &
  \multicolumn{1}{c|}{\underline{34.38}/\underline{0.844}/\underline{34.34}} &
  \multicolumn{1}{c||}{34.32/0.842/34.31} &
  \multicolumn{1}{c|}{34.13/0.843/34.10} &
  \multicolumn{1}{c }{\textbf{34.70}/\textbf{0.843}/\textbf{32.64}}
   \\
 &
  \multicolumn{1}{|c|}{30} &
  \multicolumn{1}{c|}{33.20/0.831/32.35} &
  \multicolumn{1}{c|}{35.22/0.856/35.15} &
  \multicolumn{1}{c|}{\underline{35.41}/\underline{0.857}/\underline{35.35}} &
  \multicolumn{1}{c||}{35.37/0.856/\underline{35.35}} &
  \multicolumn{1}{c|}{35.07/\textbf{0.857}/35.02} &
  \multicolumn{1}{c}{\textbf{35.61}/\textbf{0.857}/\textbf{35.54}}
   \\
 &
  \multicolumn{1}{|c|}{40} &
  \multicolumn{1}{c|}{33.95/0.840/33.14} &
  \multicolumn{1}{c|}{35.78/0.860/35.71} &
  \multicolumn{1}{c|}{\underline{36.02}/\underline{0.866}/35.95} &
  \multicolumn{1}{c||}{35.99/0.860/\underline{35.97}} &
  \multicolumn{1}{c|}{35.71/0.862/35.65} &
  \multicolumn{1}{c}{\textbf{36.14}/\textbf{0.866}/\textbf{36.05}}
   \\ 
\bottomrule
\label{tab_1}
\end{tabular}
\end{table*}

\begin{figure*}[!ht]
  \centering
  
  \subfloat[QF=10 JPEG (23.63 dB / 0.753)]{
    \includegraphics[width=0.32\textwidth]{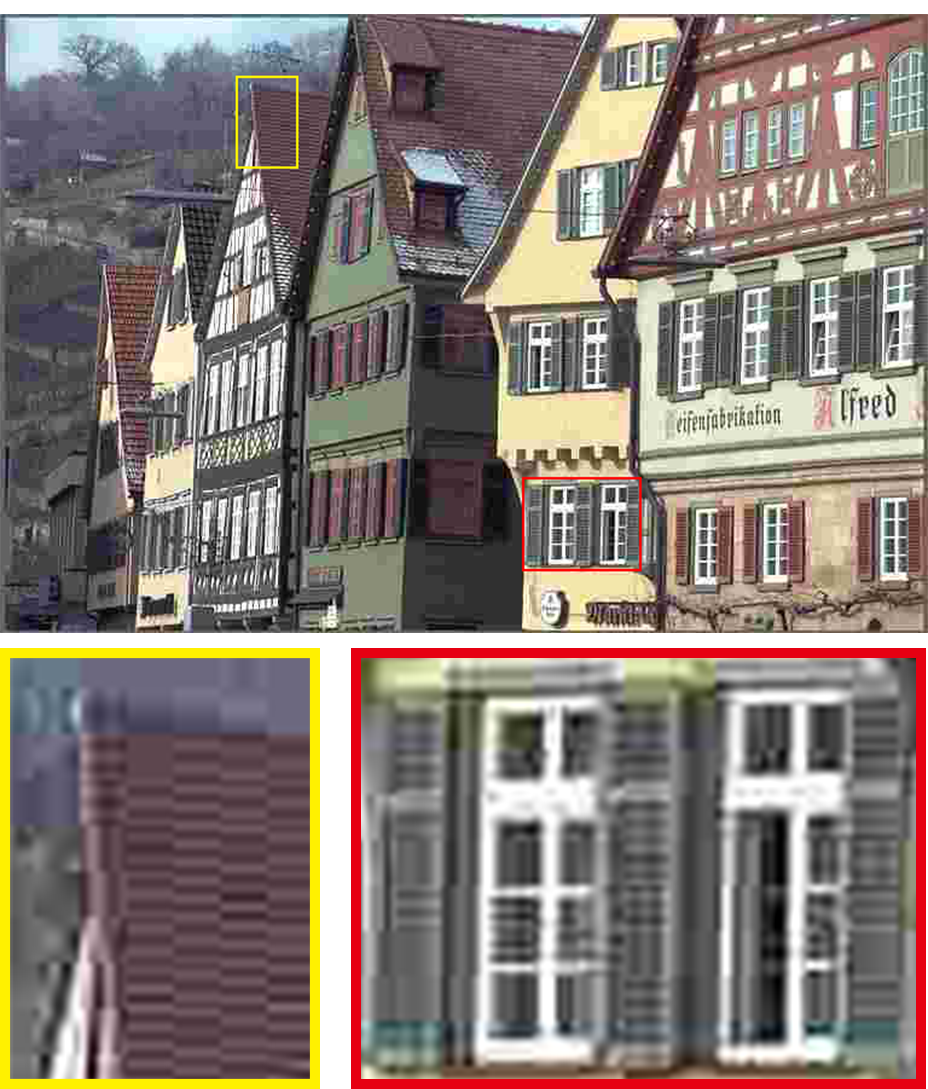}%
    \label{fig5:subfig_a}
  }
    \hspace{0.01cm}
  \subfloat[DnCNN-3 (24.93 dB / 0.790)]{
    \includegraphics[width=0.32\textwidth]{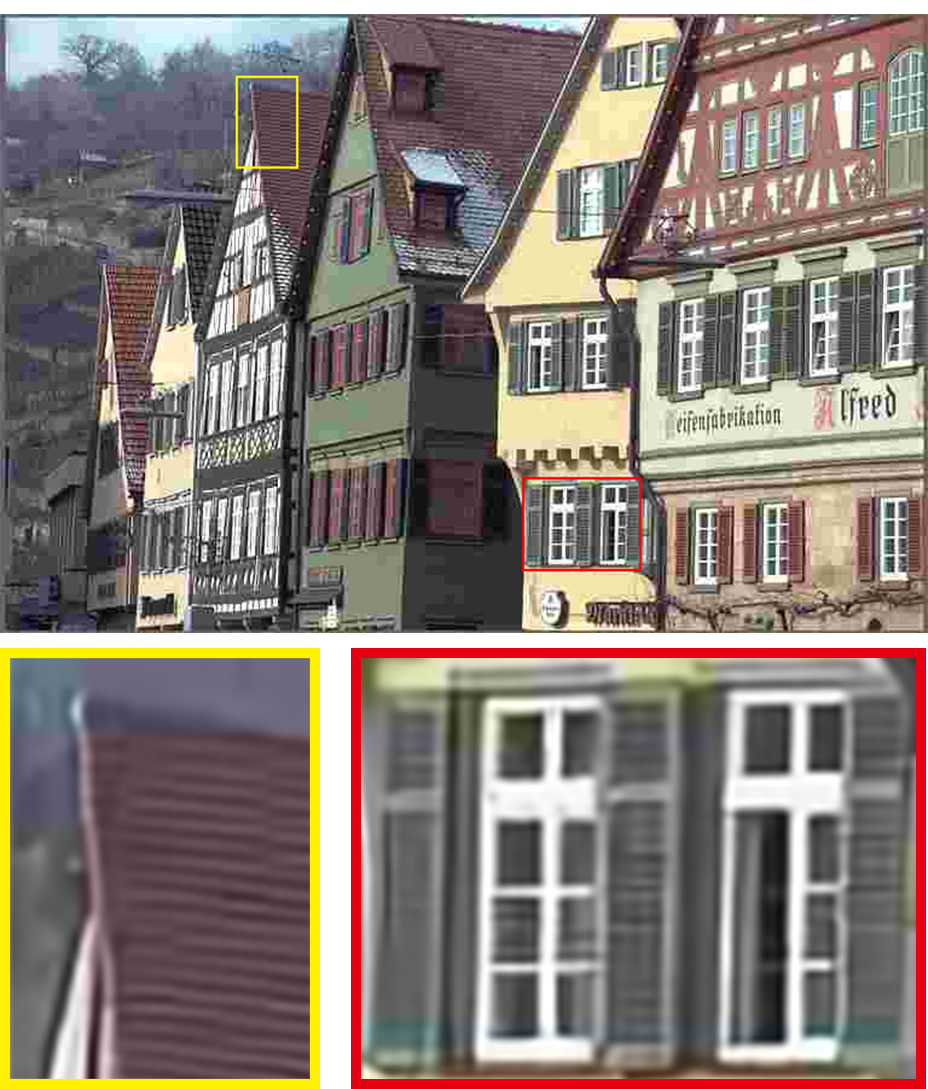}%
    \label{fig5:subfig_b}
  }
    \hspace{0.01cm}
  \subfloat[QGAC (25.78 dB / 0.815)]{
    \includegraphics[width=0.32\textwidth]{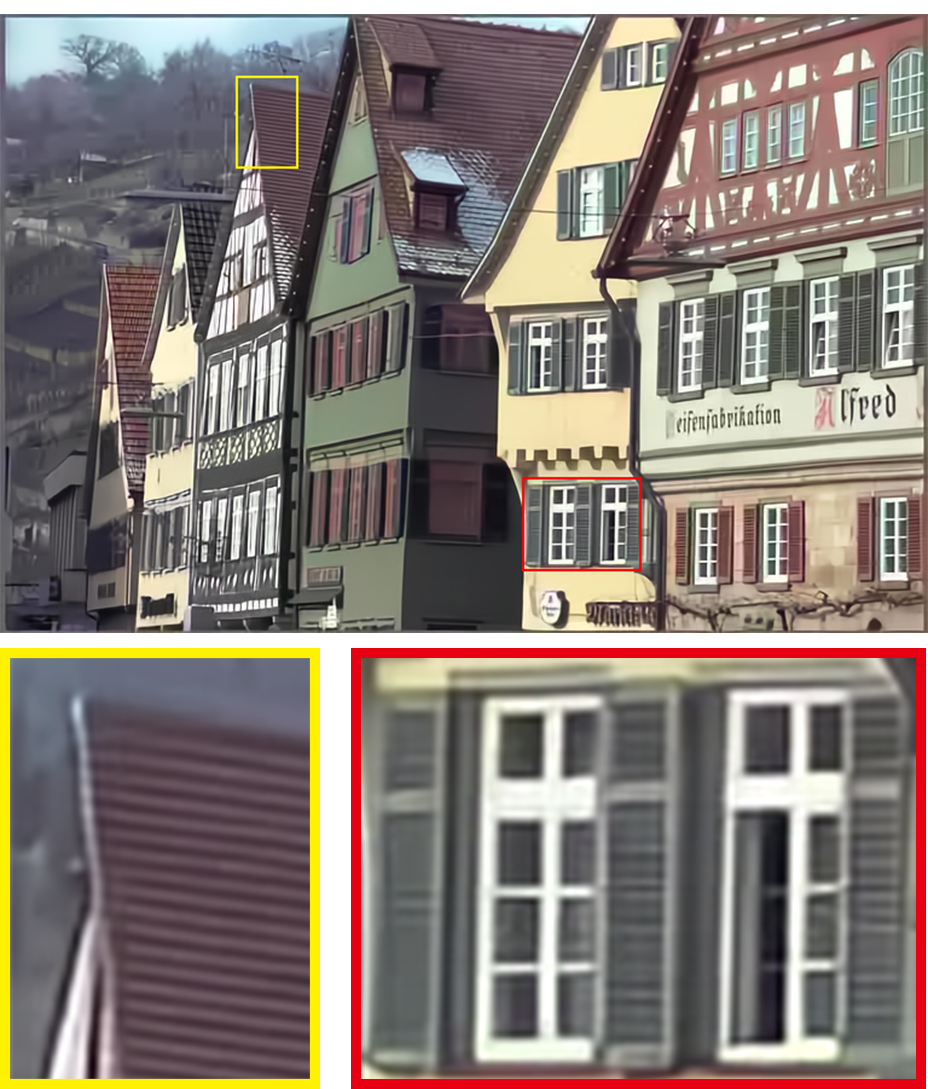}%
    \label{fig5:subfig_c}
  }
  \vspace{-7.5pt}
    \subfloat[FBCNN (25.99 dB / 0.814)]{
    \includegraphics[width=0.32\textwidth]{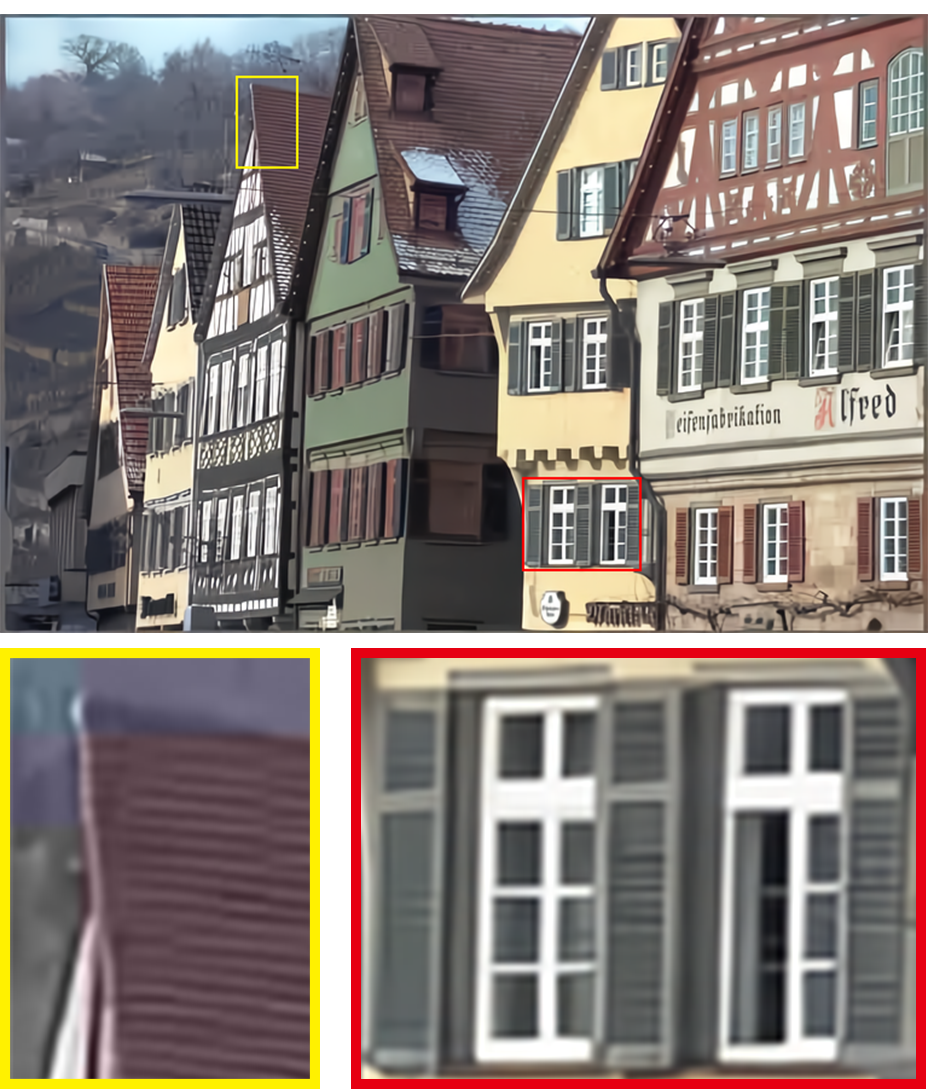}%
    \label{fig5:subfig_d}
  }
    \hspace{0.01cm}
  \subfloat[DCTransformer (26.27 dB / 0.822)]{
    \includegraphics[width=0.32\textwidth]{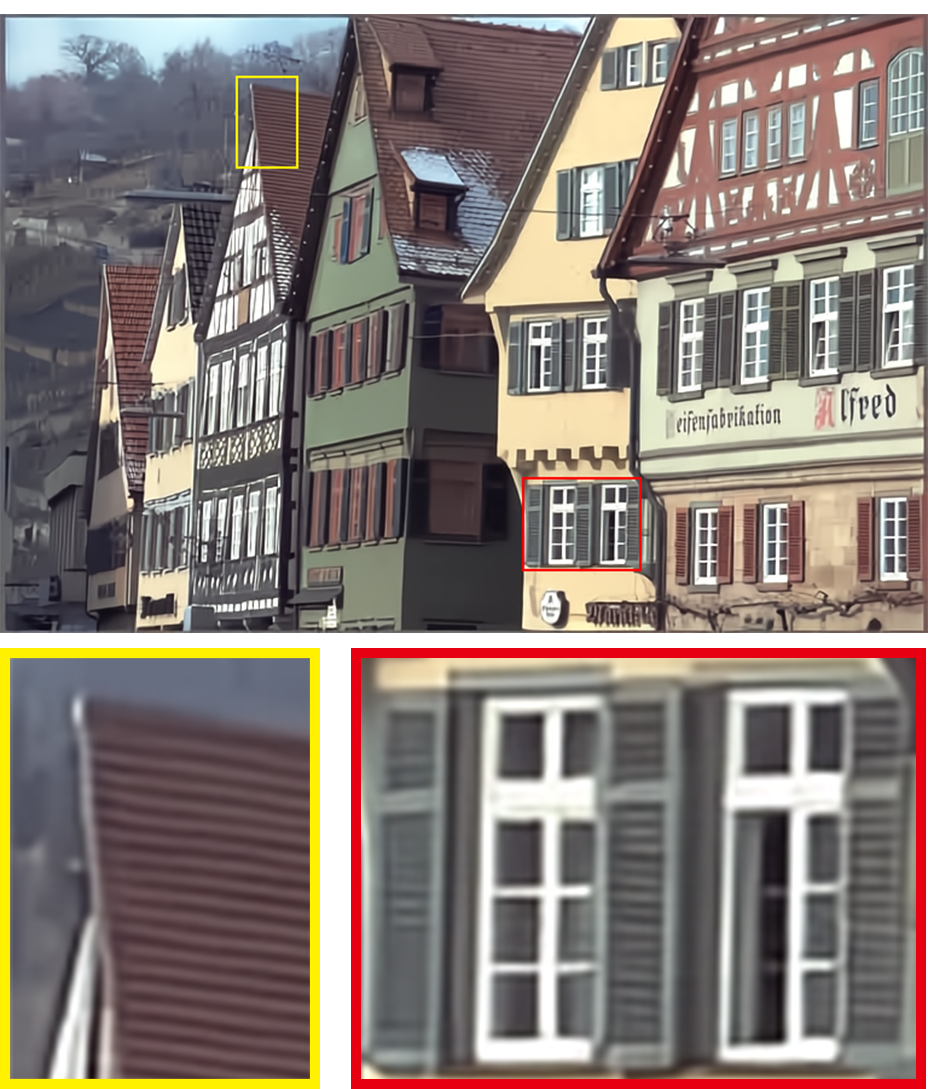}%
    \label{fig5:subfig_e}
  }
  \hspace{0.01cm}
  \subfloat[Ground Truth (PSNR / SSIM)]{
    \includegraphics[width=0.32\textwidth]{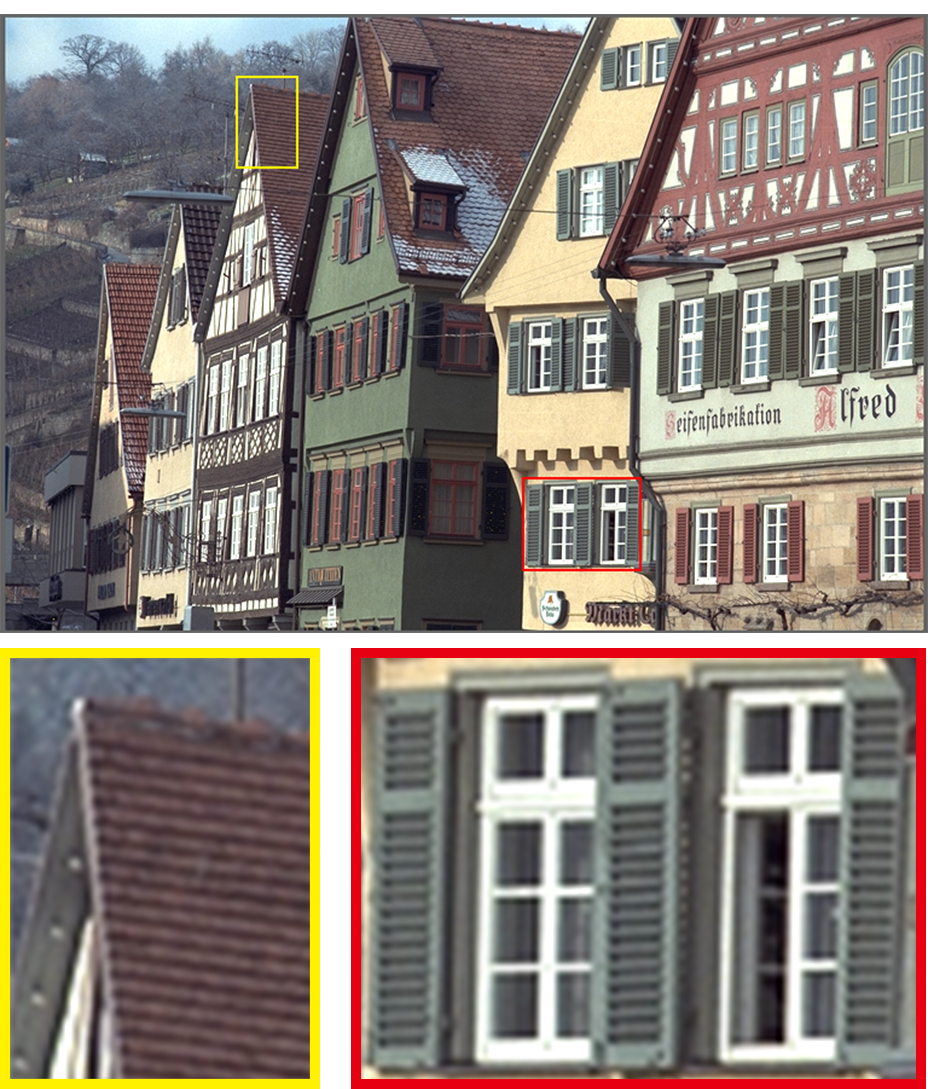}%
    \label{fig5:subfig_f}
  }
  \caption{Recovery comparisons of "LIVE1: buildings.bmp" with JPEG compression quality factor = 10. (a) compressed JPEG in 23.63 dB and SSIM 0.790. (b) DnCNN-3 \cite{DNCNN} in 24.93 dB and SSIM 0.753. (c) QGAC \cite{QGAC} method in 25.78 dB and SSIM 0.815. (d) FBCNN \cite{FBCNN} method in 25.99 dB and SSIM 0.814. (e) our DCTransformer in 26.23 dB and SSIM 0.820. (f) the original image. Note that our method provides a less smoothing result with more natural textures (on window shutters and the rooftop), and less incorrect color blurring (red-colored above the sky in (a), (b), and (d)). Please zoom in to view the details.}
  \label{fig_5}
\end{figure*}

\textit{4) Feature Concatenation in SFTB and Building DCTransformer Block: }
Low quality factors correspond to higher values in the quantization matrix, which bring many zeros (e.g. approximately 95\% are zeros at QF=10, about 90\% are zeros at QF=20) in quantized coefficients, especially more obvious for high-frequency components. Due to so many invalid tokens caused by this sparsity, we apply the feature concatenation to fuse the feature maps from two branches in SFTB. Consider $Y_{spat}$ and $Y_{freq}$ as the output feature maps of two branches. Instead of adding them together as $Y_{spat} + Y_{freq}$, we concatenate them along the channel dimension as $[Y_{spat}, Y_{freq}]$ and pass through a convolution operation. Moreover, we incorporate residual connections \cite{He_2016_CVPR} in each branch to further alleviate the sparsity problem. We can formulate the combined operations of fusion as:
\begin{equation}
Z_{out} = H_{\text{CONV}}\left([Y_{spat} + X_{in},~ Y_{freq} + X_{in}]\right),
\end{equation}
where $Z_{out}$ is the output feature map, $H_{\text{CONV}} \left( \cdot \right)$ is the $3 \times 3$ convolutional layer in SFTB, $X_{in}$, $Y_{spat}$, $Y_{freq}$ are the input, output feature maps from spatial and frequential branch, respectively. This helps both reserve the previous information and avoid too much invalid information. Moreover, utilizing a $3 \times 3$ convolution contributes to the integration of local neighboring information in two branches. Replacing the adding function with feature concatenation followed by convolution leads to obviously superior results, as verified in Section IV-E of our ablation studies. 

As shown in Figure \ref{fig_2}, our DCTransformer is comprised of a series of stacked DCTransformer blocks. Each DCTransformer block is a residual block with several RSTBs and a convolutional layer before the residual connection. Given intermediate features $F_1, F_2, \dots, F_K$ in the $i$-th DCTransformer Block, the entire DCTransformer Block with $K$ SFTBs can be expressed as:
\begin{equation}
\begin{gathered}
    F_{i,j} = \text{SFTB}_{i,j}\left( F_{i,j-1} \right),~~j=1,2,\dots,K,\\
    F_{out,i} = H_{\text{CONV}_{i}} \left( F_{i,K} \right) + F_{i,1},
\end{gathered}
\end{equation}
where $H_{\text{CONV}} \left( \cdot \right)$ is the $3 \times 3$ convolutional layer in the SFTB. Using such a convolutional layer followed by Transformer-based feature extraction can introduce the translational equivalence of the convolution operation. This composition design takes advantage of the strengths of residual learning \cite{He_2016_CVPR}, also providing more aggregation of features at different stages in the network.

\subsection{Dual-domain Loss Function}
To align the optimization objective of our DCT domain model with the pixel domain and reach a better visual effect, we propose a dual-domain loss to balance the target between pixel and frequency domains. Either a single pixel domain loss or a frequency domain loss usually leads to worse results, as verified in Section IV-E. For the pixel domain, we utilize the Charbonnier loss \cite{charbonnier1994two} expressed as:
\begin{equation}
L_{pixel}=\sqrt{{||\mathcal{F}(x)-\mathcal{F}(\hat{x})||^2 + \epsilon^2}^{\vphantom{\frac{hh}{hh}}}}, 
\end{equation}
where $\mathcal{F}(\cdot)$ represents reconstruction operation, converting the lossless coefficients $x$ and recovered coefficients $\hat{x}$ to the pixel domain images, and $\epsilon$ is a constant value set to $10^{-3}$ empirically. For recovered DCT coefficients in the frequency domain, we adopt the same L1 loss as \cite{ouyang2020towards}, which is:
\begin{equation}
L_{freq}=||x-\hat{x}||_1.
\end{equation}

To sum both up, the proposed dual-domain loss function is defined as:
\begin{equation}
L_{dual} = L_{freq} + \lambda L_{pixel},
\end{equation}
where the $\lambda$ donates a balancing ratio considering the different value ranges, which is set as 255 for our experiment.

\section{EXPERIMENTS}

\subsection{Implement Details}

{\bf{Dataset Preparation.}}
For the training stage, we use DIV2K and Flickr2k \cite{agustsson2017ntire} as training sets, which contain 800 and 2650 high-quality images. For experimental evaluation and comparisions, we use LIVE1 \cite{sheikh2006statistical} \cite{HRSheikhLIVE}, Classic-5 \cite{zeyde2012single}, BSDS500 \cite{arbelaez2010contour}, 8-bit ICB \cite{Rawzor}, and Urban100 \cite{HuangCVPR2015} datasets. For grayscale experiments, we first convert color images from RGB into YCbCr colorspace, and then we use the Y channel for the grayscale JPEG recovery. 

{\bf{Training Settings.}}
We use the Adam\cite{ADAM} optimizer with default parameters $\beta_1=0.9$ and $\beta_2=0.99$ during all training procedures. Since the input sent to the network is filled with a large part of zeros, we adopt gradient clipping with a maximum value of $0.2$ to stabilize the training procedure. The learning rate is set at $1 \times 10^{-4}$ from the start, and gradually rises to $4 \times 10^{-4}$ in the first quarter of epochs. Then we apply a cosine annealing decay strategy\cite{loshchilov2016sgdr}, decreasing the learning rate to $1 \times 10^{-5}$. For fine-tuning our augmented DCTransformer-A for double JPEG compression, we set the starting learning rate to $5 \times 10^{-5}$ and again apply the cosine annealing decaying to $1 \times 10^{-5}$. The batch size is set to 96. We implement our model on the PyTorch 1.7.1 framework \cite{paszke2019pytorch}, and train all models on six NVIDIA 3090 GPUs.

During training procedures, we generate quantized and lossless coefficients as training pairs from high-quality images. We first randomly crop the patches with the size $256 \times 256$ pixels, then encode these cropped patches into JPEG images at quality factors from 10 to 100. Random rotation and flipping are applied as augmentation methods. The encoding and decoding processes are facilitated by the OpenCV library\cite{opencv}, which internally leverages the libjpeg by Independent JPEG Group \cite{libjpeg} for efficient JPEG data handling. Since our method processes images entirely in the DCT domain, we use \textit{torchjpeg.codec.read\_coefficients} \cite{QGAC} to decode the quantized DCT coefficients and quantization matrix from compressed JPEG images.

\subsection{Pixel Domain Evaluation}

{\textit{1) Performance on color JPEG quantized coefficient recovery.}}

We compare the proposed method with other state-of-the-art methods for JPEG quantized coefficient recovery on three common test datasets LIVE1, BSDS500, and ICB. We calculate the PSNR, SSIM \cite{wang2004image}, and PSNR-B \cite{yim2010quality} for pixel domain evaluation. The quantitative results are reported in Table \ref{tab_1}. To emphasize the concept of frequency domain learning, we use marker * to refer to methods of recovering the quantization coefficients in the DCT domain. Note that all the results of our DCTransformer are predicted from the single model. Besides, the marker $^{\prime}$ specifies the method of training each model for a specific quality factor. As can be seen from Table \ref{tab_1}, our method has significantly better results than the state-of-the-art DCT domain method and moderately better results than FBCNN\cite{FBCNN} and ARCRL\cite{wang2022jpeg}, which are the state-of-the-art methods in the pixel domain. Of particular note, our deblocking results are substantially superior on the PSNR-B metric, which suggests our effectiveness in eliminating the artifacts and blockiness. We also provide a subjective comparison of the recovered images of "buildings.bmp" from the LIVE1 dataset. As presented in Figure \ref{fig_5}, our model can alleviate the visual artifacts and restore richer high-frequency textures, \textit{i.e.} on the shutter and the rooftop, while other methods suffer from over-smoothness and their recovered images have less natural details.

\begin{table*}[!t]
\caption{Pixel domain evaluation on \textbf{Grayscale} JPEG images from three benchmark datasets. Represented in \textbf{PSNR(dB)/SSIM/PSNR-B(dB)} format. The best and second best performances are \textbf{Boldfaced} and \underline{Underlined}, respectively. Please note that marker $ ^\prime$ represents quality factor specified methods, and marker $ ^*$ represents DCT domain methods.\label{tab:table_result}}
\centering
\setlength{\tabcolsep}{1pt} 
\begin{tabular}{c c c c c c c c}
\toprule
\multicolumn{1}{c|}{Dataset} & \multicolumn{1}{c|}{QF}  & \multicolumn{1}{c|}{JPEG} & \multicolumn{1}{c|}{$\text{ARCNN}^{\prime}$\cite{ARCNN}} & \multicolumn{1}{c|}{$\text{DnCNN-3}$\cite{DNCNN}} & \multicolumn{1}{c|}{$\text{MWCNN}^{\prime}$\cite{MWCNN}}  & \multicolumn{1}{c|}{DCSC\cite{DCSC}} & \multicolumn{1}{c}{$\text{RNAN}^{\prime}$\cite{zhang2019rnan}}  \\ 
\midrule
\midrule
\multicolumn{1}{c}{\multirow{4}{*}{Classic-5}}&
\multicolumn{1}{|c|}{10}&
\multicolumn{1}{c|}{27.82/0.760/25.21}&
\multicolumn{1}{c|}{29.03/0.793/28.76}&  
\multicolumn{1}{c|}{29.40/0.803/29.13}&
\multicolumn{1}{c|}{30.01/0.820/29.59}&
\multicolumn{1}{c|}{29.62/0.810/29.30}&
\multicolumn{1}{c}{29.96/0.819/29.42}
\\
&
\multicolumn{1}{|c|}{20}&
\multicolumn{1}{c|}{30.12/0.834/27.50}&
\multicolumn{1}{c|}{31.15/0.852/30.59}&  
\multicolumn{1}{c|}{31.63/0.861/31.19}&
\multicolumn{1}{c|}{32.16/{0.870}/31.52}&
\multicolumn{1}{c|}{31.81/0.864/31.34}&
\multicolumn{1}{c}{32.11/0.869/31.26}
\\
&
\multicolumn{1}{|c|}{30}&
\multicolumn{1}{c|}{31.48/0.867/28.94}&
\multicolumn{1}{c|}{32.51/0.881/31.98}&  
\multicolumn{1}{c|}{32.91/0.886/32.38}&
\multicolumn{1}{c|}{33.43/{0.893}/{32.62}}&
\multicolumn{1}{c|}{33.06/0.888/32.49}&
\multicolumn{1}{c}{33.38/0.892/32.35}
\\
&
\multicolumn{1}{|c|}{40}&
\multicolumn{1}{c|}{32.43/0.885/29.92}&
\multicolumn{1}{c|}{33.32/0.895/32.79}&  
\multicolumn{1}{c|}{33.77/0.900/33.23}&
\multicolumn{1}{c|}{34.27/{0.906}/33.35}&
\multicolumn{1}{c|}{33.87/0.902/33.30}&
\multicolumn{1}{c}{34.27/0.906/33.40}
\\
\midrule
\multicolumn{1}{c}{\multirow{4}{*}{LIVE1}}&
\multicolumn{1}{|c|}{10}&
\multicolumn{1}{c|}{27.77/0.773/25.33}&
\multicolumn{1}{c|}{28.96/0.808/28.68}&  
\multicolumn{1}{c|}{29.19/0.812/28.90}&
\multicolumn{1}{c|}{29.69/0.825/29.32}&
\multicolumn{1}{c|}{29.34/0.818/29.01}&
\multicolumn{1}{c}{29.63/0.824/29.13}
\\
&
\multicolumn{1}{|c|}{20}&
\multicolumn{1}{c|}{30.07/0.851/27.57}&
\multicolumn{1}{c|}{31.29/0.873/30.76}&  
\multicolumn{1}{c|}{31.59/0.880/31.07}&
\multicolumn{1}{c|}{32.04/\underline{0.889}/31.51}&
\multicolumn{1}{c|}{31.70/0.883/31.18}&
\multicolumn{1}{c}{32.03/0.888/31.12}
\\
&
\multicolumn{1}{|c|}{30}&
\multicolumn{1}{c|}{31.41/0.885/28.92}&
\multicolumn{1}{c|}{32.67/0.904/32.14}&  
\multicolumn{1}{c|}{32.98/0.909/32.34}&
\multicolumn{1}{c|}{33.45/{0.915}/32.80}&
\multicolumn{1}{c|}{33.07/0.911/32.43}&
\multicolumn{1}{c}{33.45/0.915/32.22}
\\
&
\multicolumn{1}{|c|}{40}&
\multicolumn{1}{c|}{32.35/0.904/29.96}&
\multicolumn{1}{c|}{33.61/0.920/33.11}&  
\multicolumn{1}{c|}{33.96/0.925/33.28}&
\multicolumn{1}{c|}{34.45/{0.930}/\underline{33.78}}&
\multicolumn{1}{c|}{34.02/0.926/33.36}&
\multicolumn{1}{c}{34.47/0.930/33.66}
\\
\midrule
\multicolumn{1}{c}{\multirow{4}{*}{BSDS500}}&
\multicolumn{1}{|c|}{10}&
\multicolumn{1}{c|}{27.80/0.768/25.10}&
\multicolumn{1}{c|}{29.10/0.804/28.73}&  
\multicolumn{1}{c|}{29.21/0.809/28.80}&
\multicolumn{1}{c|}{29.61/{0.820}/29.14}&
\multicolumn{1}{c|}{29.32/0.813/28.91}&
\multicolumn{1}{c}{29.08/0.805/28.48}
\\
&
\multicolumn{1}{|c|}{20}&
\multicolumn{1}{c|}{30.05/0.849/27.22}&
\multicolumn{1}{c|}{31.28/0.870/30.55}&  
\multicolumn{1}{c|}{31.53/0.878/30.79}&
\multicolumn{1}{c|}{{31.92}/\underline{0.885}/31.15}&
\multicolumn{1}{c|}{31.63/0.880/30.92}&
\multicolumn{1}{c}{31.25/0.875/30.27}
\\
&
\multicolumn{1}{|c|}{30}&
\multicolumn{1}{c|}{31.37/0.884/28.53}&
\multicolumn{1}{c|}{32.67/0.902/31.94}&  
\multicolumn{1}{c|}{32.90/0.907/31.97}&
\multicolumn{1}{c|}{33.30/{0.912}/{32.34}}&
\multicolumn{1}{c|}{32.99/0.908/32.08}&
\multicolumn{1}{c}{32.70/0.907/31.33}
\\
&
\multicolumn{1}{|c|}{40}&
\multicolumn{1}{c|}{32.30/0.903/29.49}&
\multicolumn{1}{c|}{33.55/0.918/32.78}&  
\multicolumn{1}{c|}{33.85/0.923/32.80}&
\multicolumn{1}{c|}{{34.27}/\underline{0.928}/\underline{33.19}}&
\multicolumn{1}{c|}{33.92/0.924/32.92}&
\multicolumn{1}{c}{33.47/0.923/32.27}
\\
\bottomrule

\\

\toprule
\multicolumn{1}{c|}{Dataset} & \multicolumn{1}{c|}{QF} & \multicolumn{1}{c|}{FBCNN\cite{FBCNN}}  & \multicolumn{1}{c||}{\text{ARCRL}\cite{wang2022jpeg}} & \multicolumn{1}{c|}{Laplacian$ ^*$\cite{price1999biased}} & \multicolumn{1}{c|}{FD-CRNet$^*$\cite{ouyang2020towards}} & \multicolumn{1}{c|}{QGAC$^*$\cite{QGAC}} & \multicolumn{1}{c}{DCTransformer$^*$} \\ 
\midrule
\midrule
\multicolumn{1}{c}{\multirow{4}{*}{Classic-5}}&
\multicolumn{1}{|c|}{10}&
\multicolumn{1}{c|}{\underline{30.12}/\textbf{0.822}/{29.80}}&
\multicolumn{1}{c||}{\textbf{30.16}/\textbf{0.822}/\textbf{29.85}}&
\multicolumn{1}{c|}{28.04/0.763/25.44}&  
\multicolumn{1}{c|}{29.54/0.806/29.14}&  
\multicolumn{1}{c|}{29.84/0.812/29.43}&
\multicolumn{1}{c}{\underline{30.12}/\textbf{0.822}/\underline{29.81}}\\
&
\multicolumn{1}{|c|}{20}&
\multicolumn{1}{c|}{{32.31}/\underline{0.872}/{31.74}}&
\multicolumn{1}{c||}{\textbf{32.37}/\textbf{0.873}/\textbf{31.84}}&
\multicolumn{1}{c|}{30.39/0.838/27.78}&
\multicolumn{1}{c|}{31.64/0.861/30.88}&
\multicolumn{1}{c|}{31.98/0.869/31.37}&
\multicolumn{1}{c}{\underline{32.33}/\underline{0.872}/\underline{31.81}}\\
&
\multicolumn{1}{|c|}{30}&
\multicolumn{1}{c|}{{33.54}/\underline{0.894}/{32.78}}&
\multicolumn{1}{c||}{\textbf{33.60}/\textbf{0.895}/\textbf{32.89}}&
\multicolumn{1}{c|}{31.83/0.871/29.30}&
\multicolumn{1}{c|}{32.84/0.885/32.03}&
\multicolumn{1}{c|}{33.22/0.892/32.42}&
\multicolumn{1}{c}{\underline{33.55}/\underline{0.894}/\underline{32.84}}\\
&
\multicolumn{1}{|c|}{40}&
\multicolumn{1}{c|}{{34.35}/\underline{0.907}/33.48}&
\multicolumn{1}{c||}{\textbf{34.43}/\textbf{0.908}/\textbf{33.58}}&
\multicolumn{1}{c|}{32.82/0.889/30.35}&
\multicolumn{1}{c|}{33.64/0.899/32.82}&
\multicolumn{1}{c|}{34.05/0.905/33.12}&
\multicolumn{1}{c}{\underline{34.37}/\underline{0.907}/\underline{33.53}}\\
\midrule

\multicolumn{1}{c}{\multirow{4}{*}{LIVE1}}&
\multicolumn{1}{|c|}{10}&
\multicolumn{1}{c|}{{29.75}/\textbf{0.827}/{29.40}}&
\multicolumn{1}{c||}{\textbf{29.80}/\textbf{0.827}/\textbf{29.44}}&
\multicolumn{1}{c|}{27.98/0.777/25.81}&
\multicolumn{1}{c|}{29.24/0.813/28.88}&  
\multicolumn{1}{c|}{29.51/0.825/29.13}&
\multicolumn{1}{c}{\underline{29.76}/\textbf{0.827}/\textbf{29.44}}\\
&
\multicolumn{1}{|c|}{20}&
\multicolumn{1}{c|}{{32.13}/\underline{0.889}/{31.57}}&
\multicolumn{1}{c||}{\textbf{32.19}/\textbf{0.890}/\textbf{31.63}}&
\multicolumn{1}{c|}{30.31/0.856/28.09}&
\multicolumn{1}{c|}{31.52/0.880/30.73}&
\multicolumn{1}{c|}{31.83/0.888/31.25}&
\multicolumn{1}{c}{\underline{32.14}/\underline{0.889}/\textbf{31.63}}\\
&
\multicolumn{1}{|c|}{30}&
\multicolumn{1}{c|}{\underline{33.54}/\underline{0.916}/32.83}&
\multicolumn{1}{c||}{\textbf{33.62}/\textbf{0.918}/\textbf{32.91}}&
\multicolumn{1}{c|}{31.67/0.890/29.43}&
\multicolumn{1}{c|}{32.84/0.908/31.96}&
\multicolumn{1}{c|}{33.20/0.914/32.47}&
\multicolumn{1}{c}{\underline{33.54}/\underline{0.916}/\underline{32.89}}\\
&
\multicolumn{1}{|c|}{40}&
\multicolumn{1}{c|}{\underline{34.53}/\underline{0.931}/33.74}&
\multicolumn{1}{c||}{\textbf{34.62}/\textbf{0.931}/\textbf{33.84}}&
\multicolumn{1}{c|}{32.66/0.911/30.54}&
\multicolumn{1}{c|}{33.76/0.923/32.92}&
\multicolumn{1}{c|}{34.16/0.929/33.36}&
\multicolumn{1}{c}{34.52/\underline{0.931}/\underline{33.79}}\\
\midrule

\multicolumn{1}{c}{\multirow{4}{*}{BSDS500}}&
\multicolumn{1}{|c|}{10}&
\multicolumn{1}{c|}{\underline{29.67}/\underline{0.821}/{29.22}}&
\multicolumn{1}{c||}{\textbf{29.70}/\textbf{0.822}/\textbf{29.27}}&
\multicolumn{1}{c|}{27.97/0.771/25.29}&
\multicolumn{1}{c|}{29.19/0.807/28.68}&
\multicolumn{1}{c|}{29.46/\underline{0.821}/28.97}&
\multicolumn{1}{c}{\underline{29.67}/\underline{0.821}/\underline{29.26}}\\
&
\multicolumn{1}{|c|}{20}&
\multicolumn{1}{c|}{{32.00}/\underline{0.885}/\underline{31.19}}&
\multicolumn{1}{c||}{\textbf{32.06}/\textbf{0.886}/\textbf{31.27}}&
\multicolumn{1}{c|}{30.31/0.852/27.49}&
\multicolumn{1}{c|}{31.40/0.876/30.42}&
\multicolumn{1}{c|}{31.73/0.884/30.93}&
\multicolumn{1}{c}{\underline{32.01}/\underline{0.885}/\textbf{31.27}}\\
&
\multicolumn{1}{|c|}{30}&
\multicolumn{1}{c|}{{33.37}/\underline{0.913}/32.32}&
\multicolumn{1}{c||}{\textbf{33.45}/\textbf{0.914}/\textbf{32.41}}&
\multicolumn{1}{c|}{31.65/0.888/28.82}&
\multicolumn{1}{c|}{32.70/0.905/31.58}&
\multicolumn{1}{c|}{33.07/0.912/32.04}&
\multicolumn{1}{c}{\underline{33.38}/\underline{0.913}/\underline{32.40}}\\
&
\multicolumn{1}{|c|}{40}&
\multicolumn{1}{c|}{\underline{34.33}/\underline{0.928}/33.10}&
\multicolumn{1}{c||}{\textbf{34.42}/\textbf{0.929}/\textbf{33.22}}&
\multicolumn{1}{c|}{32.62/0.909/29.83}&
\multicolumn{1}{c|}{33.59/0.921/32.42}&
\multicolumn{1}{c|}{34.01/0.927/32.81}&
\multicolumn{1}{c}{\underline{34.33}/\underline{0.928}/\underline{33.19}}\\
\bottomrule

\label{tab_2}
\end{tabular}
\end{table*}

\begin{figure*}[!t]
	\centering
	\subfloat[GT]{
	\begin{minipage}[b]{0.119\textwidth}
		\includegraphics[width=1\textwidth]{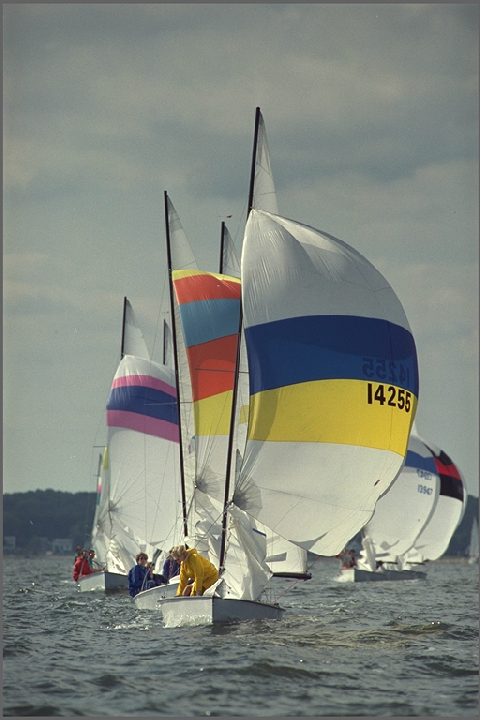}\\
        \vspace{-0.30cm}
		\includegraphics[width=1\textwidth]{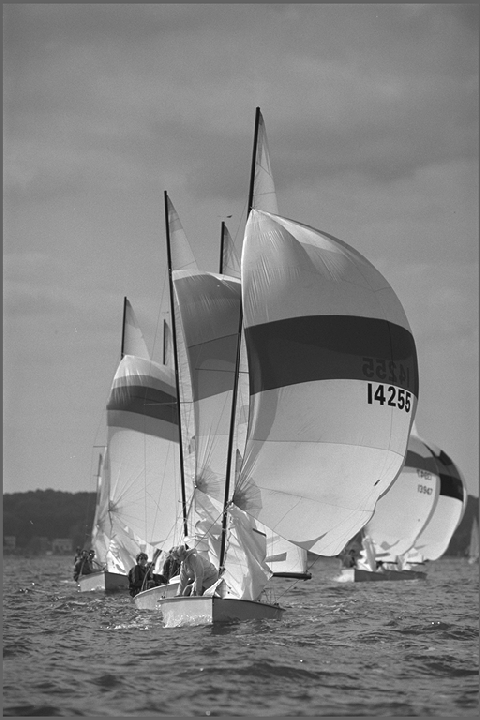}\\
        \vspace{-0.30cm}
		\includegraphics[width=1\textwidth]{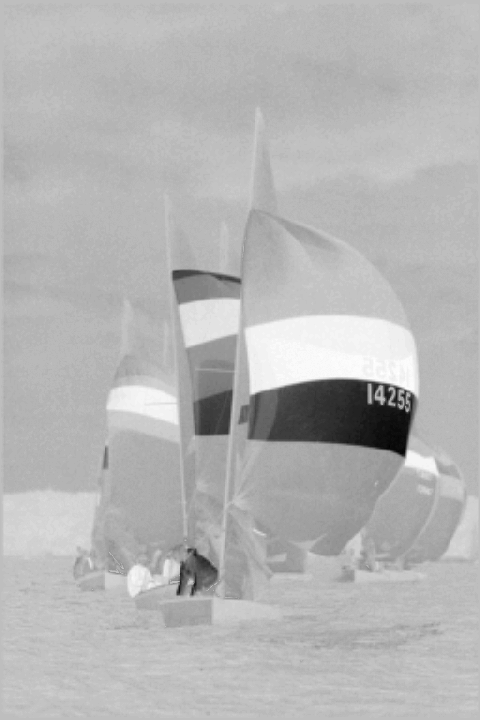}\\
        \vspace{-0.30cm}
		\includegraphics[width=1\textwidth]{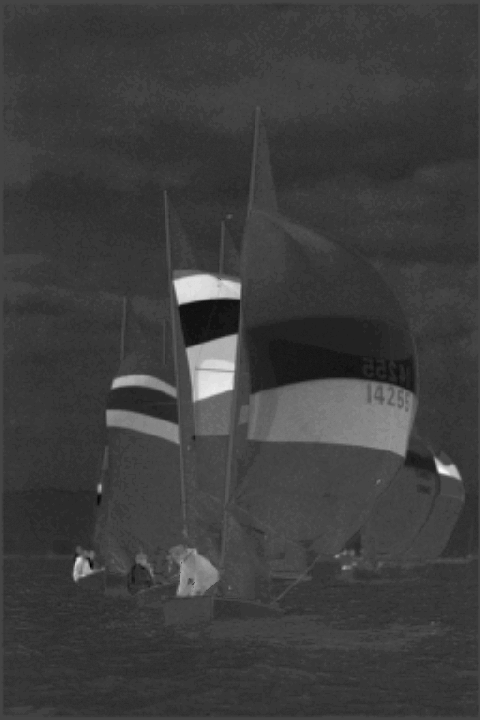}
	\end{minipage}
}
    \hspace{-0.20cm}
	\subfloat[JPEG]{
	\begin{minipage}[b]{0.119\textwidth}
		\includegraphics[width=1\textwidth]{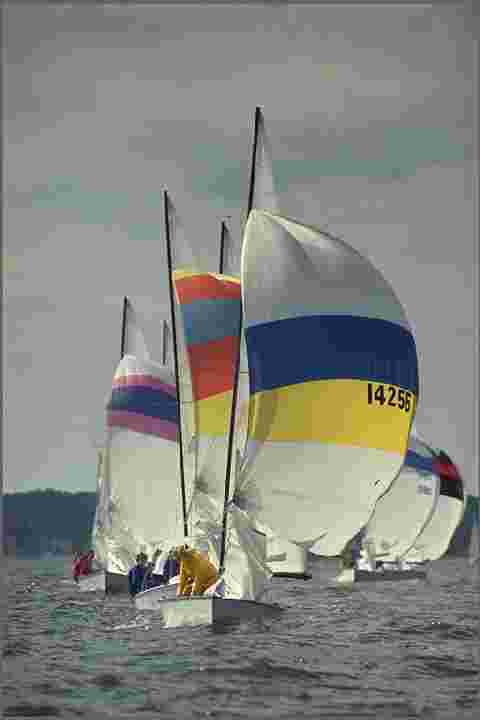}\\
        \vspace{-0.30cm}
		\includegraphics[width=1\textwidth]{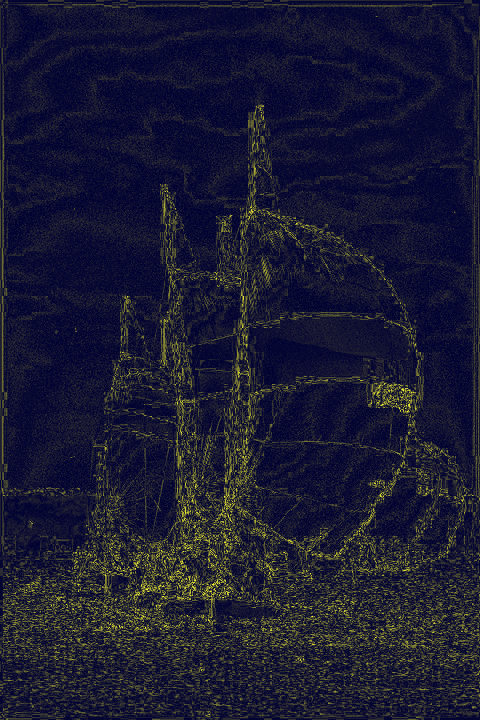}\\
        \vspace{-0.30cm}
		\includegraphics[width=1\textwidth]{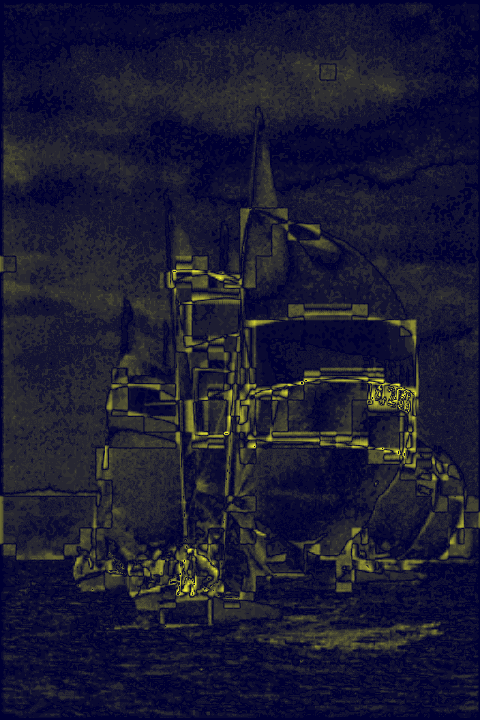}\\
        \vspace{-0.30cm}
		\includegraphics[width=1\textwidth]{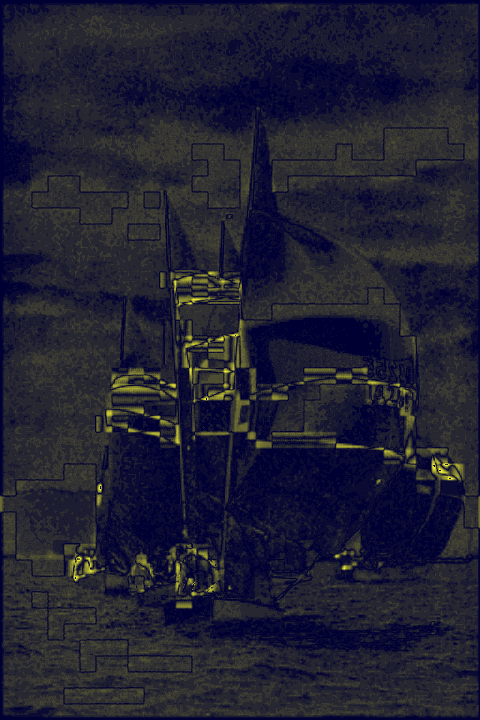}
	\end{minipage}
}
    \hspace{-0.20cm}
	\subfloat[ARCNN]{
	\begin{minipage}[b]{0.119\textwidth}
		\includegraphics[width=1\textwidth]{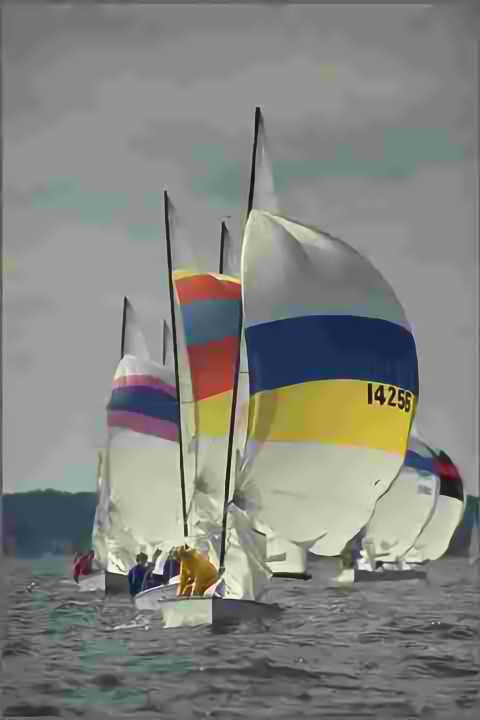}\\
        \vspace{-0.30cm}
		\includegraphics[width=1\textwidth]{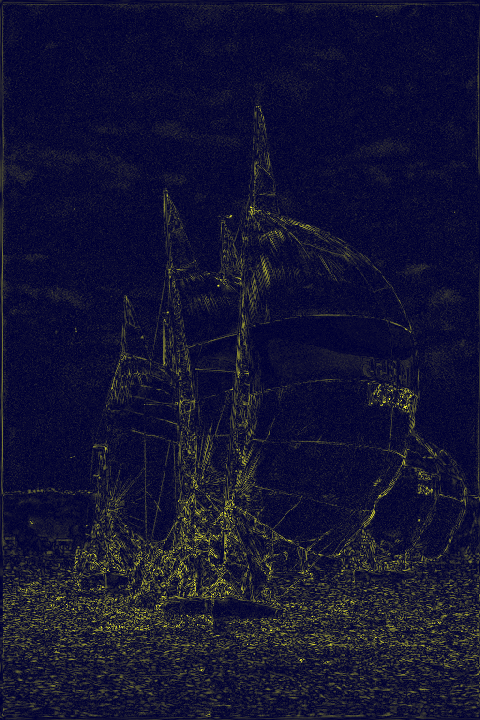}\\
        \vspace{-0.30cm}
		\includegraphics[width=1\textwidth]{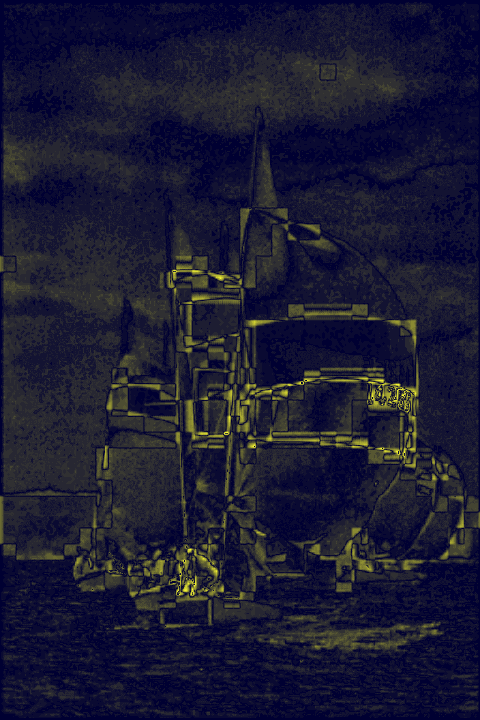}\\
        \vspace{-0.30cm}
		\includegraphics[width=1\textwidth]{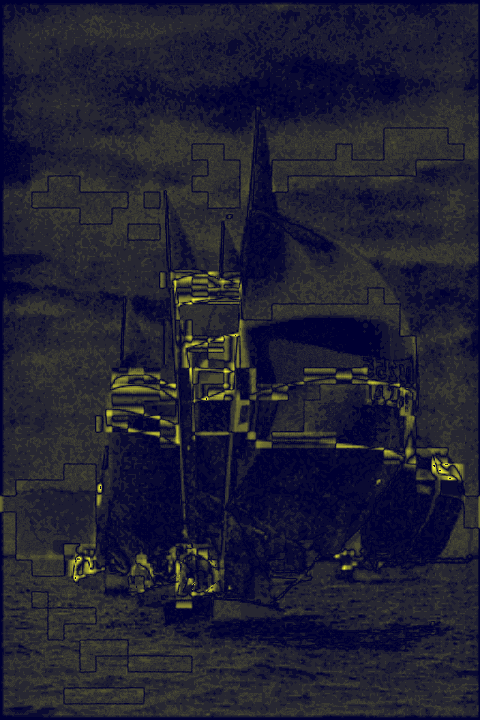}
	\end{minipage}
}
    \hspace{-0.20cm}
	\subfloat[DnCNN-3]{
	\begin{minipage}[b]{0.119\textwidth}
		\includegraphics[width=1\textwidth]{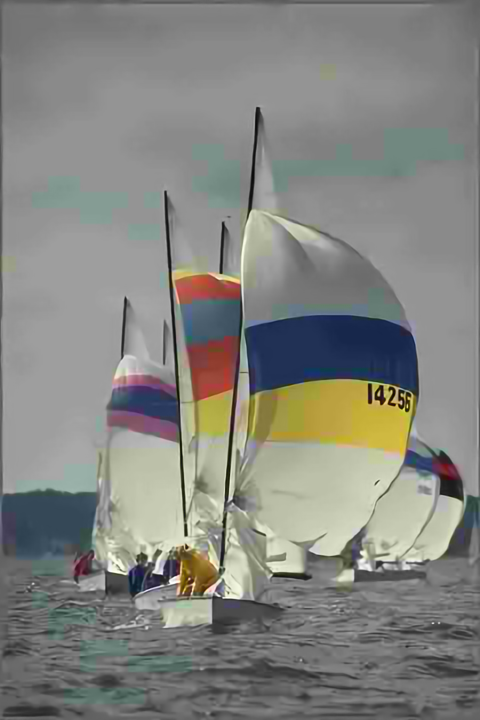}\\
        \vspace{-0.30cm}
		\includegraphics[width=1\textwidth]{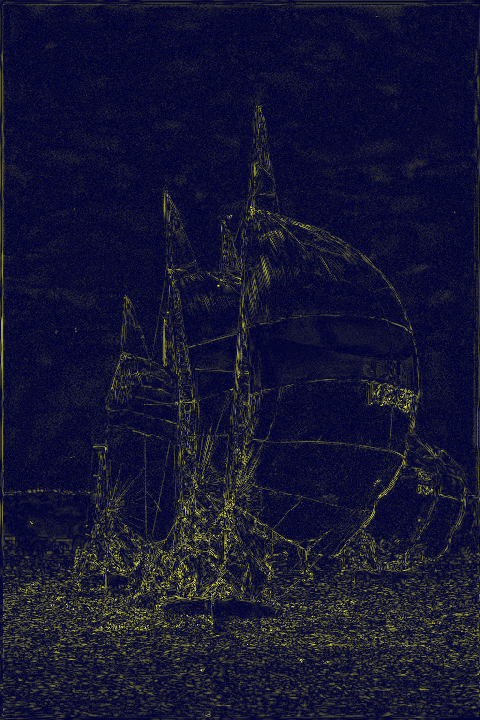}\\
        \vspace{-0.30cm}
		\includegraphics[width=1\textwidth]{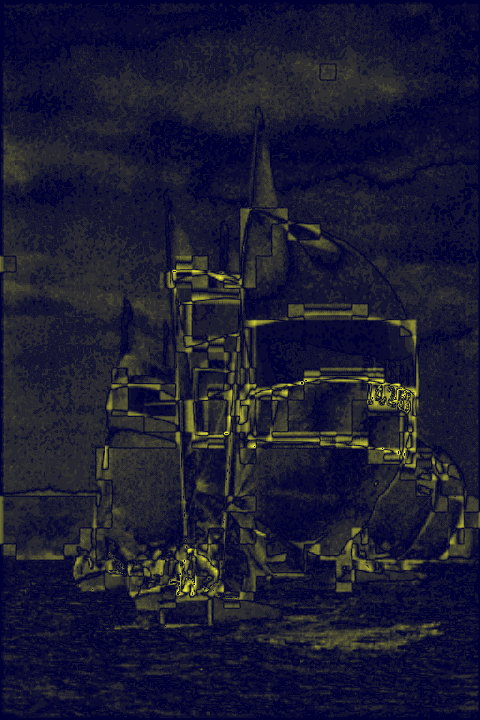}\\
        \vspace{-0.30cm}
		\includegraphics[width=1\textwidth]{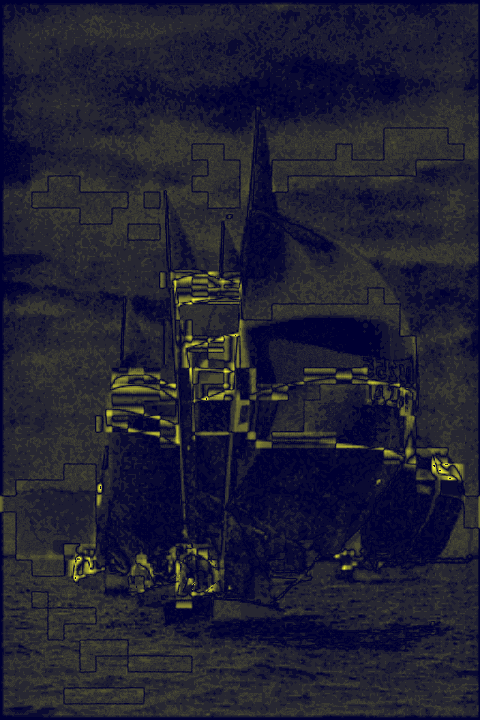}
	\end{minipage}
}
    \hspace{-0.20cm}
	\subfloat[QGCN]{
	\begin{minipage}[b]{0.119\textwidth}
		\includegraphics[width=1\textwidth]{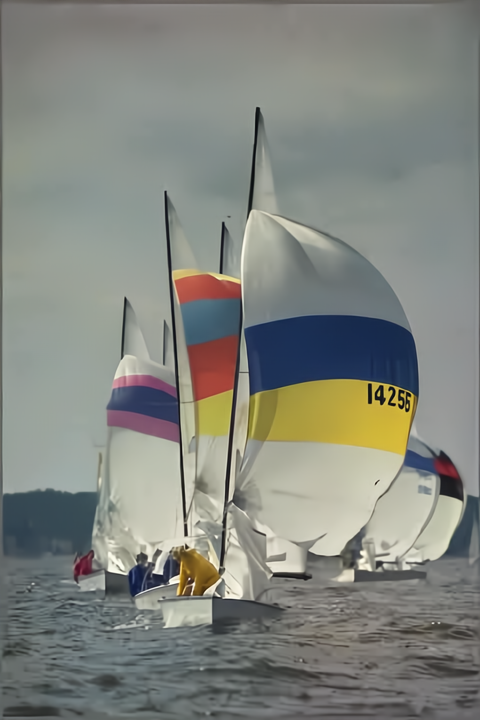}\\
        \vspace{-0.30cm}
		\includegraphics[width=1\textwidth]{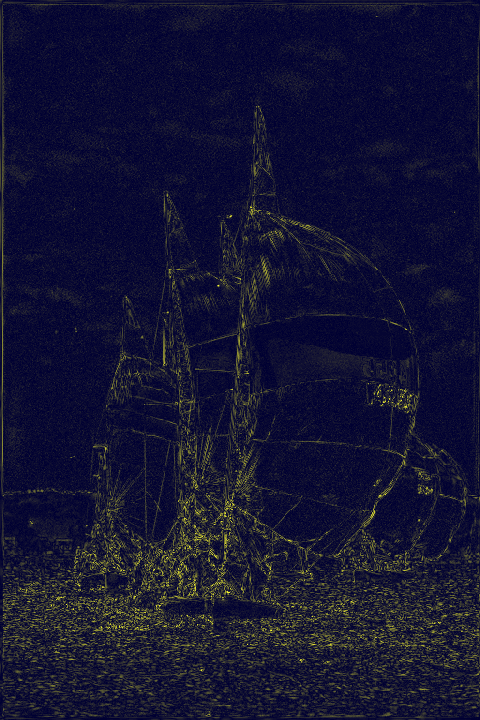}\\
        \vspace{-0.30cm}
		\includegraphics[width=1\textwidth]{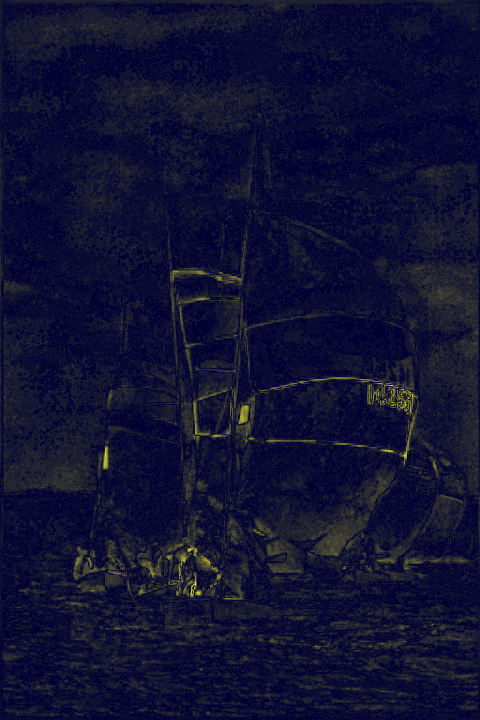}\\
        \vspace{-0.30cm}
		\includegraphics[width=1\textwidth]{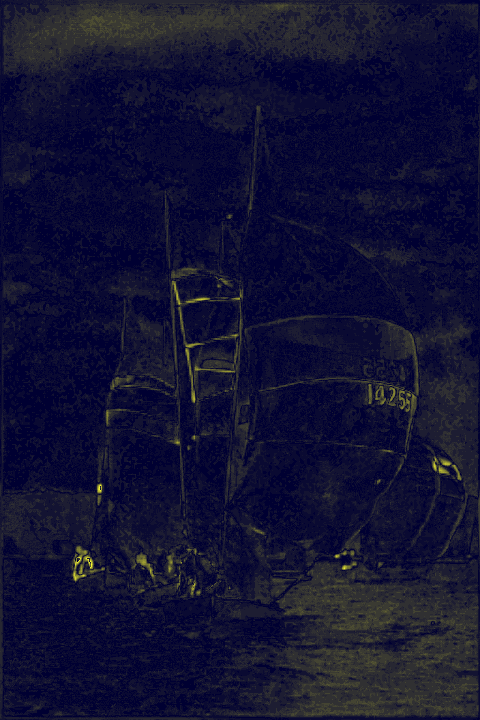}
	\end{minipage}
}
    \hspace{-0.20cm}
	\subfloat[QGAC]{
	\begin{minipage}[b]{0.119\textwidth}
		\includegraphics[width=1\textwidth]{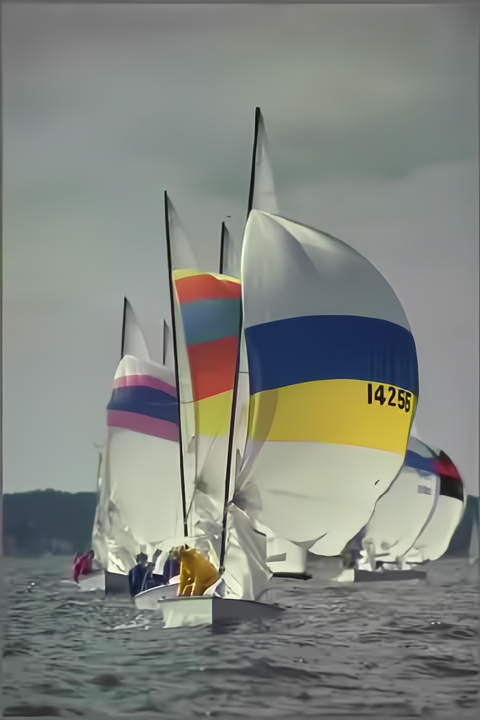}\\
        \vspace{-0.30cm}
		\includegraphics[width=1\textwidth]{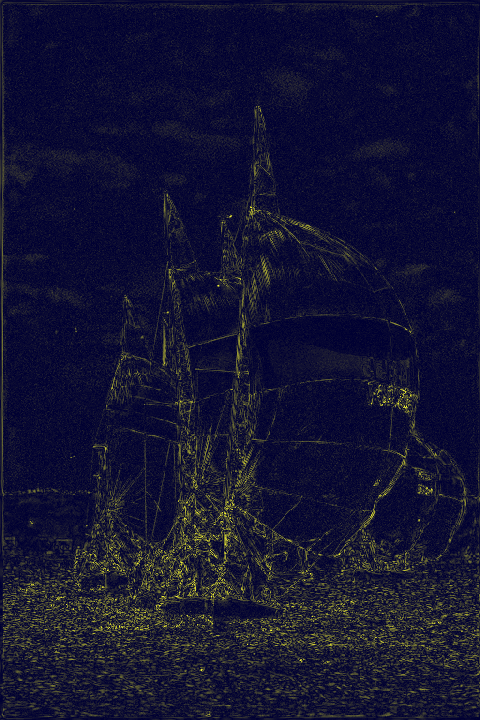}\\
        \vspace{-0.30cm}
		\includegraphics[width=1\textwidth]{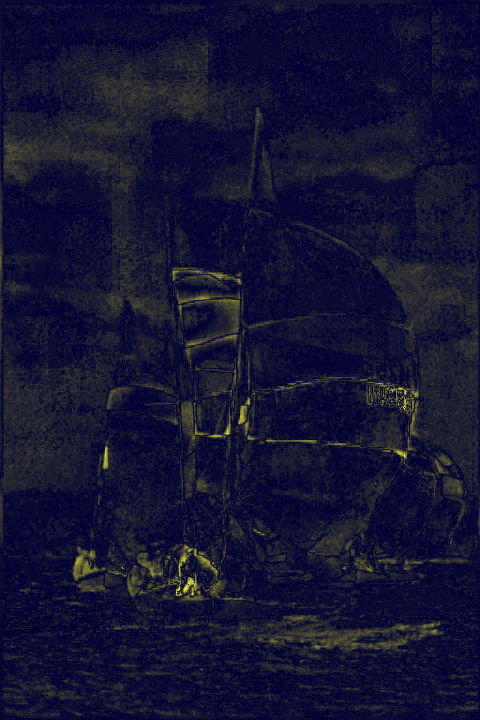}\\
        \vspace{-0.30cm}
		\includegraphics[width=1\textwidth]{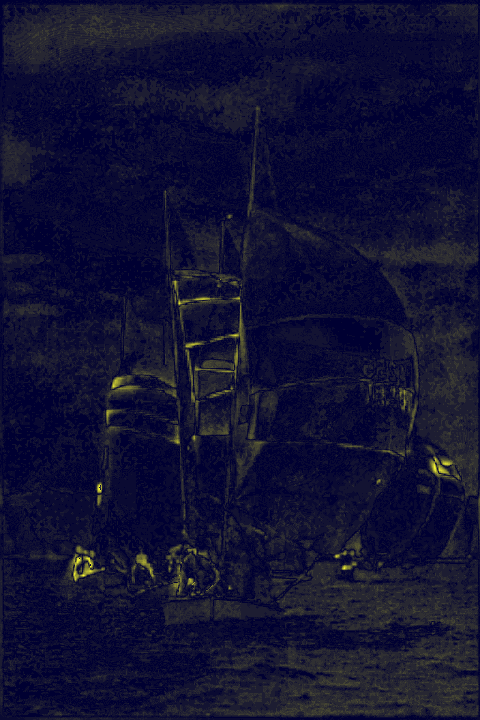}
	\end{minipage}
}
    \hspace{-0.20cm}
	\subfloat[FBCNN]{
	\begin{minipage}[b]{0.119\textwidth}
		\includegraphics[width=1\textwidth]{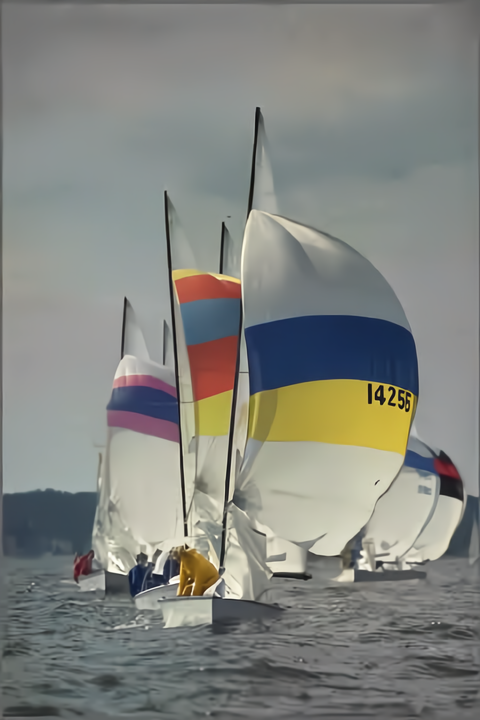}\\
        \vspace{-0.30cm}
		\includegraphics[width=1\textwidth]{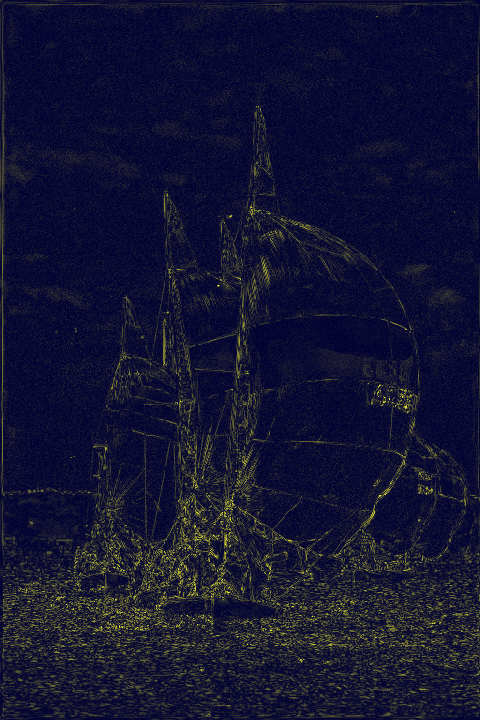}\\
        \vspace{-0.30cm}
		\includegraphics[width=1\textwidth]{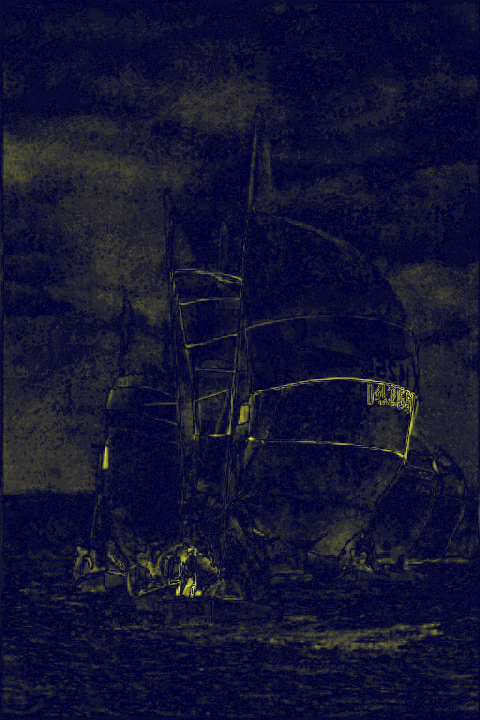}\\
        \vspace{-0.30cm}
		\includegraphics[width=1\textwidth]{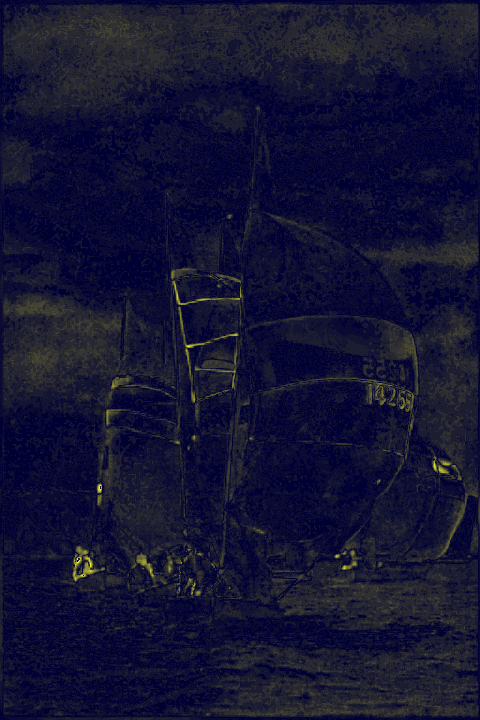}
	\end{minipage}
}
    \hspace{-0.20cm}
	\subfloat[Ours]{
	\begin{minipage}[b]{0.119\textwidth}
		\includegraphics[width=1\textwidth]{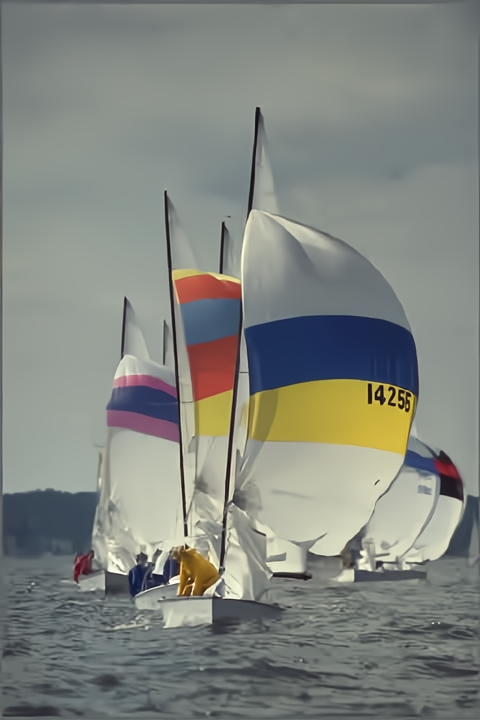}\\
        \vspace{-0.30cm}
		\includegraphics[width=1\textwidth]{dctransformer_sailing2_diff_Y.png}\\
        \vspace{-0.30cm}
		\includegraphics[width=1\textwidth]{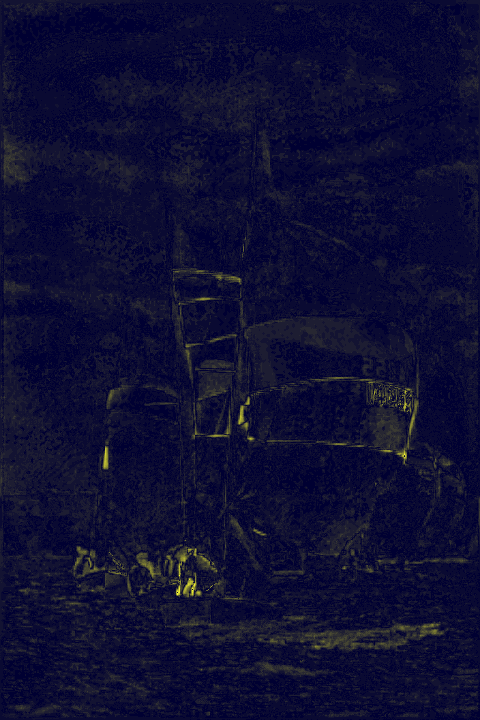}\\
        \vspace{-0.30cm}
		\includegraphics[width=1\textwidth]{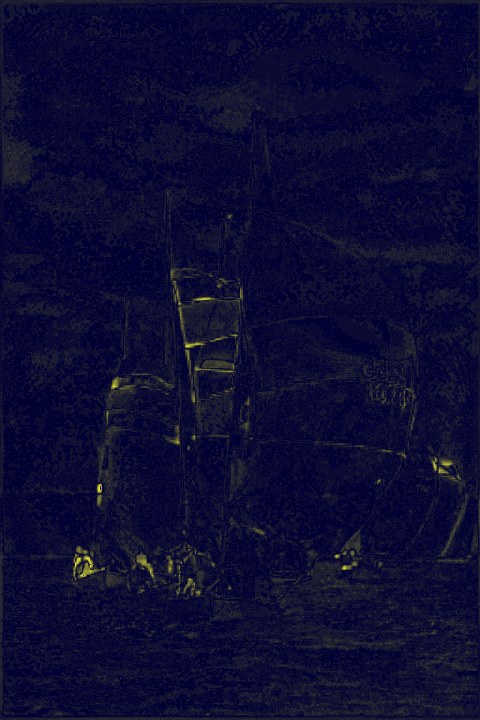}
	\end{minipage}
}
\caption{Channel residuals visualization on "LIVE1: sailing2.bmp" at the JPEG compression quality factor = 10. Each row (from top to bottom) represents the \textbf{color images}, \textbf{Y channels}, \textbf{Cb channels}, and \textbf{Cr channels}, respectively. The lighter place stands for a higher value of differences between the recovered images with ground truth. Note that although there are only slight differences in recovered Y channels, our DCTransformer performs significantly better on recovering color components (Cb and Cr), as shown in the last two rows with a darker result.}
\label{fig_6}
\end{figure*}

Figure \ref{fig_6} presents another comparison of the results for artifact removal. The channel residual refers to the difference between the recovered images and the corresponding ground truth. We convert the recovered results to YCbCr colorspace and present the residual of each channel separately. Through this example, we highlight that the proposed DCT domain method is able to handle color components quite well. It is worth emphasizing that since our method generates upsampled chrominance components in the DCT domain, the recovery result from our method shows a more consistent color with the ground truth, \textit{i.e.} less red-colored blurring in the area of the sky. That also demonstrates the design of luminance-chrominance alignment has an advantage in eliminating large area color artifacts, which eventually produces a more visually pleasing restoration.

\captionsetup[subfloat]{labelformat=empty} 
\begin{figure*}[tb]
  \centering
  \begin{minipage}{0.8\textwidth}
    \subfloat[QF=10 Compressed]{
      \includegraphics[width=0.24\textwidth]{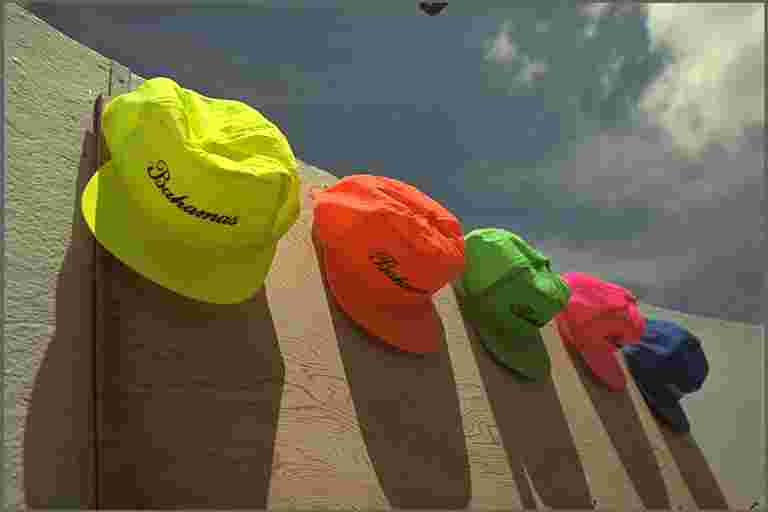}
      \label{fig7:subfig_1}
    }
    \hspace{-0.15cm}
    \subfloat[QF=20 Compressed]{
      \includegraphics[width=0.24\textwidth]{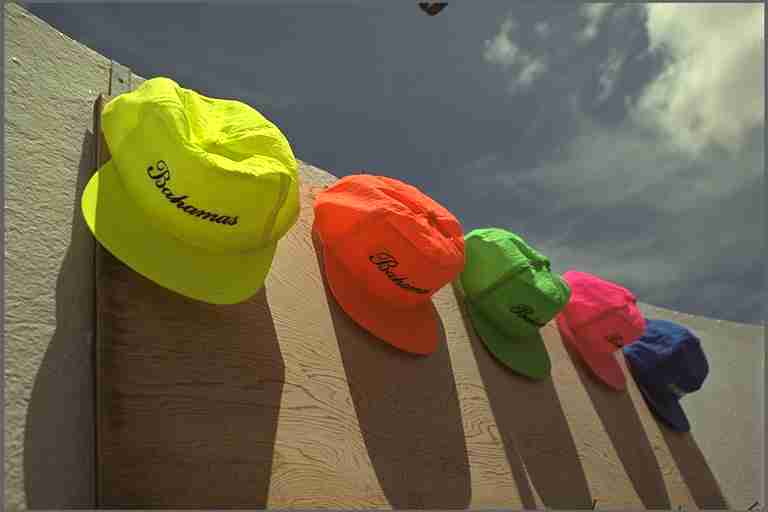}%
      \label{fig7:subfig_2}
    }
    \hspace{-0.15cm}
    \subfloat[QF=30 Compressed]{
      \includegraphics[width=0.24\textwidth]{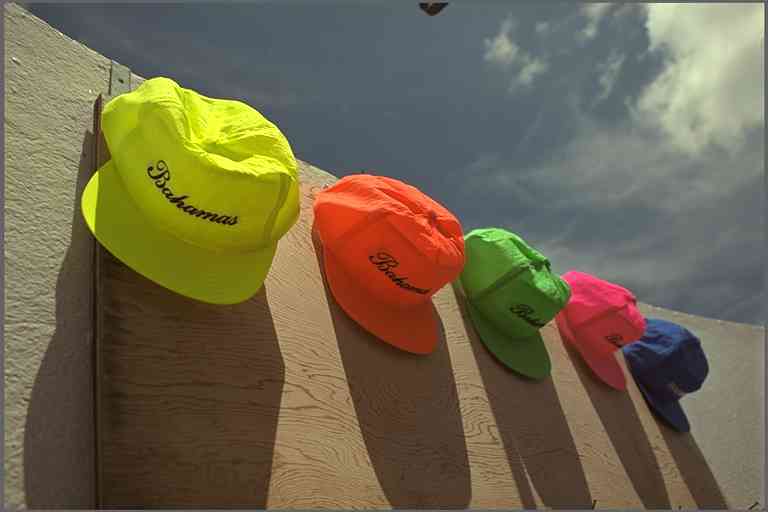}%
      \label{fig7:subfig_3}
    }
    \hspace{-0.15cm}
    \subfloat[QF=40 Compressed]{
      \includegraphics[width=0.24\textwidth]{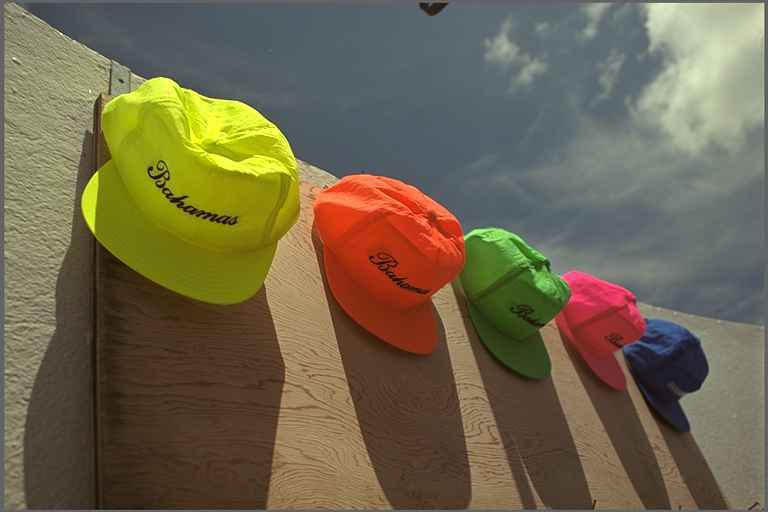}%
      \label{fig7:subfig_4}
    }
    
    \vspace{-6pt}
    
    \subfloat[QF=10 Recovered]{
      \includegraphics[width=0.24\textwidth]{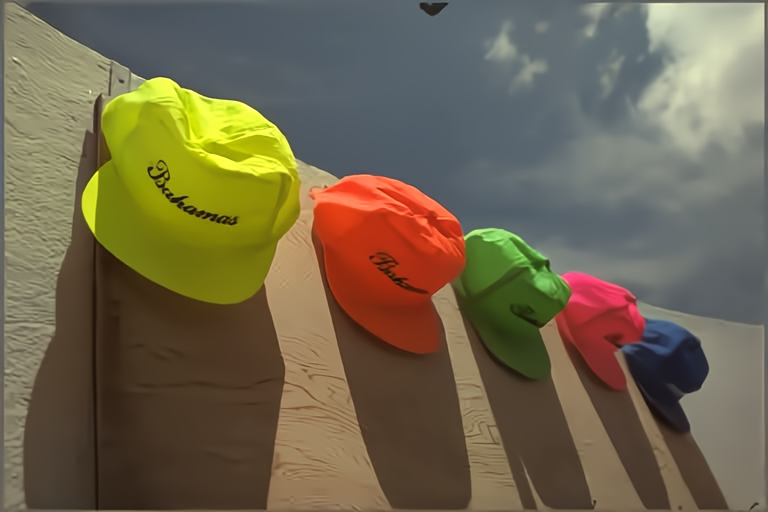}%
      \label{fig7:subfig_5}
    }
    \hspace{-0.15cm}
    \subfloat[QF=20 Recovered]{
      \includegraphics[width=0.24\textwidth]{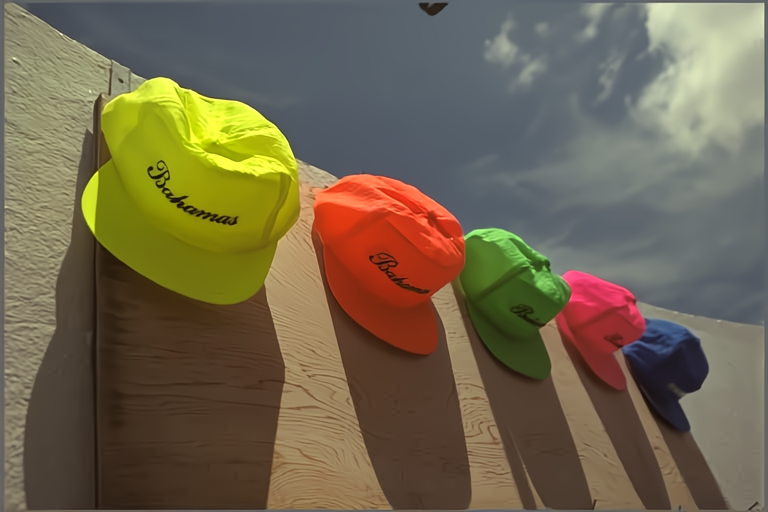}%
      \label{fig7:subfig_6}
    }
    \hspace{-0.15cm}
    \subfloat[QF=30 Recovered]{
      \includegraphics[width=0.24\textwidth]{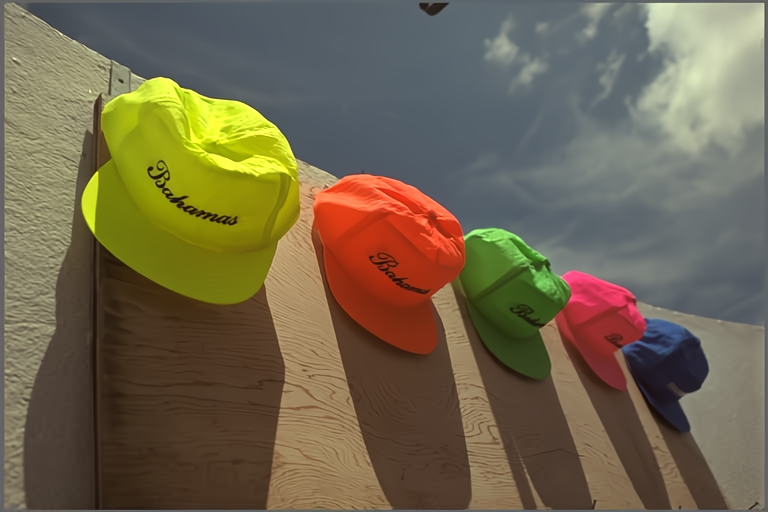}%
      \label{fig7:subfig_7}
    }
    \hspace{-0.15cm}
    \subfloat[QF=40 Recovered]{
      \includegraphics[width=0.24\textwidth]{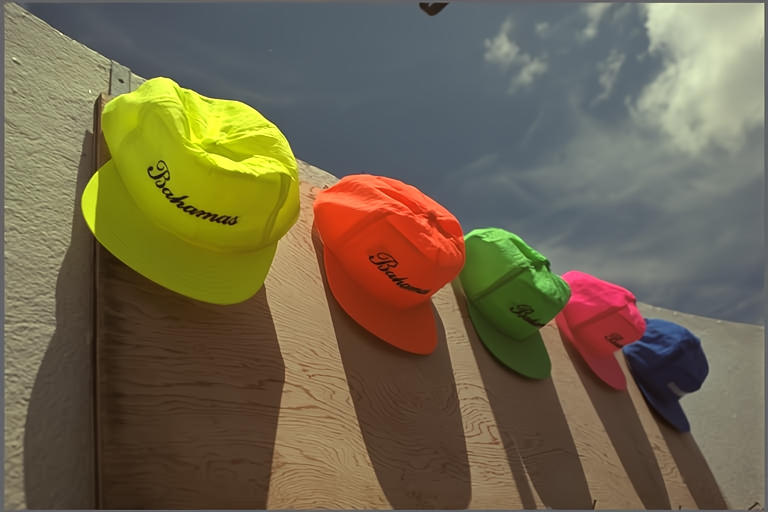}%
      \label{fig7:subfig_8}
    }
  \end{minipage}%
  \hspace{-0.20cm}
  \begin{minipage}{0.192\textwidth}
    \centering
    \subfloat[Ground Truth]{
      \includegraphics[width=\textwidth]{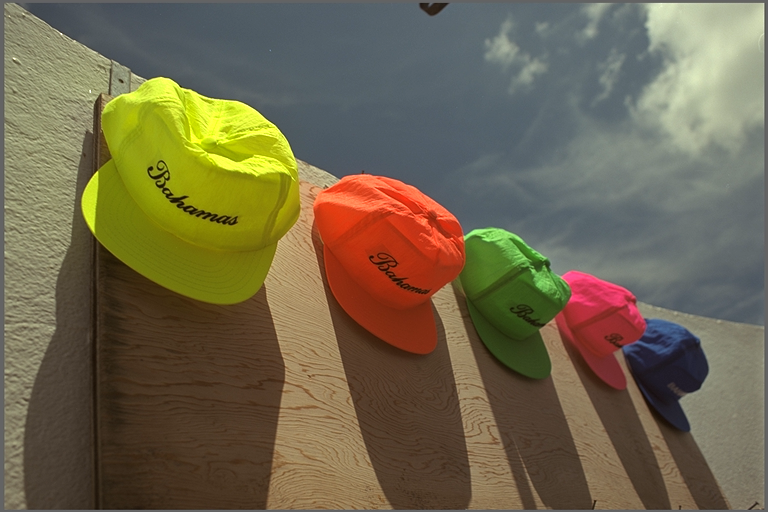}%
      \label{fig7:subfig_9}
    }
  \end{minipage}

  \vspace{-6pt}

  \begin{minipage}{0.8\textwidth}
    \subfloat[QF=10 Compressed]{
      \includegraphics[width=0.24\textwidth]{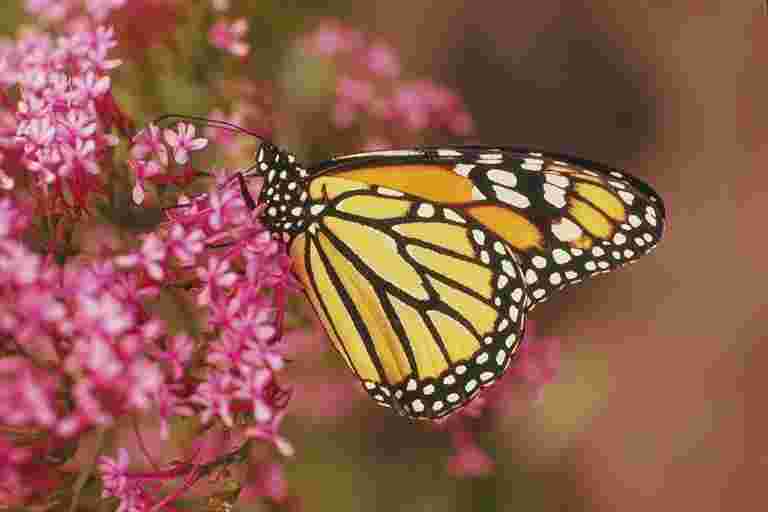}
      \label{fig:subfig_721}
    }
    \hspace{-0.15cm}
    \subfloat[QF=20 Compressed]{
      \includegraphics[width=0.24\textwidth]{compressed_color_qf10_monarch.jpg}%
      \label{fig:subfig_722}
    }
    \hspace{-0.15cm}
    \subfloat[QF=30 Compressed]{
      \includegraphics[width=0.24\textwidth]{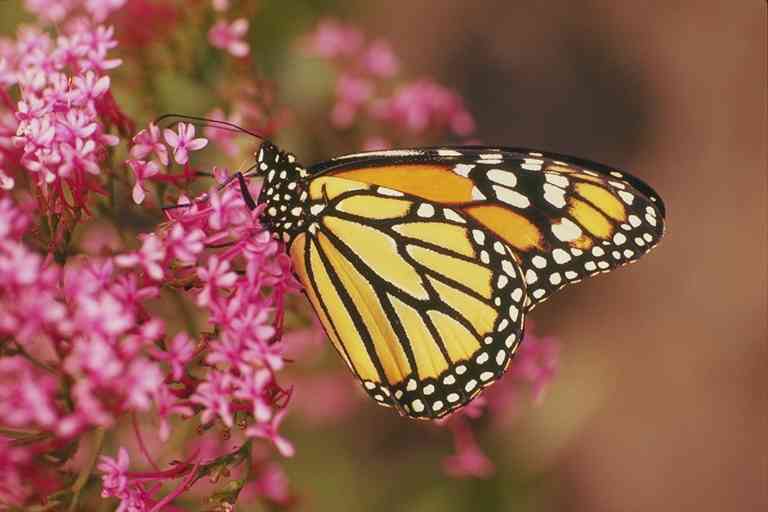}%
      \label{fig:subfig_723}
    }
    \hspace{-0.15cm}
    \subfloat[QF=40 Compressed]{
      \includegraphics[width=0.24\textwidth]{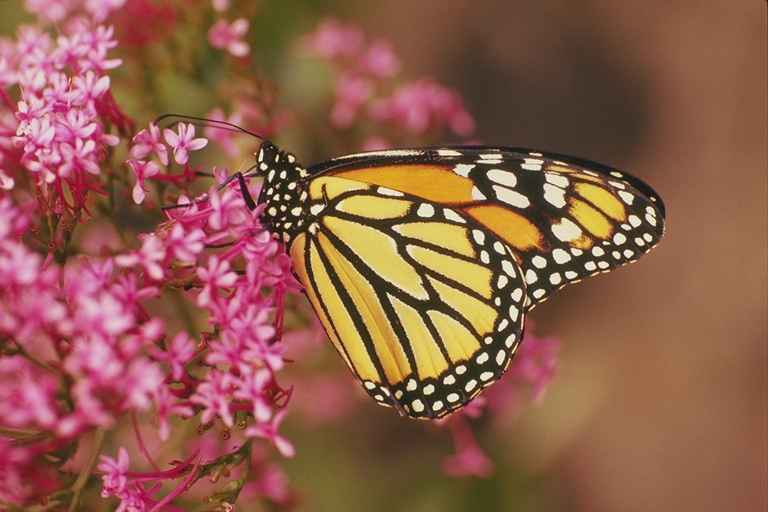}%
      \label{fig:subfig_724}
    }

    \vspace{-8pt}

    \subfloat[QF=10 Recovered]{
      \includegraphics[width=0.24\textwidth]{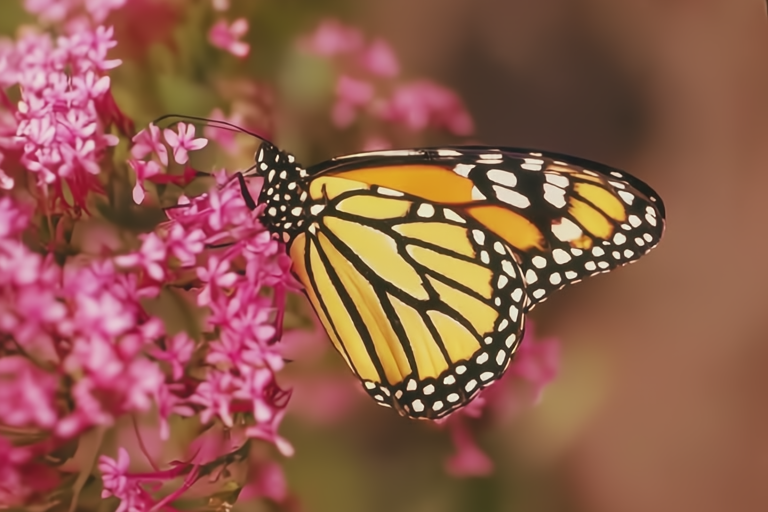}%
      \label{fig7:subfig_725}
    }
    \hspace{-0.15cm}
    \subfloat[QF=20 Recovered]{
      \includegraphics[width=0.24\textwidth]{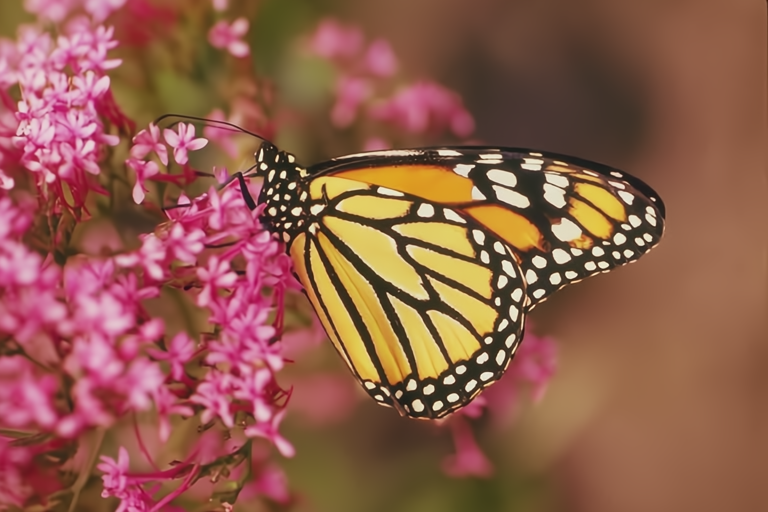}%
      \label{fig7:subfig_726}
    }
    \hspace{-0.15cm}
    \subfloat[QF=30 Recovered]{
      \includegraphics[width=0.24\textwidth]{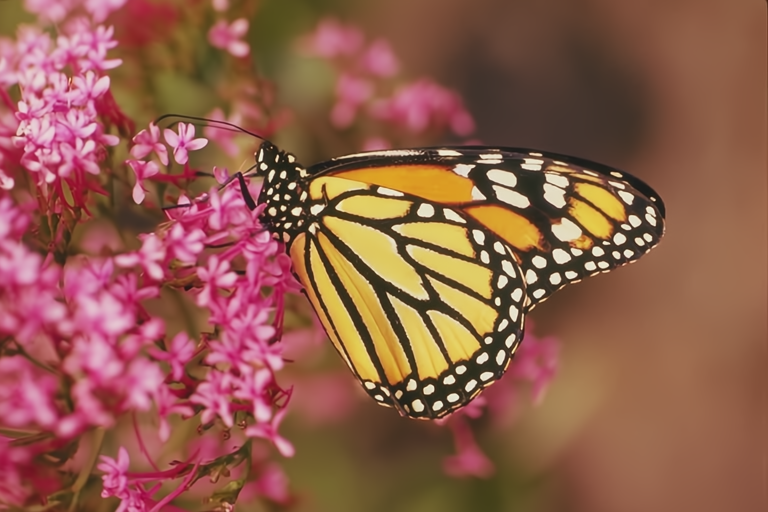}%
      \label{fig7:subfig_727}
    }
    \hspace{-0.15cm}
    \subfloat[QF=40 Recovered]{
      \includegraphics[width=0.24\textwidth]{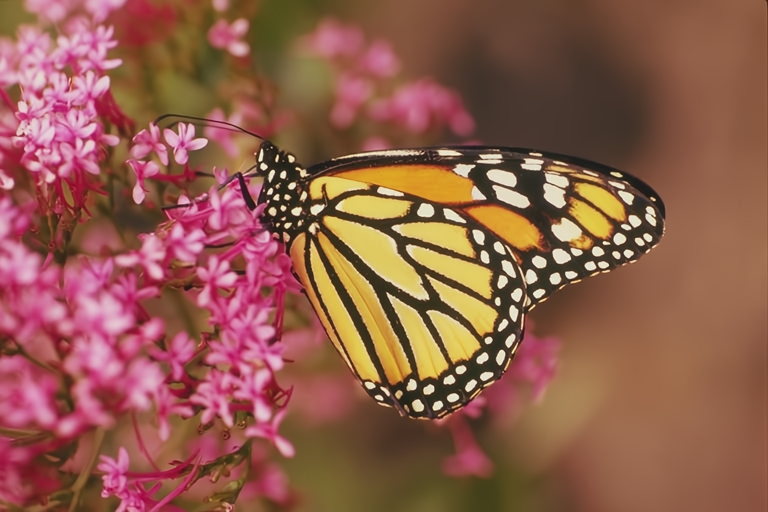}%
      \label{fig7:subfig_728}
    }
  \end{minipage}%
  \hspace{-0.20cm}
  \begin{minipage}{0.192\textwidth}
    \centering
    \subfloat[Ground Truth]{
      \includegraphics[width=\textwidth]{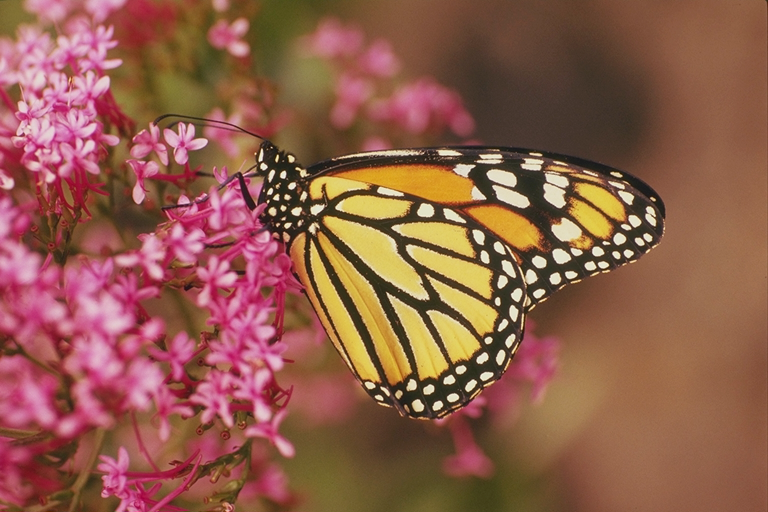}%
      \label{fig7:subfig_729}
    }
  \end{minipage}
  
  \caption{Examples of our single DCTransformer recover the compressed JPEG images of "LIVE1: caps.bmp" and "LIVE1: monarch.bmp" at different quality factors. Irrespective of the differences in input quality factors, all recovery results show consistently richer details, which demonstrates the generalization capabilities of our model to cover a wide range of quality factors.}
  \label{fig_7}
\end{figure*}

{\textit{2) Performance on grayscale JPEG quantized coefficient recovery.}}

We also trained our model with the acceptance of only Y coefficients for grayscale JPEG recovery. The luminance-chrominance alignment head is replaced by convolutional layers in our DCTransformer for grayscale experiments, with other settings remaining unchanged. Table \ref{tab_2} reports the PSNR, SSIM, and PSNR-B results on the Classic-5, LIVE1, and BSDS500. Although our model is designed for recovering both the luminance and chrominance components, we can also match the previous state-of-the-art grayscale correlation models, and consistently provides high PSNR-B results demonstrating our strength in removing blockiness.

{\textit{3) Generalization capabilities at a wide range of quality factors.}} 

Since the quality factor is flexible but unknown for JPEG compressed images, generalizing a robust method in the recovery can avoid assembling quality-specific trained models. As one of the main superiority of our approach, the generalization capabilities enable our model to handle a wide range of compression quality factors, thus allowing our model to be more practical.

To present this feature, we first show the qualitative results of two recovered images from the LIVE1 dataset in Figure \ref{fig_7}. The first row of each presented image is JPEG compressed at quality factors from 10 to 40, with a step size of 10, and the second row is the recovered images via our model. As can be seen in Figure \ref{fig_7}, while the significance of artifacts varies from the quality factors (e.g. see the sky region of "caps" and the background of "monarch"), our DCTransformer generates visually pleasing results with finer visual details.

\begin{figure}[!b]
\centering
\includegraphics[width=3.5in]{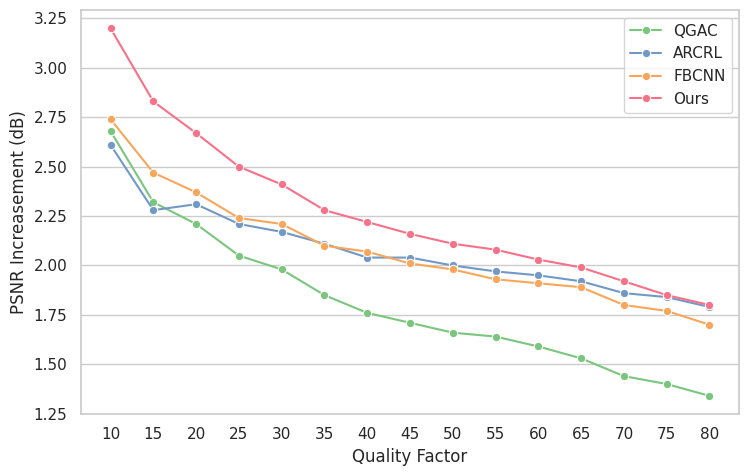}
\caption{
Generalization capabilities of our DCTransformer at quality factors from 10 to 80. Increase in PSNR of four methods on color benchmark ICB is reported. At high quality factors, less information loss during the compression leads to a drop in the PSNR increase.}
\label{fig_8}
\end{figure}

To further distinguish our model from single-quality specified methods, we also explore its generalization capabilities and compare it with previous methods at a wide range of quality factors. Images from color 8-bit ICB are used for the evaluation, with quality factors set from 10 to 80 and a step size of 5. As the result shown in Figure \ref{fig_8}, our single model maintains great performance consistently on a wide range of quality factors and shows better results over previous methods. We attribute it to the operation of quantization matrix embedding, since we do not extract features from the quantization matrix directly, but recover the original representation of coefficients in discrete cosine transformed value range by using a linear embedding. It can also be observed from the result in Figure \ref{fig_8} that the increase in PSNR is generally high at low quality factors, and declines steeply as the quality factor reaches a higher value.

\textit{4) Performance on double JPEG compression and real-world scenarios.}

Due to the widespread application of sharing and storage optimization, real-world JPEG images often undergo multiple repetitive encoding-decoding processes, and therefore result in more complex compression artifacts. To further demonstrate the generalizability of our model, we consider two types of double compression: aligned and non-aligned double JPEG compression. Specifically, in the aligned scenario, we perform two successive compressions with different quality factors. For the non-aligned scenario, after the first compression, the image is shifted towards the bottom right by (4, 4) pixels before undergoing a second compression. To formulate this degradation, the double JPEG compressed image $I_{deg}$ with a $shift$ of $(dx, dy)$ can be synthesized from high quality $I_{gt}$ via:
\begin{equation}
I_{deg} \!=\! \text{JPEG}(shift(\text{JPEG}(I_{gt}, QF1), (dx, dy)), QF2).
\end{equation}
Furthermore, we conduct experiments on 4 pairs of quality factors (donated as (QF1,QF2) respectively) in each scenario for a more general comparison. In particular, QF = (10,90), (90,10) simulate the case of a large gap in two quality factors, and the settings of QF = (75,95), (95,75) simulate the common compression ratios in real-world applications.

\begin{figure*}[!ht]
\centering
  \vspace{-0.5cm}
  \subfloat[Ground Truth]{
  \includegraphics[width=0.190\textwidth]{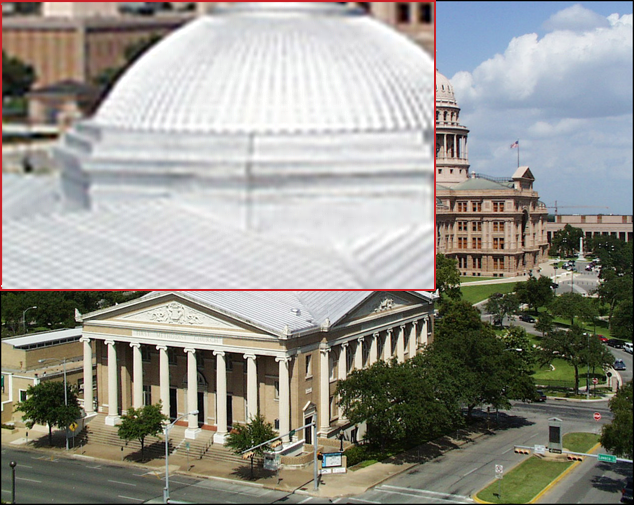}%
    \label{fig5:subfig_0}
  }
  \hspace{0.05cm}
  \subfloat[Ground Truth]{
    \includegraphics[width=0.190\textwidth]{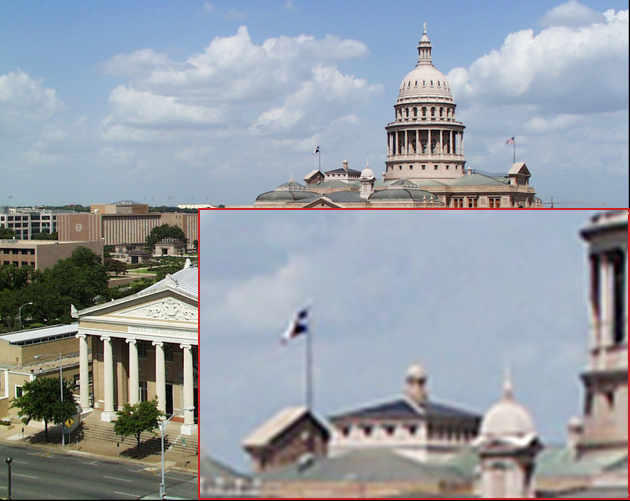}%
    \label{fig9:subfig_c}
  } 
  \hspace{0.05cm}
  \subfloat[QF=(75, 95) JPEG]{
    \includegraphics[width=0.190\textwidth]{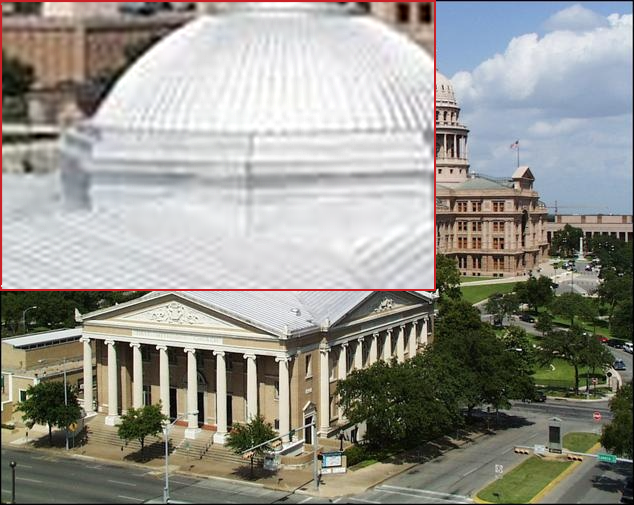}%
    \label{fig9:subfig_a}
  }
    \hspace{0.05cm}
  \subfloat[QF=(75, 95) JPEG$^*$]{
    \includegraphics[width=0.190\textwidth]{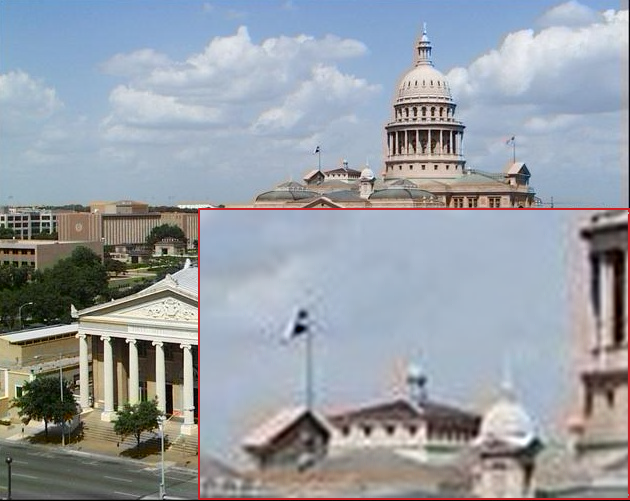}%
    \label{fig9:subfig_b}
  }
  \\  
  \vspace{-8pt}
    \subfloat[DNCNN]{
    \includegraphics[width=0.190\textwidth]{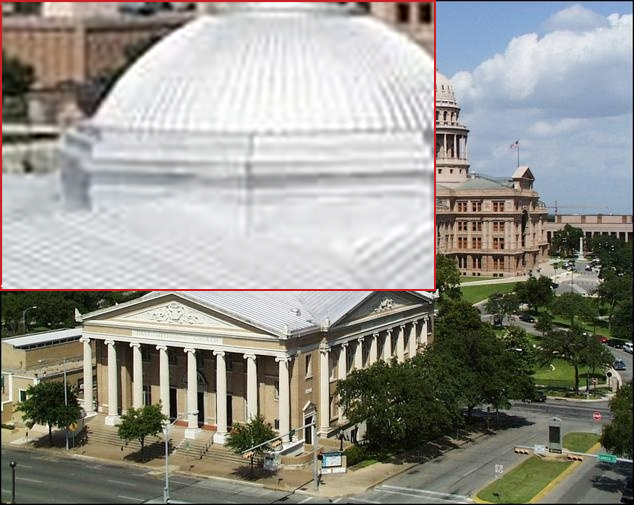}%
    \label{fig9:subfig_d}
  }
    \hspace{0.00cm}
  \subfloat[QGAC]{
    \includegraphics[width=0.190\textwidth]{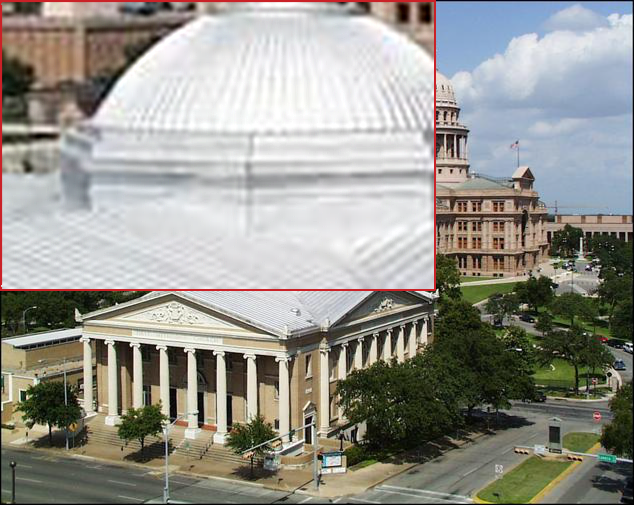}%
    \label{fig9:subfig_e}
  }
    \hspace{0.00cm}
  \subfloat[FBCNN]{
    \includegraphics[width=0.190\textwidth]{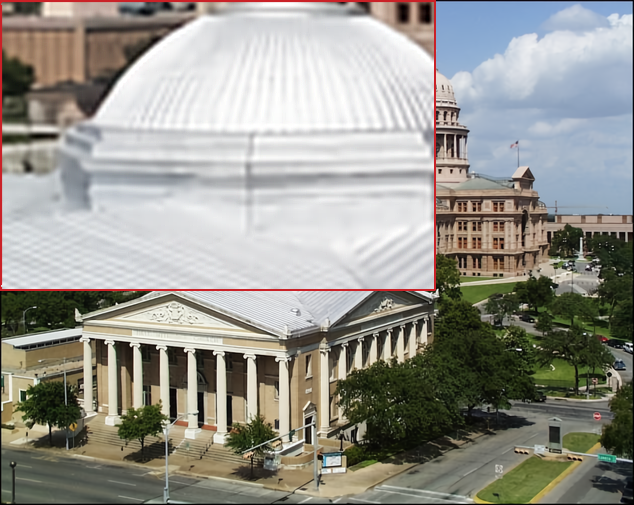}%
    \label{fig9:subfig_f}
  }
  \hspace{0.00cm}
  \subfloat[DCTransformer]{
    \includegraphics[width=0.190\textwidth]{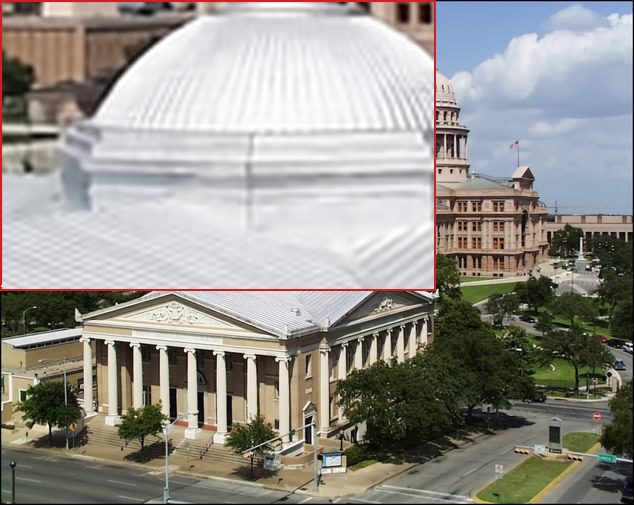}%
    \label{fig9:subfig_g}
  }
    \hspace{0.00cm}
  \subfloat[DCTransformer-A]{
    \includegraphics[width=0.190\textwidth]{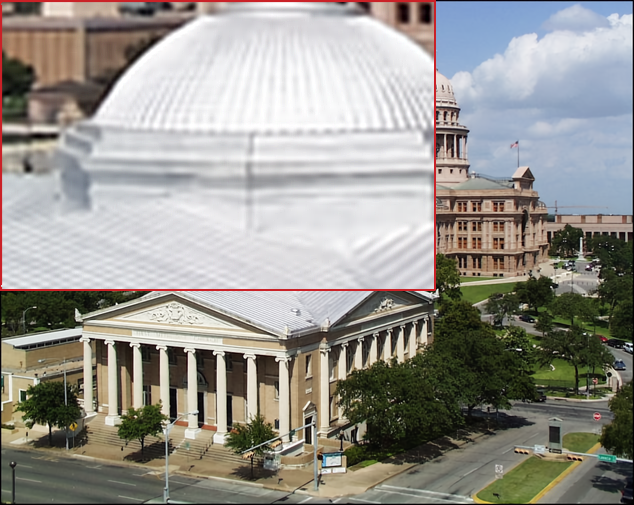}%
    \label{fig9:subfig_h}
  }
  \\ 
    \vspace{-8pt}
    \subfloat[DNCNN$^*$]{
    \includegraphics[width=0.190\textwidth]{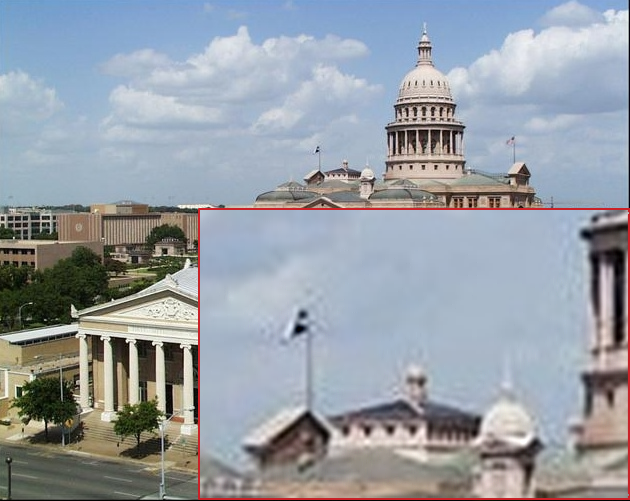}%
    \label{fig9:subfig_i}
  }
    \hspace{0.00cm}
  \subfloat[QGAC$^*$]{
    \includegraphics[width=0.190\textwidth]{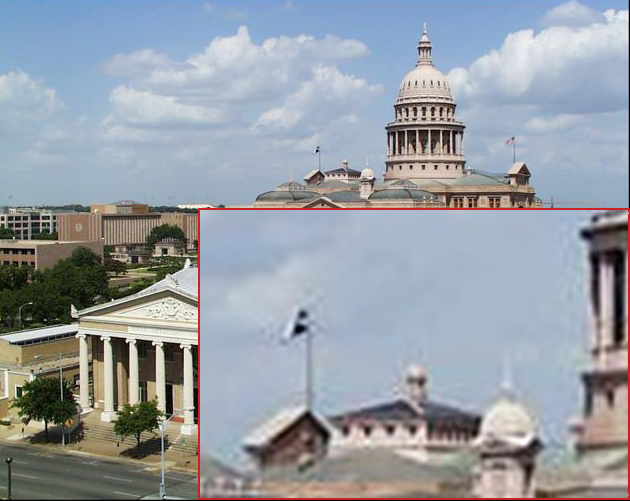}%
    \label{fig9:subfig_j}
  }
    \hspace{0.00cm}
  \subfloat[FBCNN$^*$]{
    \includegraphics[width=0.190\textwidth]{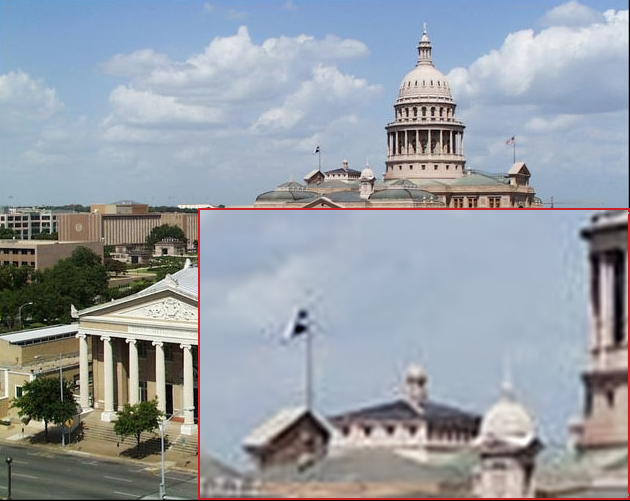}%
    \label{fig9:subfig_k}
  }
  \hspace{0.00cm}
  \subfloat[DCTransformer$^*$]{
    \includegraphics[width=0.190\textwidth]{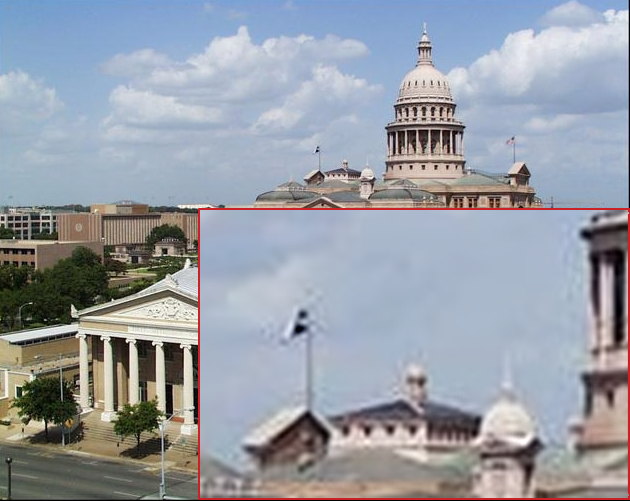}%
    \label{fig9:subfig_l}
  }
    \hspace{0.00cm}
  \subfloat[DCTransformer-A$^*$]{
    \includegraphics[width=0.190\textwidth]{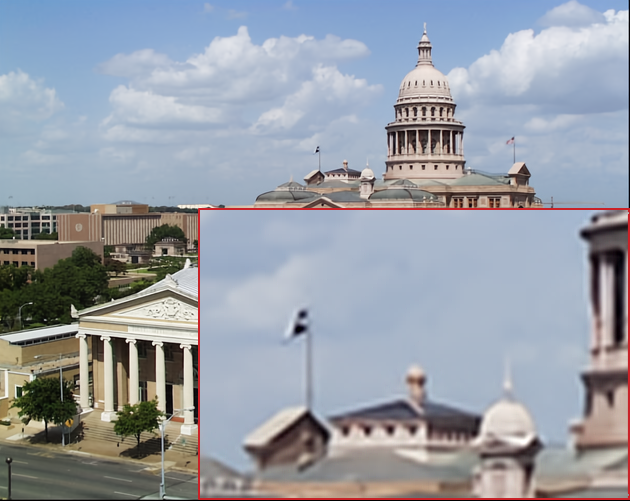}%
    \label{fig9:subfig_m}
  }
  \caption{
    Visual examples of aligned/non-aligned double JPEG compressions and corresponding recovery results. Note that QF=(QF1, QF2) represents the image is firstly compressed with QF1 and then compressed with QF2, and marker $^*$ donates the non-aligned compression with a (4,4) pixel shift in between. The restored images of different methods are shown in the second and third rows. Our DCTransformer restores rich texture details (the rooftop, second row) and alleviates more artifacts (skies around the flag, third row). Please zoom in for the best view.}
  \label{fig_9}
\end{figure*}

The quantitative results are shown in Table \ref{tab_doublejpeg}. Following previous literature setups \cite{FBCNN, 10080951}, we obtained the augmented model DCTransformer-A by fine-tuning our DCTransformer via a randomly double compression training pipeline. As shown in Table \ref{tab_doublejpeg}, although all the methods show a significant performance drop, our DCTransformer remains relatively effective in recovering the lost coefficients and outperforms the other methods in most cases. More importantly, with further fine-tuning, our augmented model can be extended to handle those complex degradation scenarios, as shown by superior metrics achieved regardless of the compression order. We can also observe from Table \ref{tab_doublejpeg} that for QF=(10,90), (75,95) in the non-align scenario, which is QF1<QF2, the recovery performance of the general model is very limited. Our fine-tuned DCTransformer-A is able to handle this complex degradation and significantly outperforming the existing methods on evaluation metrics. Compared to the second best method, the PSNR of our model is +1.99dB and +1.19dB higher in these two cases. This demonstrates the robustness of our proposed model and the possibility of further extensions. For the visual comparison, we illustrate this generalizability by presenting the results of double compressed JPEG and recovered images in Figure \ref{fig_9}. We can see that the proposed DCTransformer and DCTransformer-A obtain better subjective quality with significantly suppressed artifacts and richer texture details. 


\begin{table*}[ht]
\caption{
Pixel domain evaluation on \textbf{Color} double compressed JPEG images on the LIVE1 dataset. Represented in \textbf{PSNR(dB)/SSIM/PSNR-B(dB)} format. The best and second best performances are \textbf{Boldfaced} and \underline{Underlined}, respectively. Note that the marker $ ^*$ represents DCT domain methods.}
 
\centering
\setlength{\tabcolsep}{0.75pt} 
\begin{tabular}{cccccccc}
\toprule
\multicolumn{1}{c|}{Type} & \multicolumn{1}{c|}{QF(1,2)} & \multicolumn{1}{c|}{JPEG} & \multicolumn{1}{c|}{$\text{DnCNN-3}$\cite{DNCNN}} & \multicolumn{1}{c||}{FBCNN\cite{FBCNN}} & \multicolumn{1}{c|}{QGAC$ ^*$\cite{QGAC}} & \multicolumn{1}{c|}{DCTransformer$^*$} & \multicolumn{1}{c}{DCTransformer-A$^*$} \\ 
\midrule
\midrule
\multicolumn{1}{c}{\multirow{4}{*}{Aligned}} &
  \multicolumn{1}{|c|}{(10,90)} &
  \multicolumn{1}{c|}{25.75/0.745/24.25} &
  \multicolumn{1}{c|}{25.80/0.747/24.33} &
  \multicolumn{1}{c||}{27.77/0.802/27.51} &
  \multicolumn{1}{c|}{25.76/0.747/24.31} &
  \multicolumn{1}{c|}{\underline{27.95}/\underline{0.807}/\underline{27.69}} &
  \multicolumn{1}{c}{\textbf{28.00}/\textbf{0.808}/\textbf{27.75}}
   \\
 &
  \multicolumn{1}{|c|}{(90,10)} &
  \multicolumn{1}{c|}{25.67/0.743/24.18} &
  \multicolumn{1}{c|}{25.73/0.745/24.27} &
  \multicolumn{1}{c||}{27.75/0.803/27.50} &
  \multicolumn{1}{c|}{27.62/0.804/27.42} &
  \multicolumn{1}{c|}{\underline{28.02}/\underline{0.807}/\underline{27.77}} &
  \multicolumn{1}{c }{\textbf{28.03}/\textbf{0.808}/\textbf{27.80}}
   \\
 &
  \multicolumn{1}{|c|}{(75,95)} &
  \multicolumn{1}{c|}{33.25/0.931/32.30} &
  \multicolumn{1}{c|}{33.40/0.933/32.54} &
  \multicolumn{1}{c||}{35.18/0.949/34.57} &
  \multicolumn{1}{c|}{33.34/0.933/32.48} &
  \multicolumn{1}{c|}{\underline{35.23}/\underline{0.951}/\underline{34.68}} &
  \multicolumn{1}{c}{\textbf{35.28}/\textbf{0.951}/\textbf{34.75}}
   \\
 &
  \multicolumn{1}{|c|}{(95,75)} &
  \multicolumn{1}{c|}{33.18/0.930/32.26} &
  \multicolumn{1}{c|}{33.34/0.932/32.51} &
  \multicolumn{1}{c||}{35.15/0.949/34.54} &
  \multicolumn{1}{c|}{34.81/0.948/34.27} &
  \multicolumn{1}{c|}{\underline{35.29}/\textbf{0.951}/\underline{34.74}} &
  \multicolumn{1}{c }{\textbf{35.32}/\underline{0.951}/\textbf{34.79}}  %
   \\ \midrule
   &
  \multicolumn{1}{|c|}{(10,90)} &
  \multicolumn{1}{c|}{25.70/0.742/25.70} &
  \multicolumn{1}{c|}{25.74/0.744/25.74} &
  \multicolumn{1}{c||}{25.81/0.750/25.80} &
  \multicolumn{1}{c|}{\underline{25.88}/\underline{0.753}/\underline{25.87}} &
  \multicolumn{1}{c|}{25.74/0.747/25.74} &
  \multicolumn{1}{c}{\textbf{27.87}/\textbf{0.803}/\textbf{27.82}}
   \\
  \multicolumn{1}{c}{Non-}  &
  \multicolumn{1}{|c|}{(90,10)} &
  \multicolumn{1}{c|}{25.70/0.744/24.14} &
  \multicolumn{1}{c|}{25.76/0.747/24.23} &
  \multicolumn{1}{c||}{27.76/0.802/27.45} &
  \multicolumn{1}{c|}{27.64/0.803/27.39} &
  \multicolumn{1}{c|}{\underline{28.03}/\underline{0.807}/\underline{27.71}} &
  \multicolumn{1}{c}{\textbf{28.04}/\textbf{0.807}/\textbf{27.74}}
   \\
  \multicolumn{1}{c}{Aligned} &
  \multicolumn{1}{|c|}{(75,95)} &
  \multicolumn{1}{c|}{33.04/0.927/32.91} &
  \multicolumn{1}{c|}{33.20/0.929/33.06} &
  \multicolumn{1}{c||}{\underline{33.91}/\underline{0.938}/\underline{33.74}} &
  \multicolumn{1}{c|}{33.14/0.930/32.87} &
  \multicolumn{1}{c|}{33.54/0.935/33.32} &
  \multicolumn{1}{c}{\textbf{35.10}/\textbf{0.949}/\textbf{34.83}}
   \\
 &
  \multicolumn{1}{|c|}{(95,75)} &
  \multicolumn{1}{c|}{33.09/0.929/31.92} &
  \multicolumn{1}{c|}{33.25/0.931/32.16} &
  \multicolumn{1}{c||}{35.10/0.949/34.15} &
  \multicolumn{1}{c|}{34.77/0.947/33.90} &
  \multicolumn{1}{c|}{\underline{35.20}/\underline{0.950}/\underline{34.30}} &
  \multicolumn{1}{c}{\textbf{35.27}/\textbf{0.950}/\textbf{34.41}}
   \\ 
\bottomrule
\label{tab_doublejpeg}
\end{tabular}
\end{table*}

\subsection{DCT Domain Evaluation}

To analyze the performance of our model in the DCT domain, we adopt Jensen-Shannon divergence and Bhattacharyya distance as frequency quantitative evaluation metrics. These comparisons are conducted among DCT domain methods since we directly compute the metrics from coefficients. For each frequency component $c$ of the recovered DCT coefficients, we first compute the histogram of the coefficients into $n$ bins, then normalize it to form a probability distribution $P_{c}(i) = \left\{ p_{c}(1),~p_{c}(2),~\dots,~p_{c}(n) \right\}$ and $Q_{c}(i) = \left\{ q_{c}(1),~q_{c}(2),~\dots,~q_{c}(n) \right\}$ to compute the metrics. All the indicators are calculated on 64 frequencies separately and averaged as the distance between two DCT coefficient pairs. Given the DCT coefficient pairs $X$ and $Y$, the Jensen-Shannon divergence between their sampled probability distributions P and Q is defined as:

\begin{equation}
    D_{JS}\!\left(\! X,Y \!\right)\! = \frac{1}{64} \sum_{c=1}^{64} \frac{D_{KL}(P_c \| M_c) \! + \! D_{KL}(Q_c \| M_c)}{2} ,
\end{equation}

where $D_{KL}$ represents the Kullback-Leibler divergence referring to \cite{kullback1951information}, $M_c$ is the average distribution of $P_c$ and $Q_c$, and is defined as $M_c = \frac{1}{2}(P_c + Q_c)$. 

The Bhattacharyya distance between two coefficients can be formulated as follows:
\begin{equation}
    D_{B}\left( X,Y \right) = \frac{1}{64} \sum_{c=1}^{64} \left( -\log\left(\sum_{i=1}^{n} \sqrt{P_{c}(i) Q_{c}(i)}\right) \right),
\end{equation}
where $c$ goes from 1 to 64, donating the frequency index; the $P_c(i)$ and $Q_c(i)$ are the probability distribution of the coefficients $X$ and $Y$, respectively. 


Table \ref{tab_dct_eval} reports the DCT domain evaluation result on the grayscale Classic-5 dataset at quality factors from 10 to 50. The smaller value indicates a more similar distribution to the lossless DCT coefficients and, correspondingly, higher-quality images can be reconstructed from these coefficients. Compared to QGAC, the state-of-the-art frequency domain method, our results have a significantly smaller JS divergence and Bhattacharyya distance. The results indicate that our method achieves a better performance in recovering the quantized coefficients from the DCT domain perspective.

\begin{table}[!hb]
\caption{DCT Domain Evaluation of \textbf{Jensen-Shannon Divergence} and \textbf{Bhattacharyya Distance} on Grayscale Classic-5 Dataset. The Best Results are \textbf{Boldfaced}.}
\centering
\setlength{\tabcolsep}{1pt} 
\begin{tabular}{c| cc | cc | cc | cc}
\toprule
\multirow{2}{*}{QF} & \multicolumn{2}{c|}{JPEG} & \multicolumn{2}{c|}{FD-CRNet} & \multicolumn{2}{c|}{QGAC} & \multicolumn{2}{c}{Ours} \\
                   & JS $\downarrow$ &  Bha $\downarrow$ & JS $\downarrow$ & Bha $\downarrow$ & JS $\downarrow$ & Bha $\downarrow$ & JS $\downarrow$ & Bha $\downarrow$\\
\midrule
10                 &  0.2223  &  0.3589  & 0.0124  & 0.0134  &  0.0055  &  0.0060  &  \textbf{0.0044}  &  \textbf{0.0048} \\
20                 &  0.2022  &  0.3228  & 0.0095  & 0.0101  &  0.0034  &  0.0036  &  \textbf{0.0022}  &  \textbf{0.0024} \\
30                 &  0.1854  &  0.2900  & 0.0093  & 0.0099  &  0.0024  &  0.0025  &  \textbf{0.0017}  &  \textbf{0.0018} \\
40                 &  0.1756  &  0.2713  & 0.0091  & 0.0096  &  0.0018  &  0.0019  &  \textbf{0.0013}  &  \textbf{0.0013} \\
50     	           &  0.1675  &  0.2577  & 0.0087  & 0.0093  &  0.0013  &  0.0014  &  \textbf{0.0009}  &  \textbf{0.0009} \\
\bottomrule 
\end{tabular}
\label{tab_dct_eval}
\end{table}

\subsection{Comparison of Number of Parameters, Runtime, and Maximum GPU Memory Consumption}

Table \ref{tab_runtime} reports the number of parameters, maximum GPU memory consumption, and runtime comparison with three state-of-the-art methods (\textit{i.e.}, FBCNN, ARCRL, QGAC) in both pixel and DCT domains. The experiment is conducted on a single V100 GPU. All the tested models performed inference on 100 color images from the Urban100 dataset with QF=10, and averaged as the reported time. The size of images in the Urban100 dataset is around (1024, 768), which matches the real-world application scenarios.

\begin{table}[!h]
\centering
\caption{Comparisons of the \textbf{Number of Parameters}, \textbf{Maximum GPU Memory consumption}, and \textbf{Inference Time} on the Urban100 dataset. The average inference time per image is reported.}
\setlength{\tabcolsep}{2pt} 
\begin{tabular}{c|cccc}
\toprule
Experiment &  Model   & Number of   & Max GPU    &  Inference \\
Model      &  Domain  &  Param [M]  & Memory [M] &  Time [ms] \\
\midrule
FBCNN \cite{FBCNN} & Pixel & 71.922 & 1542.89 & 217.88 \\
\midrule
ARCRL \cite{wang2022jpeg} & Pixel & \textbf{13.884} & 7617.18 & 972.02 \\ 
\midrule
QGAC \cite{QGAC} & DCT & 259.409 & 4108.32 & 2074.23 \\
\midrule
DCTransformer & DCT & 21.101 & \textbf{600.41} & \textbf{125.49} \\
\bottomrule
\end{tabular}
\label{tab_runtime}
\end{table}

As shown in Table \ref{tab_runtime}, our DCTransformer demonstrates superior efficiency in terms of the max memory and inference time compared with previous state-of-the-art methods, and ranks the second best in terms of the number of parameters. In particular, although our approach has more parameters than ARCRL, it still has a significant advantage in memory consumption (7\% of theirs, 600M vs. 7617M) and inference speed (13\% of theirs, 125ms vs. 972ms). Note that excessive memory consumption can make high-resolution image inference unaffordable, e.g., FBCNN, ARCRL, and QGAC require $\geq$40GB of GPU memory for inference on the ICB benchmark. In contrast, the GPU memory consumption of our model remains almost unchanged because of the fixed sequence length. Besides, compared to our most similar work, QGAC, which also operates in the DCT domain, the number of parameters of our model (21M) is significantly less than theirs (259M). This comparison underlines the efficiency of the proposed DCTransformer and its advantage as a strong and memory-efficient model for quantized coefficient recovery.

\subsection{Ablation Studies}

We test a set of architectures of our model to investigate the contribution of each component. Our ablation models are trained exclusively on the DIV2k dataset ($\approx$ 20\% of the training set) and for 50K iterations ($\approx$ 10\% of the full training). We primarily investigate our model based on four aspects as below while ensuring that the number of parameters remains approximately the same for fair comparisons. The results of the ablation analysis are presented in Table \ref{tab_ablation}.

\textit{a) The effectiveness of dual-branch architecture.} To verify whether our SFTB module models the correlations between coefficients effectively, we explore three variations of the proposed architecture. In the comparisons, parallel spatial branches and parallel frequential branches refer to two spatial or frequential breaches in parallel, replacing the original spatial-frequential branches. Successive branches imply the spatial and frequential branches are cascaded in succession. As shown in Table \ref{tab_ablation}, the effectiveness of the dual-branch design is substantiated by the performance of our full-setting model.

\textit{b) The effectiveness of feature fusion methods.} For the comparison of fusion methods, we utilize the adding operation between the output from two branches to replace the channel-wise feature concatenation. We also explore the contribution brought by the convolutional layer after the concatenation. The results are shown as “Add Fusion + Conv.” and "Concat Fusion + No Conv." in Table \ref{tab_ablation}. We can find that feature concatenation considerably improves the performance than adding operation, which fails to learn. It is worth noting that although only a small number of parameters are reduced, the performance on PSNR drops significantly by up to 0.08 when the convolutional layer after the fusion operation is removed.

\begin{table}[!t]
\caption{Ablation Analysis on Color LIVE1 Dataset. \textbf{PSNR(dB)} and \textbf{SSIM} is Reported and the Best Results are \textbf{Boldfaced}. Please Refer to Section IV-E for Detailed Settings of Each Model.\label{tab:table_ablation}}
\centering
\setlength{\tabcolsep}{1pt} 
\begin{tabular}{l|c|cc|cc}
\toprule
\multirow{2}{*}{Ablation Method} & Params & \multicolumn{2}{c|}{QF=20} & \multicolumn{2}{c}{QF=40} \\
                            & [M]     &  PSNR    & SSIM        & PSNR        & SSIM        \\ 
\midrule
Parallel Spatial Branch     & 21.02  &  29.77         & \textbf{0.864} & 31.90          & 0.908          \\
Parallel Frequential Branch & 21.17  &  29.75         & 0.863          & 31.88          & 0.907          \\
Successive Branch           & 19.87  &  29.58         & 0.856          & 31.65          & 0.900          \\ 
\midrule
Add Fusion + Conv.            & 21.10  &  29.73         & 0.861          & 30.85          & 0.905          \\  
Concat Fusion + No Conv.      & 19.87  &  29.75         & 0.862          & 31.89          & 0.907          \\ 
\midrule
W/o Quantization Matrix   & 21.10  & 22.12          & 0.620           & 24.37          & 0.683 \\
Concat Quantization Matrix    & 21.10  & 29.40          & 0.849           & 31.46          & 0.898 \\
\midrule
Standard Pixel Loss           & 21.10  &  29.76         & 0.863          & 31.86          & 0.907          \\
Standard Frequency Loss       & 21.10  &  29.72         & 0.862          & 31.82          & 0.906          \\ 
\midrule
Full-Setting DCTransformer            & 21.10  & \textbf{29.80} & \textbf{0.864} & \textbf{31.94} & \textbf{0.910} \\ 
\bottomrule
\end{tabular}
\label{tab_ablation}
\end{table}

\textit{c) The effectiveness of quantization matrix embedding.} We perform an ablation study on how the quantization matrix embedding contributes to our modeling. We first remove the quantization matrix from our network, which finally leads the model to fail in recovery at any quality factors. Concatenate the quantization matrix with input coefficients is also tested. We suggest the proposed embedding helps handle different compression qualities, as the results reported in Table \ref{tab_ablation}.

\textit{d) The effectiveness of loss functions.} We adopt a common loss function in different domains to validate the effectiveness of our dual-domain loss. L1 is one of the most widely used losses in vision tasks, and we adopt it as the standard loss to train the same model for comparisons. We implement the L1 loss in pixel and frequency domains to compute the loss within image pairs and DCT coefficient pairs, respectively. As shown in Table \ref{tab_ablation}, the model trained with dual-domain loss (as "Full-setting DCTransformer") attains better performance than implementing any of the single-domain losses.

\section{CONCLUSION}

In this work, we propose a novel DCT domain spatial-frequential Transformer (DCTransformer) for JPEG quantized coefficient recovery. Incorporating the quantization matrix embedding and luminance-chrominance alignment module, our proposed DCTransformer has proven robustness in handling a wide range of quality factors and recovering the compressed luminance and chrominance components. As fully operating in the DCT domain, our single model is comparable to the state-of-the-art JPEG artifact removal methods on both pixel and frequency domain metrics. Extensive experiments also demonstrate the generalizability and effectiveness of our design. We suggest our model demonstrates the potential of DCT domain learning for JPEG artifact removal, which can possibly broaden the scope of image processing and computer vision fields.

\bibliography{ms}

\end{document}